\documentclass[]{aastex631}
\usepackage{tabularx,rotating}
\usepackage{xcolor}
\usepackage{tablefootnote}
\usepackage{ulem} 
\usepackage{soul}
\usepackage{cancel}
\usepackage{wasysym} 
\usepackage{graphicx}

\newcommand{\sw}{{\it Swift}}
\newcommand{\cha}{{\it Chandra}}
\newcommand{\xmm}{XMM-{\it Newton}}
\newcommand{\ergss}{~ergs~s$^{-1}$}
\newcommand{\kms}{~km~s$^{-1}$}

\newcommand{\msun}{{M$_{\odot}$}}

\newcommand{\am}{Paper~I}

\shorttitle{A \sw~X-ray view of the SMS4 sample - II.}
\shortauthors{Maselli et al.}
\graphicspath{{./}{figures/}}

\begin{document}






\title{A \sw~X-ray view of the SMS4 sample - II: X-ray properties of 17 bright radio sources.}

\correspondingauthor{Alessandro Maselli}
\email{alessandro.maselli@inaf.it}

\author[0000-0003-3760-1910]{Alessandro Maselli} 
\affiliation{INAF-Osservatorio Astronomico di Roma \\
via Frascati 33, I-00078, Monte Porzio Catone (Roma), Italy}
\affiliation{ASI-Space Science Data Center \\
via del Politecnico snc, I-00133, Roma, Italy}

\author[0000-0002-9478-1682]{William R. Forman}
\affiliation{Harvard-Smithsonian Center for Astrophysics \\
60 Garden Street, Cambridge, MA-02138, USA}

\author[0000-0003-2206-4243]{Christine Jones}
\affiliation{Harvard-Smithsonian Center for Astrophysics \\
60 Garden Street, Cambridge, MA-02138, USA}

\author[0000-0002-0765-0511]{Ralph P. Kraft}
\affiliation{Harvard-Smithsonian Center for Astrophysics \\
60 Garden Street, Cambridge, MA-02138, USA}

\author[0000-0003-3613-4409]{Matteo Perri}
\affiliation{INAF-Osservatorio Astronomico di Roma \\
via Frascati 33, I-00078, Monte Porzio Catone (Roma), Italy}
\affiliation{ASI-Space Science Data Center \\
via del Politecnico snc, I-00133, Roma, Italy}


\begin{abstract}
Based on a proposal to observe 18 bright radio sources from the SMS4 catalog with the Neil Gehrels \sw~Observatory (hereafter \sw), we obtained X-ray observations of 17 targets (one target was not observed). 
Following up our first paper that discussed 31 sources (see Maselli et~al.~2022; 20 sources detected as point sources and one very extended source), we present results for this final sample of 17 radio sources, that previously lacked dedicated, pointed narrow FOV X-ray observations.
One of these 17 sources, undetected by \sw~due to a very short exposure, was instead detected by eROSITA, and given in the Data Release 1 (DR1) Catalog.
No 1eRASS source was found in the DR1 for the remaining source, unobserved by \sw.
The new \sw~observations led to eleven X-ray source detections in the 0.3--10~keV band and six upper limits.
We investigated the extent of the X-ray emission, the hardness ratio, and when statistics allowed we carried out a spectral analysis. 
The X-ray emission of eight sources is consistent with point-like emission, while three sources show clear evidence of extent, each with peculiar properties.  
We used the X-ray determined positions and uncertainties of the twelve detected sources to establish associations with infrared and optical sources from the AllWISE and the GSC~2.4.2 catalogs.
Requiring a detection in both the infrared and the optical bands to establish a candidate counterpart for our X-ray detections, we identify counterparts for all twelve sources.
Following this X-ray based approach to derive the position of the active nucleus, we are able to confirm the same IR counterparts previously proposed by White et~al.~(2020a,b) for eight sources, and provide four new IR candidates.
In the optical, we identify counterparts that match the candidates previously given by Burgess \& Hunstead (2006a,b) for all sources.
We discuss the interesting structure of MRC~B0344$-$345 and PKS~B2148$-$555, two of the six extended X-ray sources that we detected in both our \sw~campaigns, and suggest they are very promising for further X-ray and radio investigations.
For the 38 SMS4 sources that lack pointed, narrow FOV X-ray telescope observations, after our \sw~campaigns, we list 18 likely counterparts from the eROSITA DR1 catalog.

\end{abstract}



\section{Introduction} 
\label{sec:1}

Bright extragalactic radio sources have been among the most remarkable astronomical targets. 
For example, Cygnus~A, Virgo~A, Centaurus~A, and Perseus~A, have revolutionized our views of how outflows of matter and radiation from supermassive black holes (SMBH) impact the evolution of their host galaxies and their environments, from the first identifications to recent investigations (e.g., Cygnus A - \citealp{1954ApJ...119..206B}; Virgo A - \citealp{1954ApJ...119..215B,2001ApJ...554..261C}; Centaurus A - \citealp{1981ApJ...251..523S,1954ApJ...119..215B,1998A&ARv...8..237I}; Perseus A - \citealp{1970ApJ...159L.151L,2000A&A...356..788C,2000MNRAS.318L..65F,2006MNRAS.366..417F}).
As a next step, the first catalogs of radio sources provided an enduring set of targets with key samples for early evolutionary studies (e.g., \citealp{1984MNRAS.211..833L,1983MNRAS.204..151L,1969MNRAS.145..121L,1969MNRAS.145..309L}) as well as individual objects that have yielded new insights for many decades (e.g., \citealp{1983Natur.301..488L,1963Natur.197.1040S,Chiaberge_2015,2019ApJ...875L...1E,2016ApJS..225...12H}).

As \cite{2006AJ....131..100B} pointed out (see also \citealp{2012ApJ...745..172D} and \citealp{2023ApJS..265...32M}), there is a lack of samples in the southern hemisphere, the site of the newest ground-based telescopes, compared to the flux limited 3C sample \citep{1959MmRAS..68...37E} and its revisions (3CR, \citealp{1962MmRAS..68..163B}; 3CRR, \citealp{1983MNRAS.204..151L}) of a few hundred objects in the northern hemisphere.  
To remedy this situation, \cite{2006AJ....131..100B} built the SMS4 sample by extracting sources with flux density $S_{408}$ higher than 4~Jy from the Molonglo Reference Catalogue (MRC; \citealp{1981MNRAS.194..693L}). 
Extrapolating to 178~MHz and matching the 3CRR flux density threshold, they produced a catalog of 137 bright radio sources very similar to 3CRR.

Subsequently, based on observations from the Galactic and Extragalactic All-sky Murchison Widefield Array (MWA; \citealp{2013PASA...30....7T}) survey (GLEAM; \citealp{2015PASA...32...25W}), \cite{2020PASA...37...18W,2020PASA...37...17W} extracted from the GLEAM Extra Galactic Catalogue (EGC; \citealp{2017MNRAS.464.1146H}) a flux limited sample in the southern hemisphere, the G4Jy, including sources with actual flux density $S_{151}$ higher than 4~Jy. 
More recently, from the G4Jy catalog, \cite{2023ApJS..265...32M} extracted a subset of 264 sources, the G4Jy-3CRE (3CR Equivalent),
having a flux limit comparable to that of the 3CR sample.

To characterize bright radio-selected sources from the SMS4 in the X-rays, in 2015 we embarked on an X-ray survey \citep{2022ApJS..262...51M} using the Neil Gehrels Swift Observatory (hereafter \sw, \citealp{2004ApJ...611.1005G}) for all those objects that did not already have good quality X-ray images. 
Note that our original X-ray program probed the SMS4 radio sample to slightly lower fluxes than the 3C sample.

In Maselli et al.~(2022; hereafter \am) we reported on \sw~observations of 31 radio sources selected from the SMS4.
In the current paper, we describe an additional sample of 17 targets observed by \sw~with the X-Ray Telescope (XRT, \citealp{2005SSRv..120..165B}).
With the XRT, we detect X-ray emission from 11 sources: then, we characterize their extent and the hardness ratio, also carrying out a spectral analysis when statistics allowed. 
We give upper limits for the remaining six undetected sources, but find for one of these six a 1eRASS source in the DR1 catalog (\citealp{2024arXiv240117274M}) obtained from the first scan of eROSITA data.

As in \am, we utilize X-ray observations to provide accurate locations for the central AGN engine of our sources, that were all classified as quasars or radio galaxies in \cite{2006AJ....131..114B}. 
We searched in the AllWISE and GSC~2.4.2 catalogs for sources within the positional uncertainty of the twelve X-ray detected sources, to identify their IR/optical counterparts.
We require a detection in both the IR and the optical bands to establish a qualified counterpart, and show their position in $W1$ AllWISE and $r$ DSS2 images. 
We find an IR/optical counterpart for all the twelve X-ray detected sources, and compare these with candidates previously given in the literature, mainly by \cite{2020PASA...37...18W,2020PASA...37...17W} (hereafter W20) in the infrared and by \cite{2006AJ....131..100B,2006AJ....131..114B} (hereafter BH06) in the optical.
With our campaigns, we obtained observations of $\sim$1/3 SMS4 sources with \sw, significantly increasing the \sw~coverage (see also \citealp{2023ApJS..268...32M}) of powerful radio sources in the southern hemisphere in the last few years.

Throughout this paper we use CGS units, unless otherwise stated.
For consistency with \am, we also assume a flat cosmology with $H_0 = 72$ km s$^{-1}$ Mpc$^{-1}$, $\Omega_M = 0.26$, and $\Omega_{\Lambda} = 0.74$ \citep{2009ApJS..180..306D}.


\begin{table*}
\footnotesize
\begin{center}
\caption{\small{Correspondences between the 17 SMS4 sources in our sample and G4Jy sources.}}
\label{tab:radio}
\begin{tabular}{cccc|cccccc}
\hline
SMS4 Name     & S$_{178}$ & LAS     &    $z$                  & IAU Name           & G4Jy ID  &    l    &     b    & Flux Density   & Morphology \\ 
              &   (Jy)   & (arcsec) &                         &                    &          &  (deg)  &   (deg)  &     (Jy)       &            \\  
   (1)        &   (2)    & (3)      &    (4)                  &   (5)              &   (6)    &   (7)   &    (8)   &      (9)       &    (10)    \\   
\hline
B0103$-$453~~ & 19.0     & 140      &    (0.71)               &   J010521$-$450527  &   120    & 295.04 & $-$71.82 & 17.62$\pm$0.02 &      t     \\ 
B0202$-$765~~ & 19.0     &  20      &     0.38925             &   J020213$-$762006  &   217    & 297.55 & $-$40.04 & 14.94$\pm$0.01 &      s     \\
B0242$-$514~~ & 19.0     &  53      &     0.791$^{\dagger}$   &   J024344$-$511231  &   290    & 269.20 & $-$57.93 & 13.61$\pm$0.02 &      s     \\
B0344$-$345~~ & 18.0     & 264      &     0.0536$^{\dagger}$  &   J034631$-$342238  &   381    & 234.94 & $-$51.99 & 14.61$\pm$0.03 &      t     \\
B0427$-$366~~ & 18.0     &  14      &     1.565$^{\dagger}$   &   J042940$-$363050  &   464    & 238.77 & $-$43.32 & 14.73$\pm$0.02 &      s     \\ 
B1017$-$426~~ & 23.0     &  10      &     1.28~               &   J102003$-$425130  &   837    & 275.58 & $+$11.83 & 21.29$\pm$0.02 &      s     \\
B1526$-$423~~ & 44.0     &  50      &    (0.50)               &   J153014$-$423146  &  1262    & 331.62 & $+$11.32 & 34.61$\pm$0.05 &      s     \\
B1754$-$597~~ & 20.0     &  21      &    (0.8)                &   J175906$-$594655  &  1453    & 333.77 & $-$17.02 & 25.16$\pm$0.04 &      s     \\
B1814$-$519~~ & 24.0     &  10      &    (0.48)               &   J181806$-$515801  &  1471    & 342.35 & $-$16.24 & 23.55$\pm$0.03 &      s     \\
B1817$-$391~~ & 19.0     &  16      &    (0.91)               &   J182035$-$390925  &  1474    & 354.51 & $-$11.26 & 15.66$\pm$0.04 &      s     \\
B1817$-$640~~ & 28.0     &  31      &     0.67~               &   J182216$-$635915  &  1477    & 330.73 & $-$21.12 & 21.41$\pm$0.05 &      s     \\
B1953$-$425~~ & 33.0     &  10      &    (0.12)               &   J195715$-$422218  &  1588    & 357.49 & $-$29.57 & 15.90$\pm$0.03 &      s     \\
B2032$-$350~~ & 27.0     &  26      &    (0.56)               &   J203547$-$345403  &  1640    &   7.85 & $-$35.58 & 23.80$\pm$0.04 &      d     \\
B2041$-$604~~ & 27.0     &  32      &     1.464               &   J204520$-$601902  &  1646    & 336.12 & $-$37.32 & 20.83$\pm$0.03 &      s     \\
B2140$-$434~~ & 19.0     &  56      &     0.648$^{\dagger}$   &   J214333$-$431245  &  1717    & 357.25 & $-$49.01 & 14.70$\pm$0.03 &      d     \\
B2140$-$817~~ & 26.0     &  44      &    (0.64)               &   J214724$-$813213  &  1723    & 310.17 & $-$32.82 & 12.73$\pm$0.04 &      s     \\          
B2331$-$416~~ & 30.0     &  19      &     0.907               &   J233426$-$412520  &  1840    & 345.83 & $-$68.71 & 23.69$\pm$0.03 &      s     \\
\hline                                                                                                                   
B1827$-$360$^{\ast}$ & 33.0 &  10   &    (0.12)               &   J183059$-$360229  &  1487    & 358.29 & $-$11.78 & 30.15$\pm$0.04 &      s     \\
\hline                                                                                                                   
\end{tabular}
\end{center}
\tablecomments{The columns show (1) the name in SMS4, according to the MRC designation; (2) the extrapolated flux density $S_{178}$; (3) the largest angular size of the radio source at 843~MHz; (4) the redshift values taken from \cite{2006AJ....131..114B} and from \cite{2024ApJS..271....8G}, with photometric estimates in parentheses; (5) the International Astronomical Union (IAU) name in G4Jy, according to the GLEAM designation; (6) the G4Jy identifier; Galactic longitude (7) and latitude (8) of the G4Jy source; (9) the actual flux density $\overline{S}_{181}$ for G4Jy; (10) the radio source morphology, following W20 (s=single; d=double; t=triple).\\
$\ast$: this source, included in our original \sw~proposal, is not part of the sample discussed in this paper since, as of May 2024, it was not observed.\\ 
$\dagger$: this value comes from the spectroscopic campaign described in \cite{2024ApJS..271....8G}.} 
\end{table*}

\section{Description of our sample of SMS4 sources} 
\label{sec:2}

Our sample is based on the SMS4 sample, assembled by BH06, and was selected from the 56 SMS4 bright radio sources with no \cha, \sw, or \xmm~observations as of September 2021, that we published in \am~(see Table~8 therein).
For each of the SMS4 sources in that list, a G4Jy source was found, with an unambiguous match.
Sorting that list for $\overline{S}_{181}$, the flux density at 181~MHz derived from W20, we selected and successfully proposed with \sw~for the 18 brightest radio sources at that frequency, in the range between 12.7~Jy and 34.6~Jy.
However, as of May 2024, one source was not yet observed by \sw~ (see Section~\ref{sec:3}): thus, our sample comprises 17 SMS4 sources.

Using data from the Sydney University Molonglo Sky Survey (SUMSS, \citealp{2003MNRAS.342.1117M}) and the NRAO VLA Sky Survey (NVSS, \citealp{1998AJ....115.1693C}), at higher frequency than the GLEAM survey, a morphological classification was provided by W20 for all the G4Jy sources, distinguishing single (s), double (d), triple (t), and complex (c) morphology.
The main difference between the double and the triple morphology is the capability of detecting what is likely the core of the radio galaxy, in addition to the lobes; see W20 for further details on their morphology classification criteria.
Of the 17 SMS4 sources in our sample, all with a G4Jy counterpart, the radio morphology is triple for 2 sources and is double for 2 sources, with all the remaining sources having a single (compact, substantially symmetric) morphology.
These SMS4-G4Jy correspondences are listed in Table~\ref{tab:radio}: quantities in columns 1--4 are taken from BH06, while those in columns 5--10 come from W20.

We used maps from the SUMSS survey, available for all 17 sources, to derive radio flux density contours at 843~MHz.
To compare the morphology of the radio source and the underlying X-ray emission, we overlay these contours on the \sw-XRT X-ray maps (see Figure~\ref{fig:xrtmaps}), generated as described in Section~\ref{sec:3}.
Furthermore, we highlight the distinct SUMSS components encompassed within a circle of radius given by the corresponding Largest Angular Size (LAS) value, reported in Table~\ref{tab:radio}.   
For each source with multiple components, we assign a letter to each component, sorting them by decreasing flux density (A corresponding to the brighter component).      
We note that for B0344$-$345, with a triple morphology (see column~10 in Table~\ref{tab:radio}), we find four SUMSS components within the LAS circle; for B2140$-$434, with double morphology, three SUMSS components are found; conversely, for B2032$-$350 just one SUMSS component is found, although the radio source was characterized by a double morphology.

In BH06, for all 17 SMS4 sources making up our sample, an optical counterpart was given using either the plates of the UK Schmidt Southern Sky Survey or R-band CCD images from the Anglo-Australian Telescope (AAT).
The spectroscopic redshifts of seven sources, collected from literature, were also reported.
For an additional nine sources, photometric redshifts were computed from the magnitudes of the corresponding optical counterparts, using Eq.~2 and Eq.~3 (see their Section~3.8).
No redshift measurement was provided for B0427$-$366.

More recently, \cite{2024ApJS..271....8G} published results from a spectroscopic campaign on G4Jy-3CRE sources, including spectra and redshifts for four of our 17 sources.
One of these sources is B0427$-$366, filling the gap of BH06; the other three sources are B0242$-$514, B0344$-$345, and B2140$-$434.
The analysis of \cite{2024ApJS..271....8G} basically confirmed the spectroscopic redshifts already reported by BH06 for B0344$-$345 ($z=0.0538$) and B2140$-$434 ($z=0.65$), as well as the BH06 photometric estimate for B0242$-$514 ($z$=0.72).   
As a result, we have redshifts for all 17 sources in our sample, nine of which are spectroscopic.
All these redshift values are reported in Table~\ref{tab:radio}.


\section{\sw-XRT Data reduction and analysis} 
\label{sec:3}

The \sw~campaign described in this paper was organized following a \sw~proposal (see \am) in which we were granted 6~ks exposures, for a total of 108~ks.
For each source, the first \sw~observation started in 2022, between May and December, excluding B1814$-$519 that was observed in February 2023 for the first time.  
As of May 2024, the observational campaign was nearly concluded: B1827$-$360 was the only target still pending any observations, and B1754$-$597 had a very short exposure. 
However, all the other sources had a minimum exposure of 2.7~ks, with six sources exceeding the requested exposure time of 6~ks, up to 12~ks for B1817$-$640.

The X-ray data reduction and procedures adopted in the present analysis are analogous to those already described in \am~and references therein.
Below, we briefly summarize these and include additional details, suited for the present analysis.
We give full observational details for detections in Table~\ref{tab:det} and upper limits in Table~\ref{tab:und} for undetected sources.

X-ray data from the \sw-XRT were retrieved from the \sw~archive and processed with the XRTDAS software package (v.3.7.0), developed at the Space Science Data Center (SSDC) of the Italian Space Agency (ASI) and distributed by the NASA High Energy Astrophysics Archive Research Center (HEASARC) within the HEASoft package (v.6.30.1), including a collection of {\sc ftools} to manipulate and analyze FITS files.
All the XRT observations were carried out in the most sensitive Photon Counting (PC) readout mode. 
Event files were calibrated and cleaned by applying standard filtering criteria with the {\sc xrtpipeline} task and using the latest calibration files available in the \sw~CALDB distributed by HEASARC. 
In particular, X-ray events are classified according to the charge distribution in a 3x3 matrix of pixels on the detector, following a library of 32 grades: grade 0 is assigned to events with the whole charge concentrated in a single pixel, while increasing grades are due to events with charge progressively spreading to neighbouring pixels, according to a coded scheme\footnote{For further details, see the \sw~XRT Data Reduction Guide (\url{https://swift.gsfc.nasa.gov/analysis/xrt_swguide_v1_2.pdf})}.
Events in the 0.3--10 keV energy range, with grades 0--12, were used in our analysis, and exposure maps were also created with {\sc xrtpipeline}.
For sources with several visits, event files and exposure maps from each observation were accumulated with the {\sc xselect} and {\sc ximage} tools, respectively, to build a single event file and a single exposure map for each source.
Fig.~\ref{fig:xrtmaps} shows the X-ray maps for all 17 SMS4 sources included in our sample, built from the corresponding event files.

\begin{figure*}
\gridline{
\fig{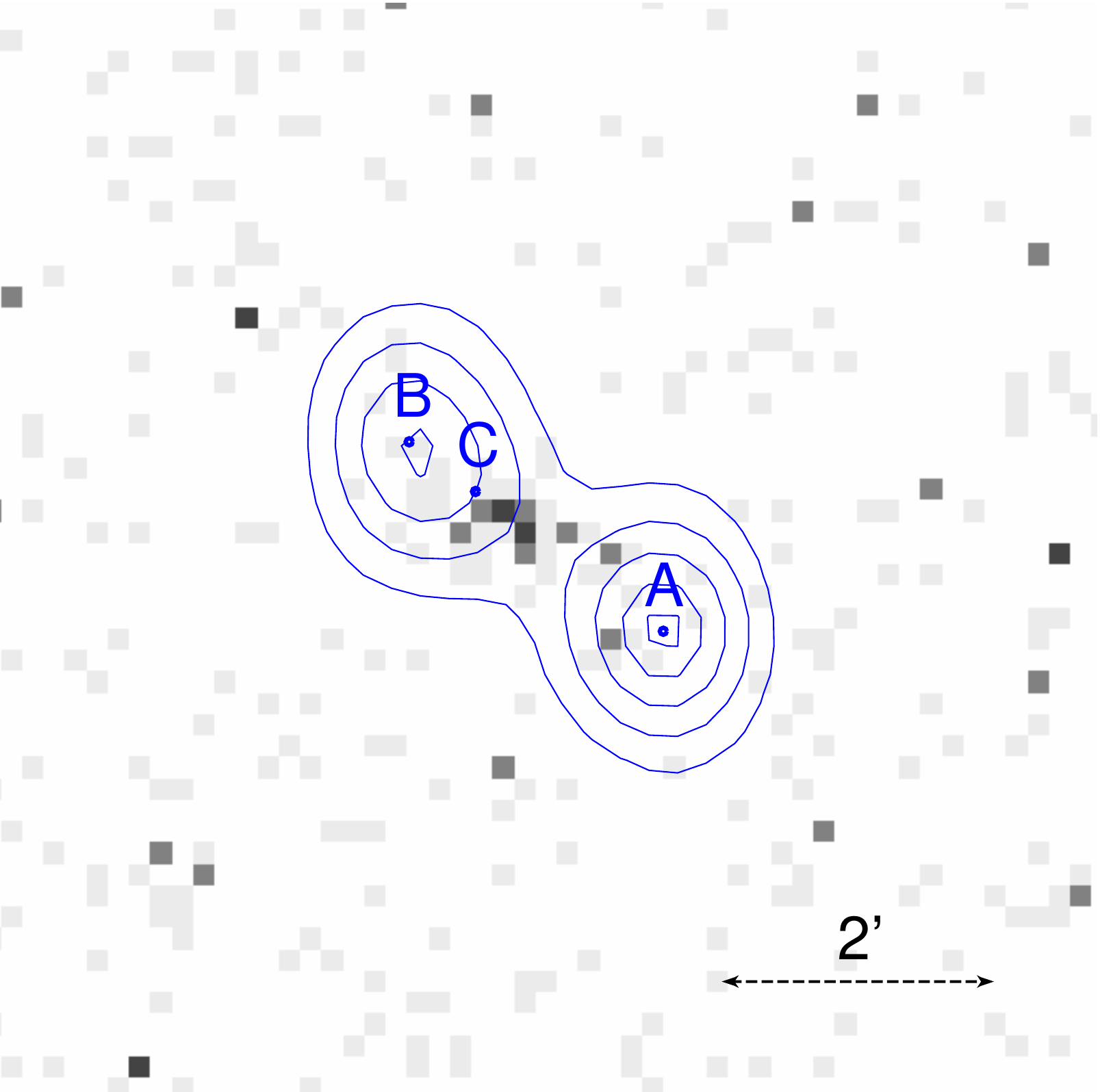}{0.30\textwidth}{(MRC~B0103$-$453 / G4Jy 120)}
\fig{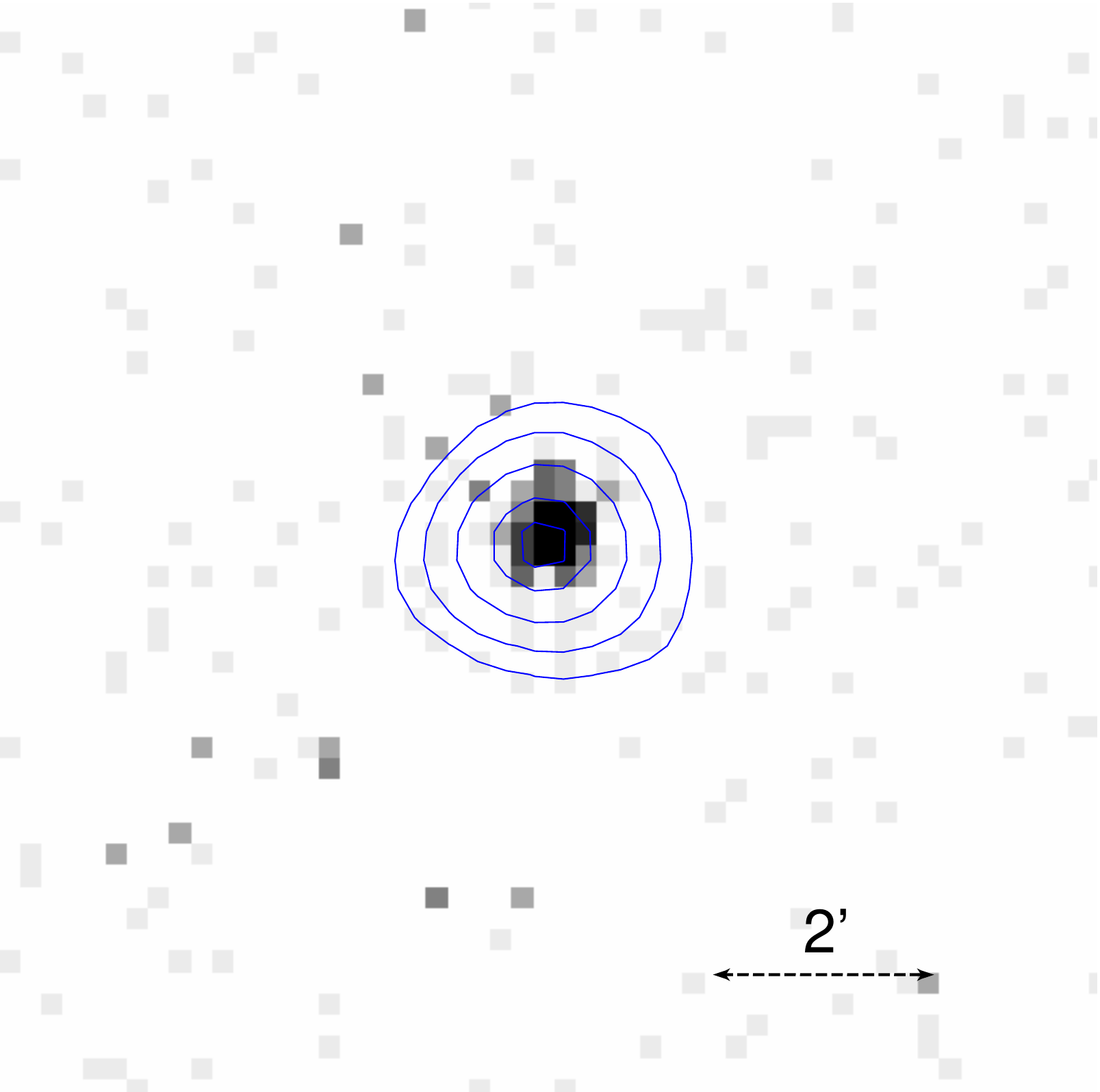}{0.30\textwidth}{(MRC~B0202$-$765 / G4Jy 217)}
\fig{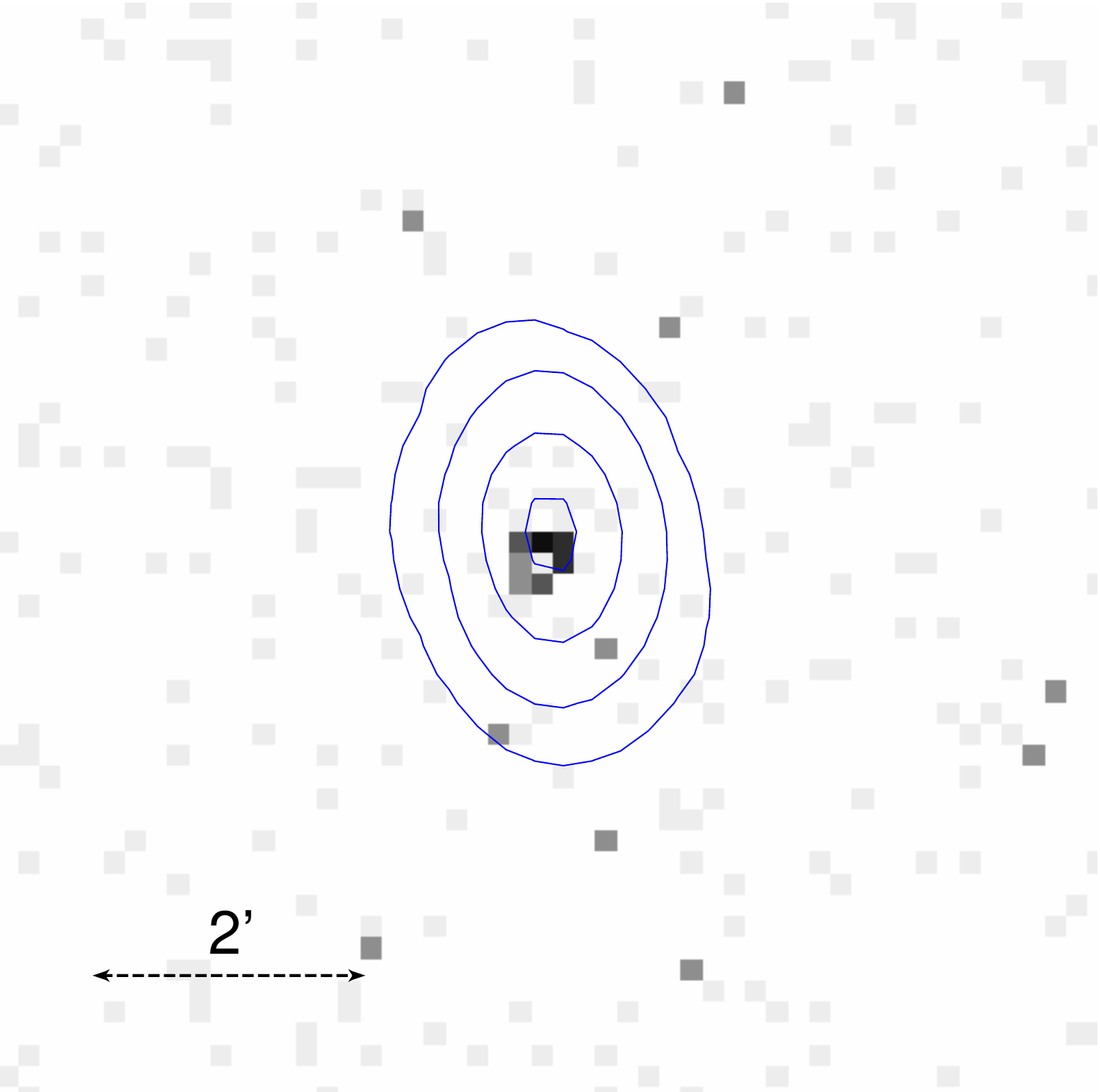}{0.30\textwidth}{(MRC~B0242$-$514 / G4Jy 290)}
}
\gridline{
\fig{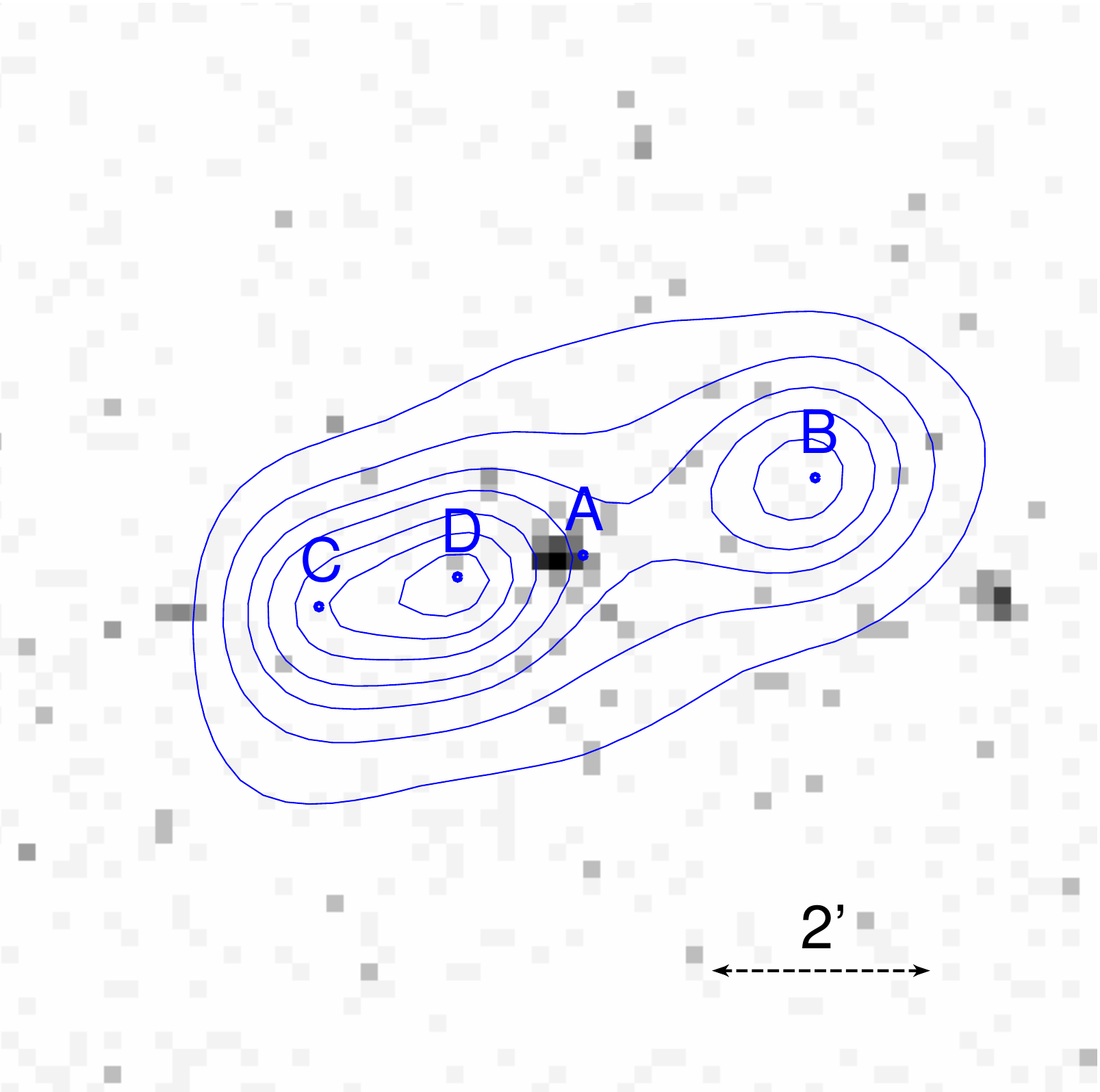}{0.30\textwidth}{(MRC~B0344$-$345 / G4Jy 381)}
\fig{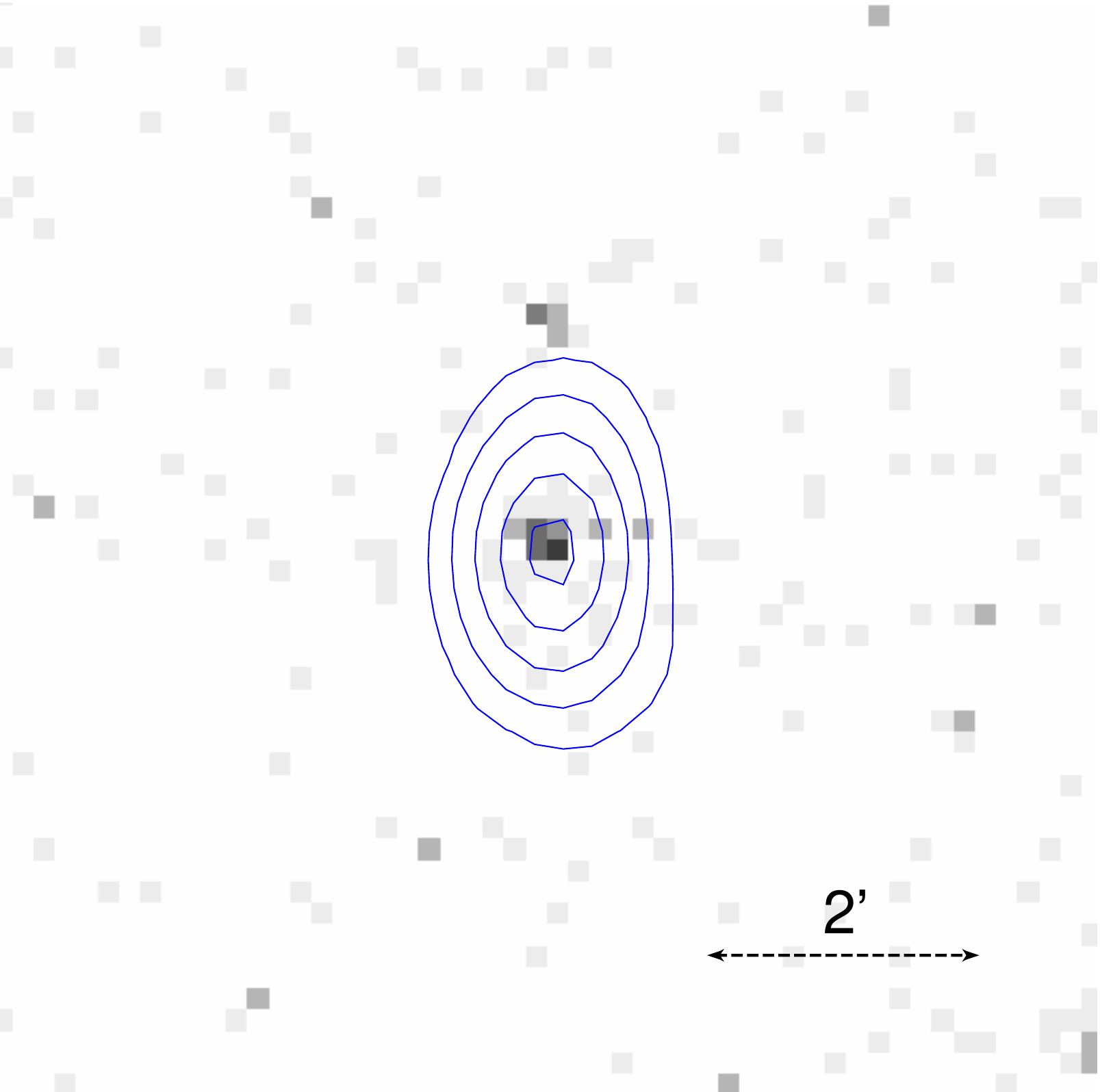}{0.30\textwidth}{(MRC~B0427$-$366 / G4Jy 464)}
\fig{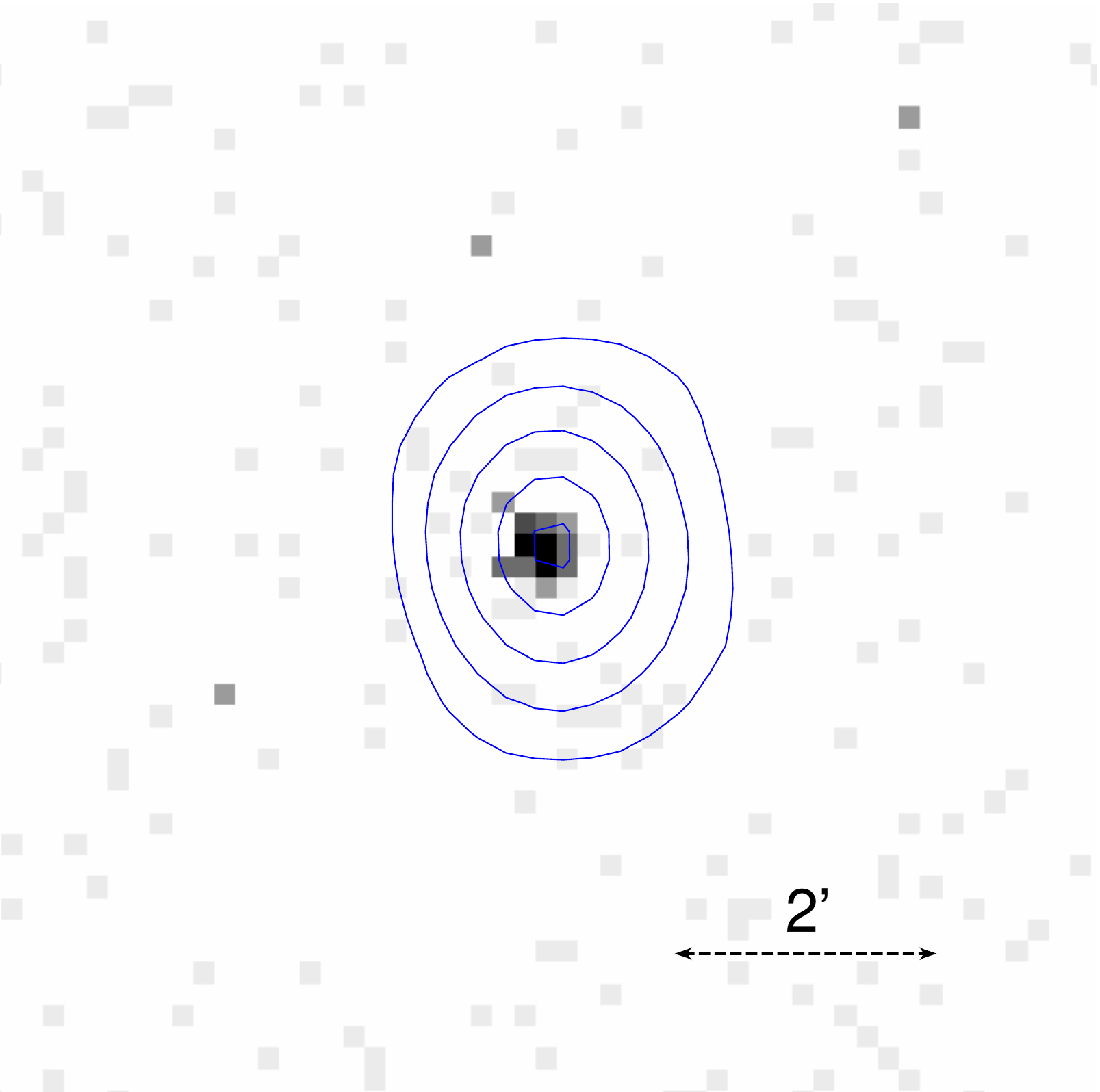}{0.30\textwidth}{(MRC~B1017$-$426 / G4Jy 837)}
}
\gridline{
\fig{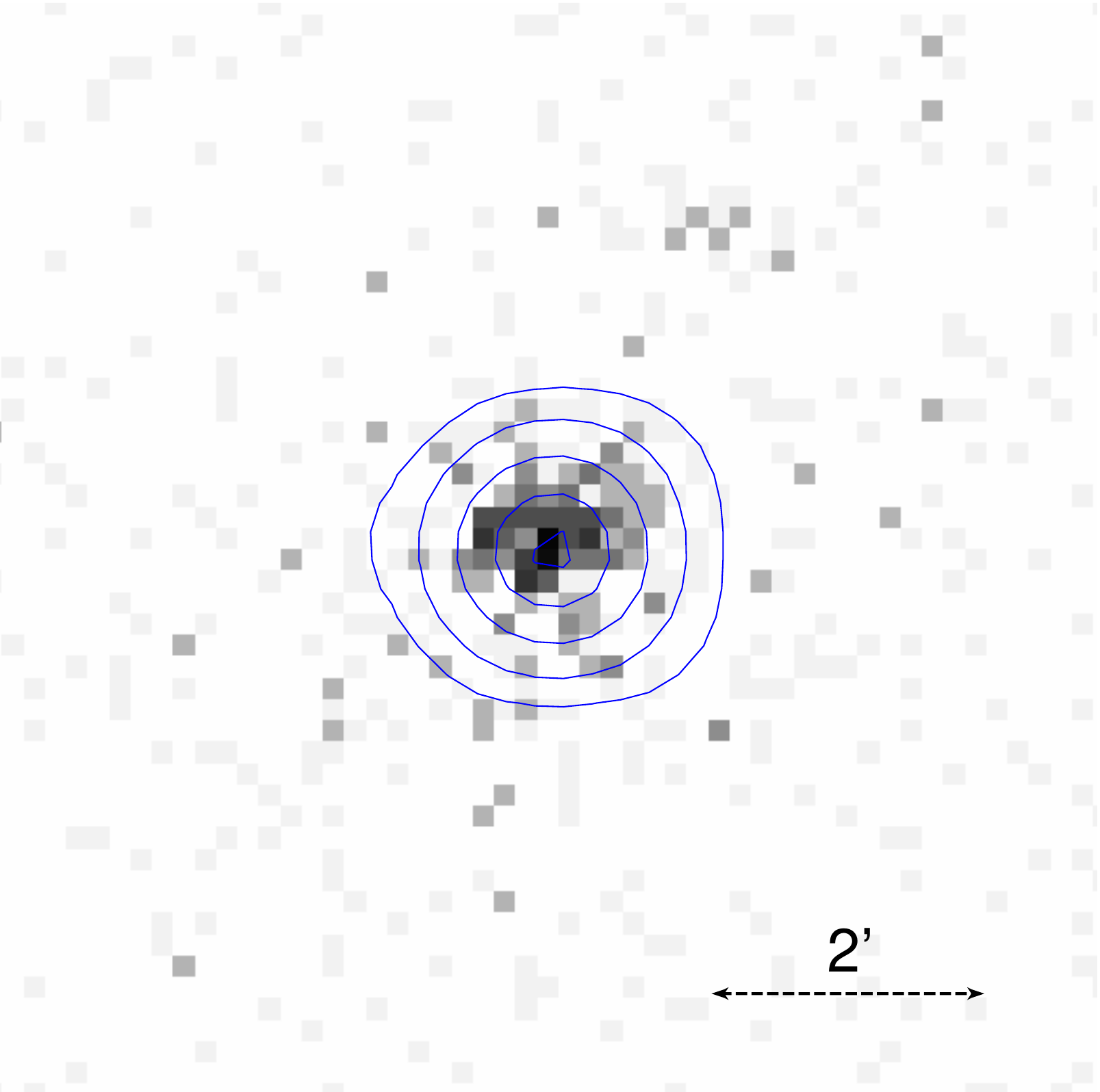}{0.30\textwidth}{(MRC~B1526$-$423 / G4Jy 1262)}
\fig{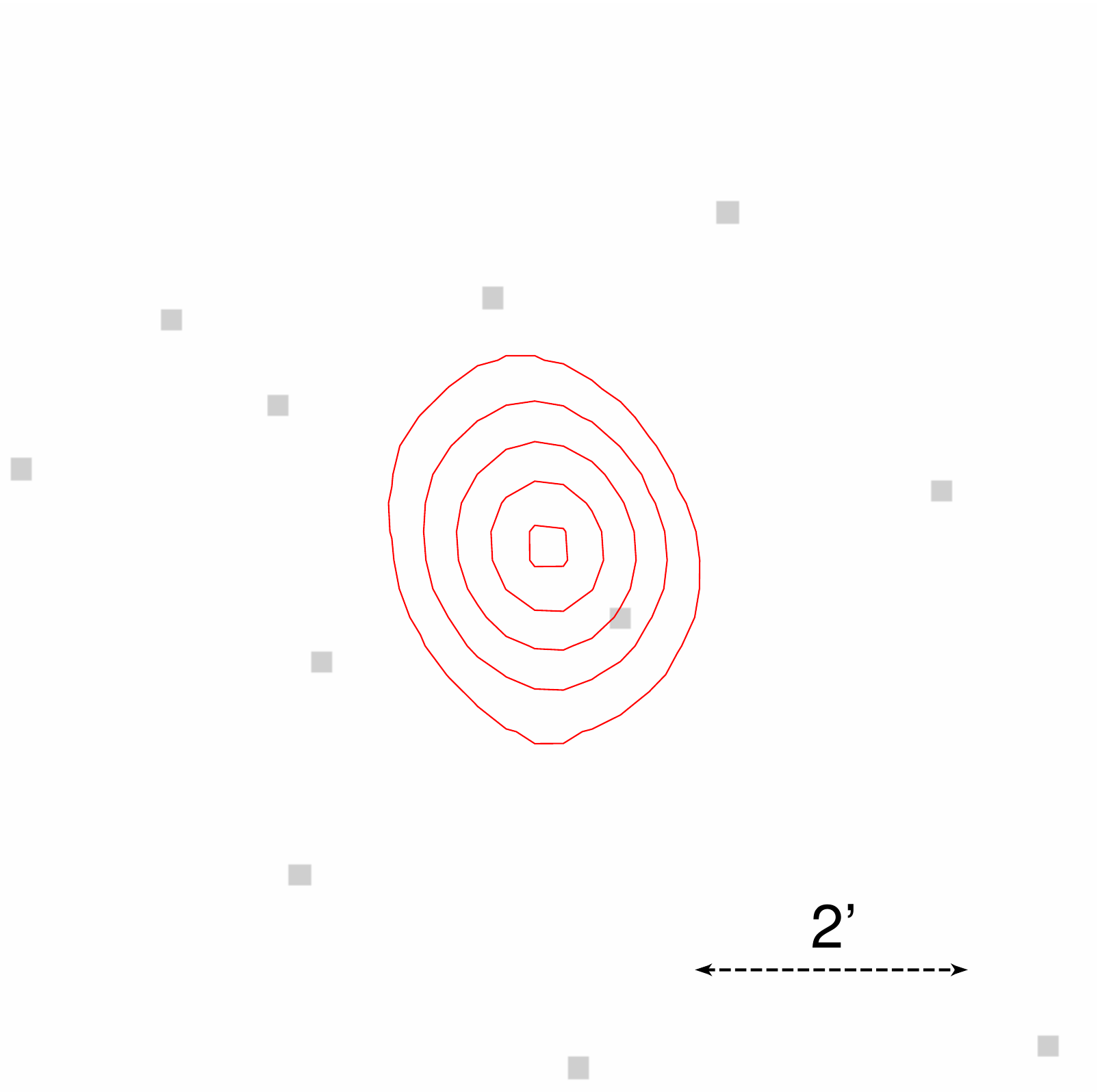}{0.30\textwidth}{(MRC~B1754$-$597 / G4Jy 1453)}
\fig{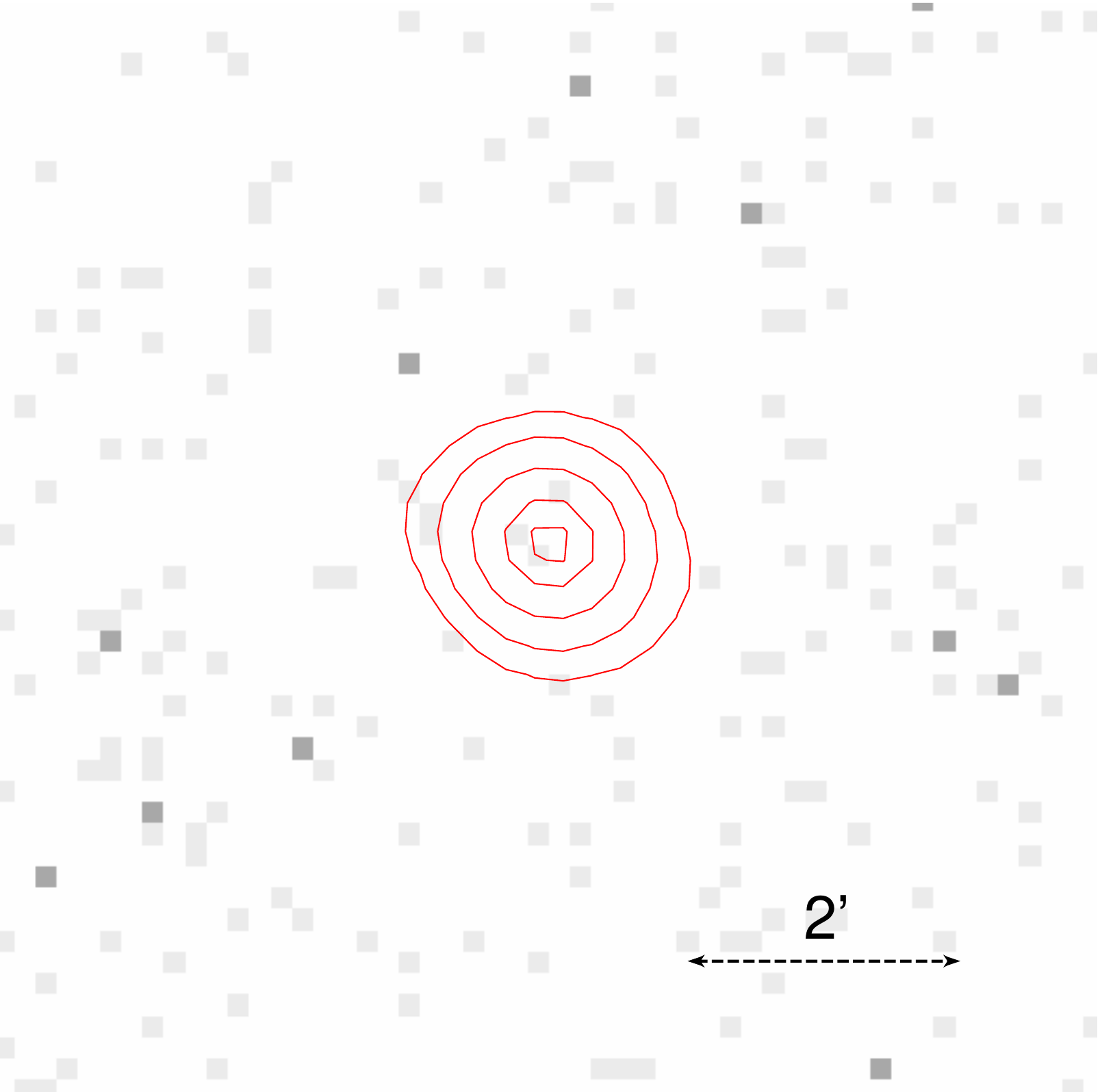}{0.30\textwidth}{(MRC~B1814$-$519 / G4Jy 1471)}
}
\caption{\sw~X-ray maps (North is up, and East to the left) in the 0.3--10.0 keV band for the 17 sources comprising our sample. All maps, centered at the positions of the SMS4 radio sources, are binned by 4x4 pixels and are 8\arcmin~x~8\arcmin~except for B0344$-$345, shown at 10\arcmin~x~10\arcmin. Radio flux density contours from the SUMSS survey, carried out at 843~MHz, overlay the X-ray maps and have been selected, for each source, to best display the shape of the radio emission. Following \am, blue contours distinguish sources detected in the X-rays by \sw~in our campaign from not detected sources, whose contours are marked in red. B1754$-$597, not detected in our analysis due to the short exposure, was detected by eROSITA in the 1eRASS scan.}
\label{fig:xrtmaps}
\end{figure*}

\setcounter{figure}{0}
\begin{figure*}
\gridline{
\fig{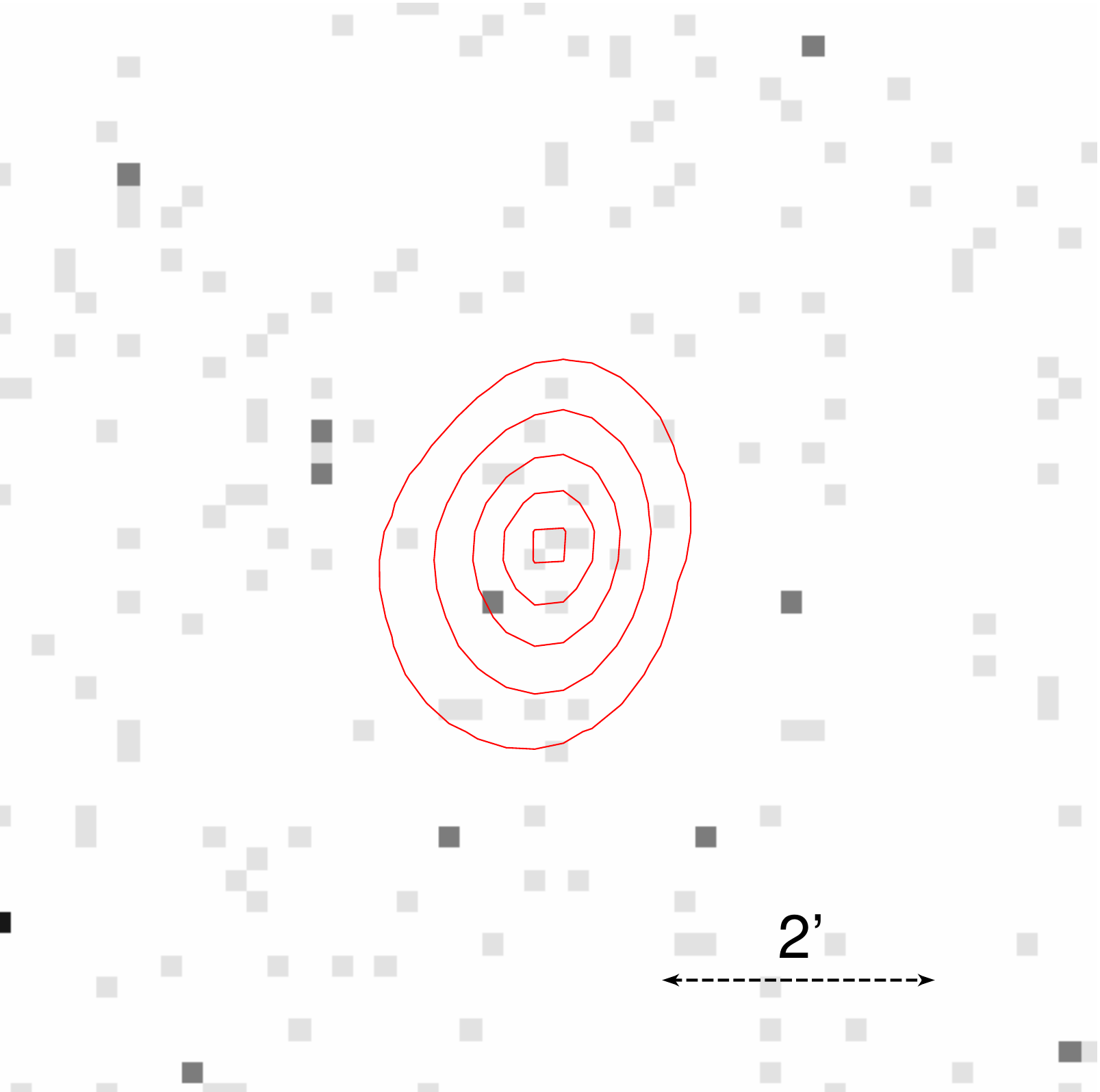}{0.30\textwidth}{(MRC~B1817$-$391 / G4Jy 1474)}
\fig{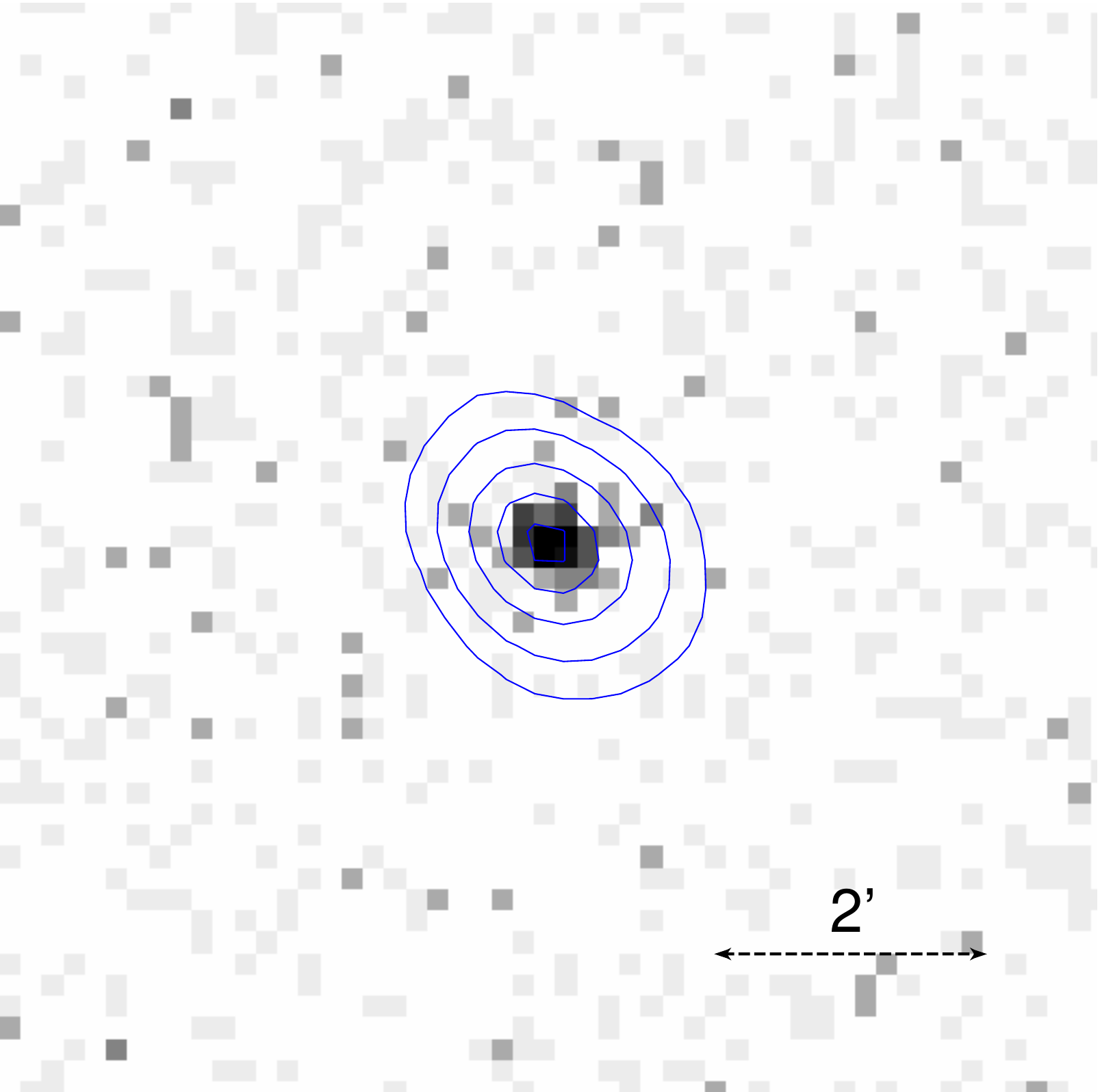}{0.30\textwidth}{(MRC~B1817$-$640 / G4Jy 1477)}
\fig{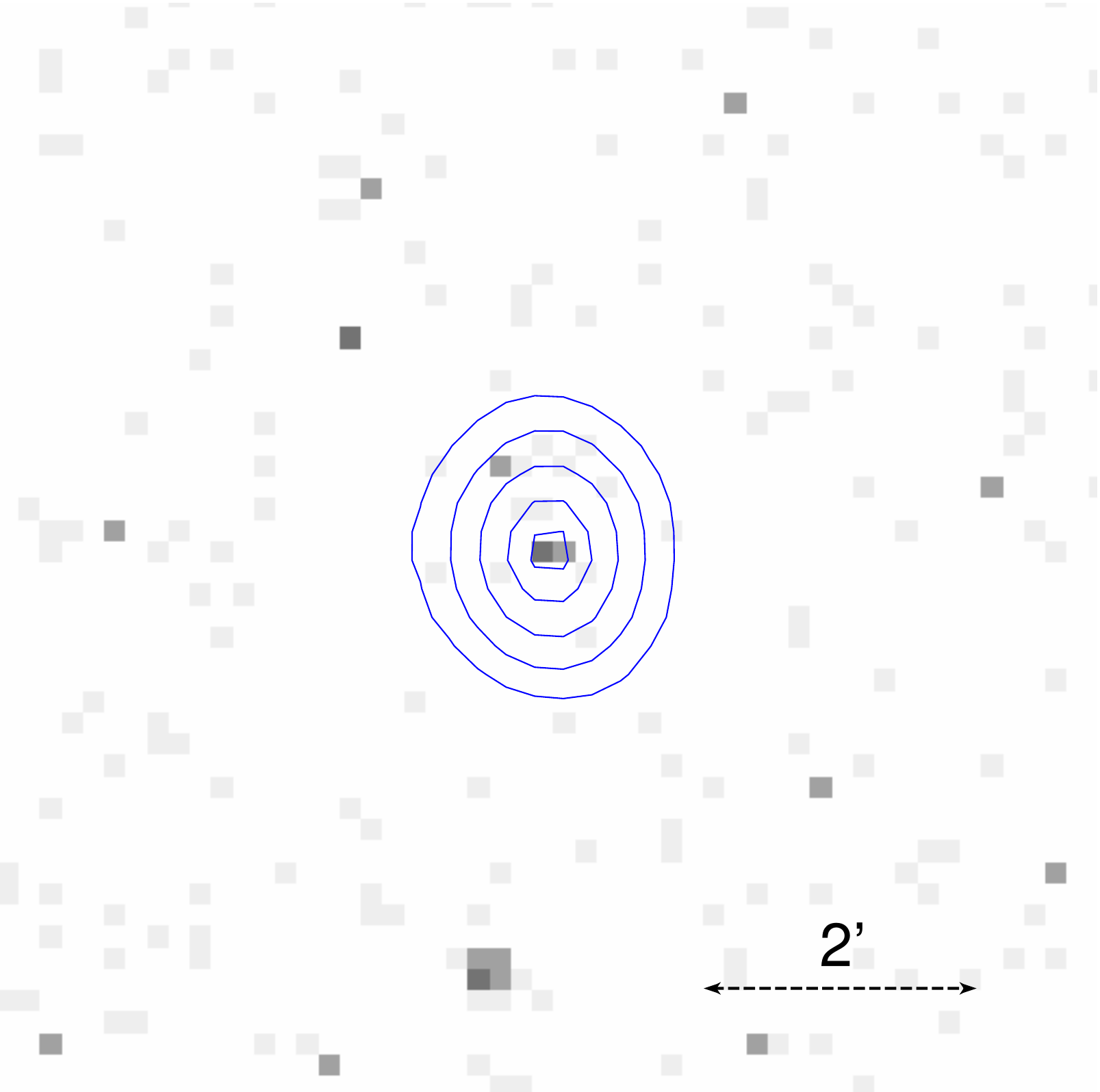}{0.30\textwidth}{(MRC~B1953$-$425 / G4Jy 1588)} 
}
\gridline{
\fig{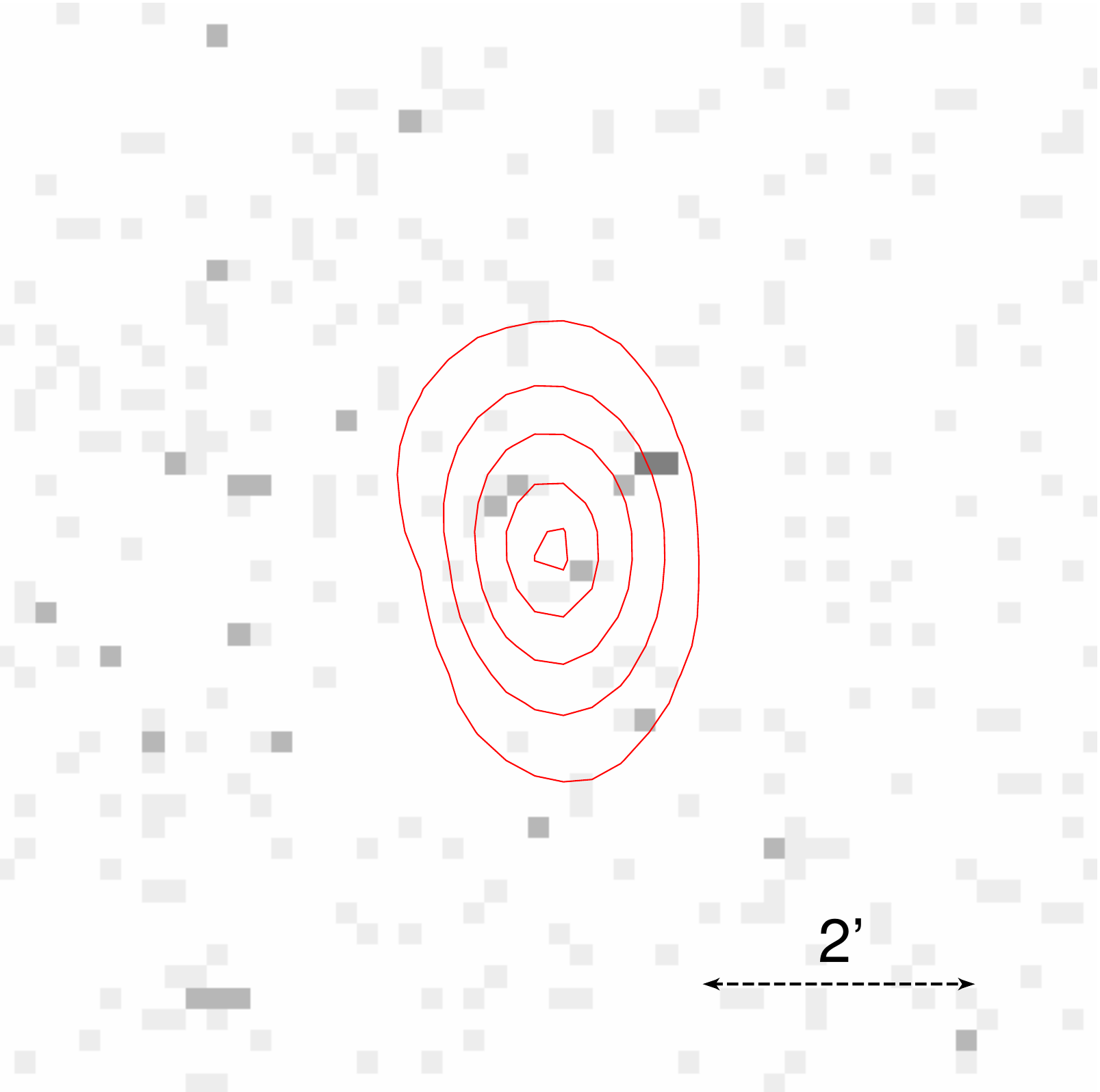}{0.30\textwidth}{(MRC~B2032$-$350 / G4Jy 1640)}
\fig{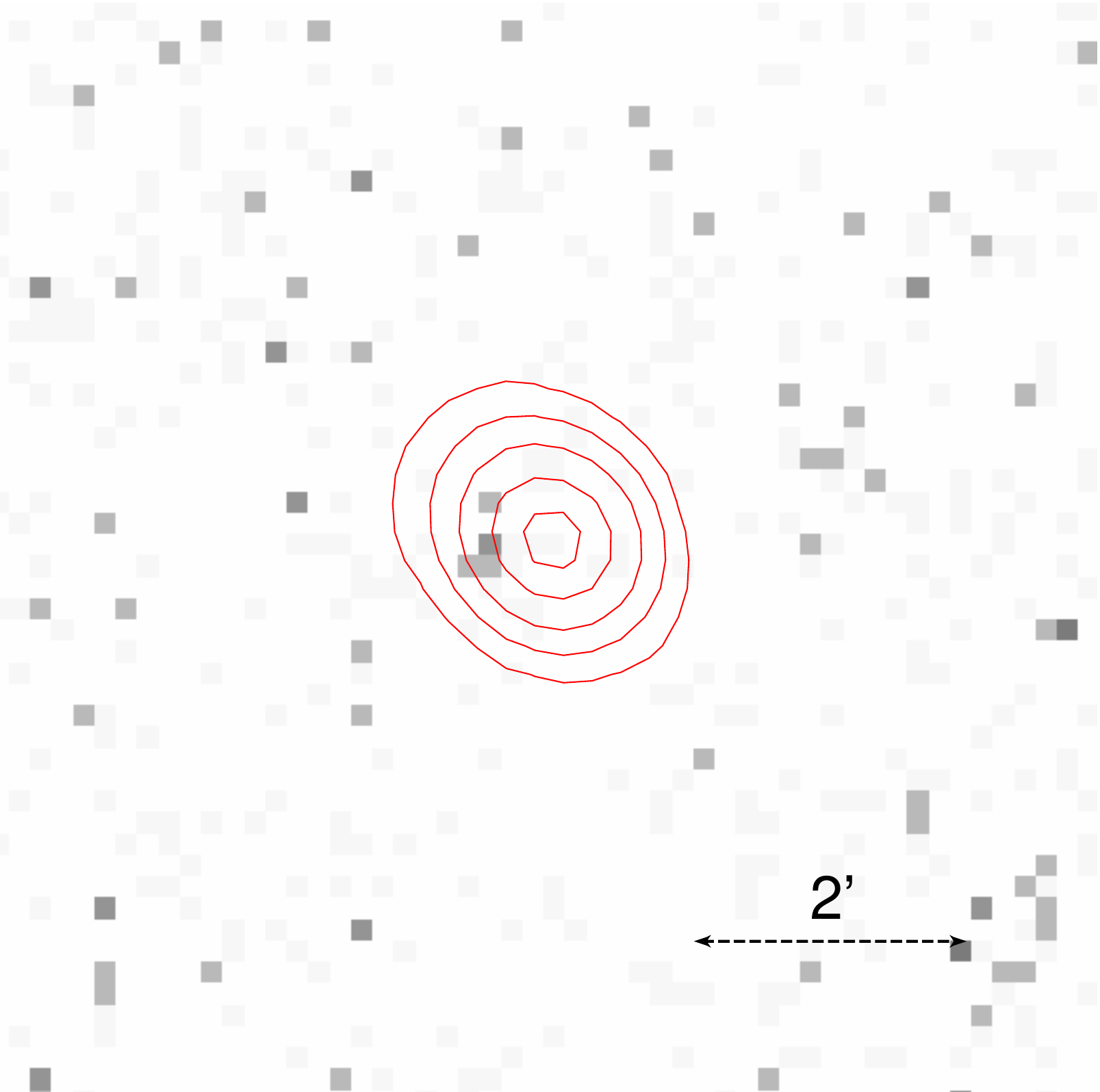}{0.30\textwidth}{(MRC~B2041$-$604 / G4Jy 1646)} 
}
\gridline{
\fig{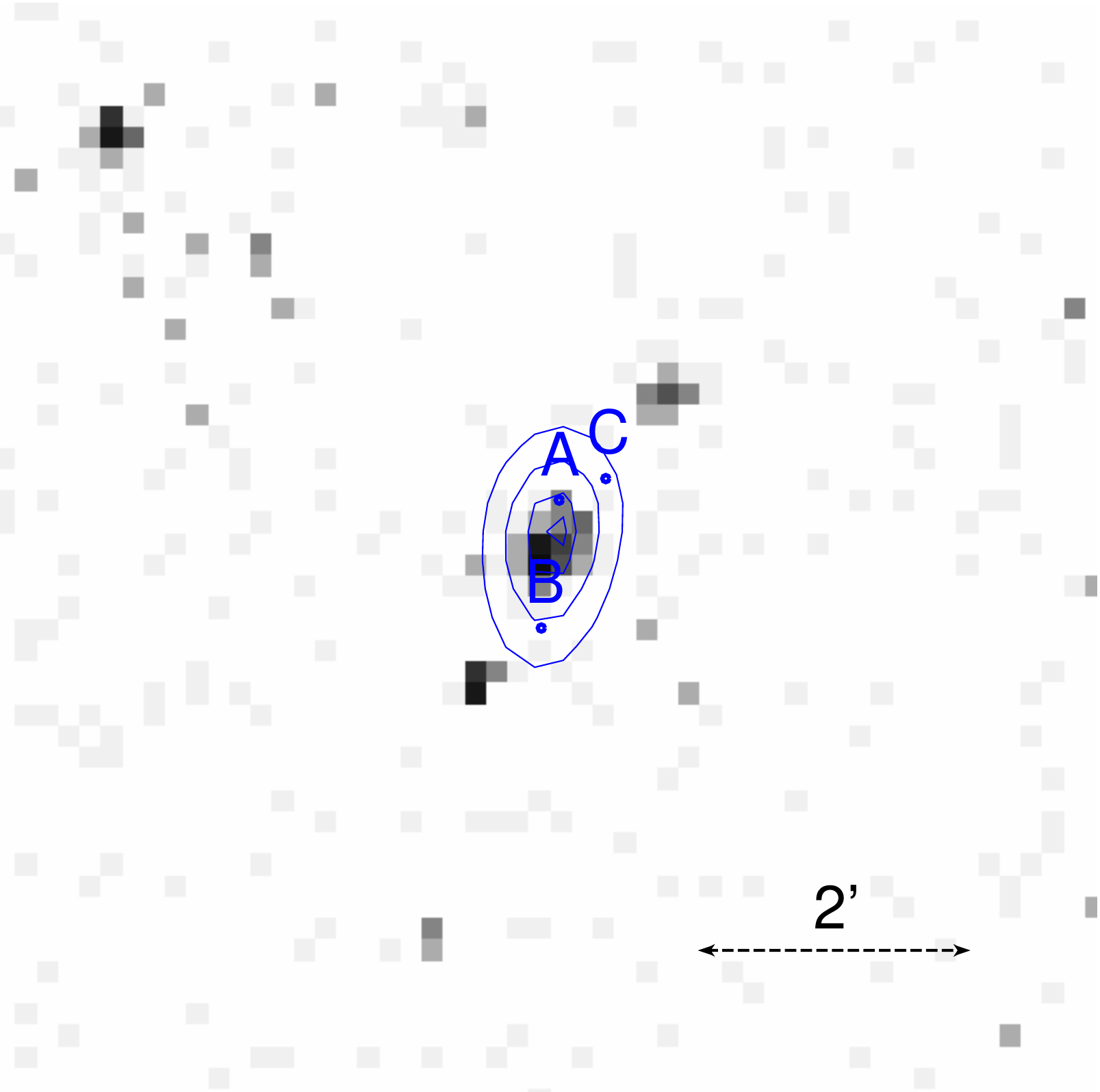}{0.30\textwidth}{(MRC~B2140$-$434 / G4Jy 1717)}
\fig{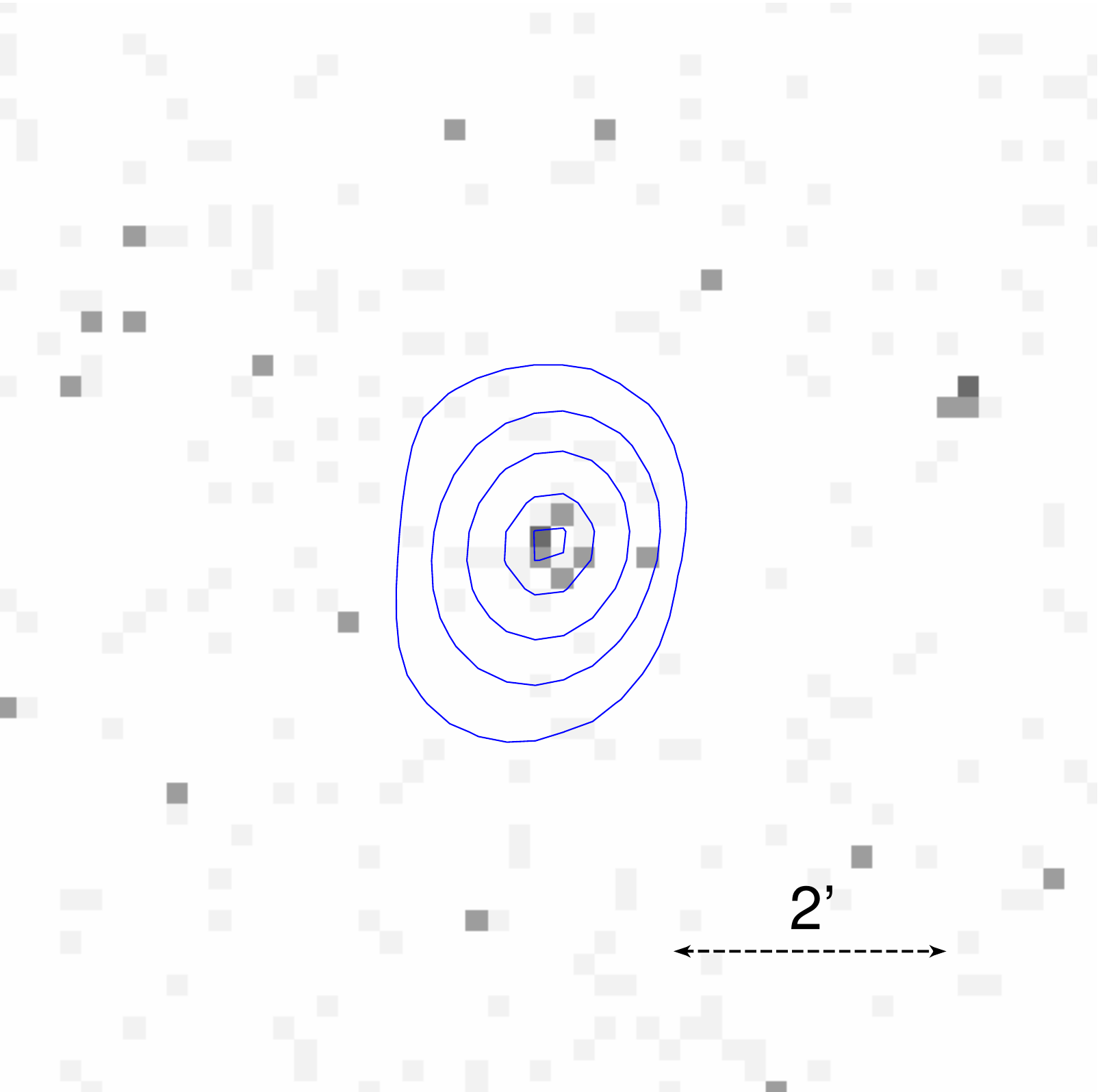}{0.30\textwidth}{(MRC~B2140$-$817 / G4Jy 1723)}
\fig{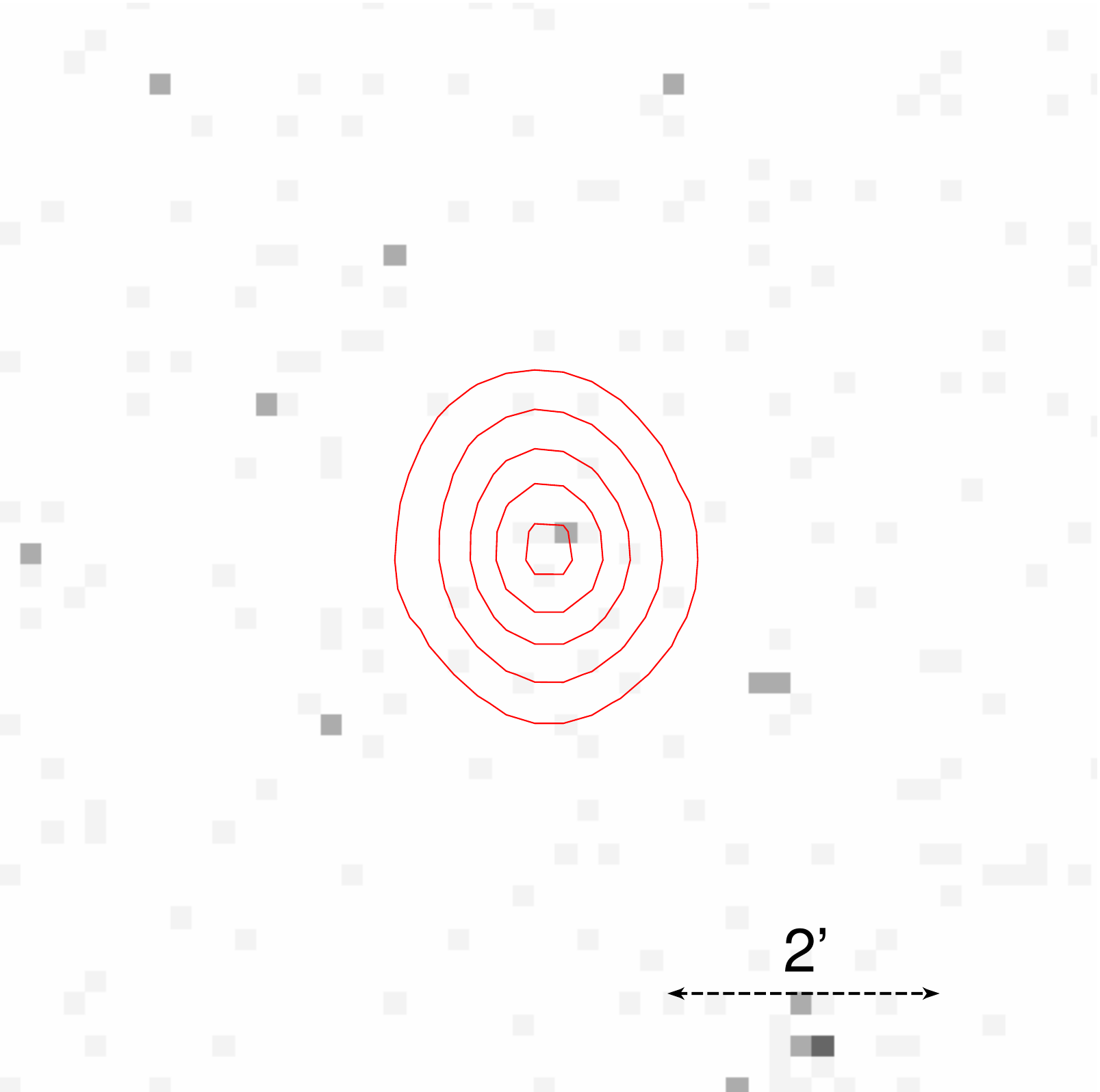}{0.30\textwidth}{(MRC~B2331$-$416 / G4Jy 1840)}
}
\caption{{\it (continued)}}
\end{figure*}

\begin{table*}
\scriptsize
\caption{Results from \sw-XRT observations for the X-ray detected sources of our sample.}
\label{tab:det}
\begin{center}
\begin{tabular}{cccccccccc}
\hline
R.A. (J2000) & Decl. (J2000)                &    $r_c$     &        Count Rate       &        $P$         & First Obs. & Latest Obs.& Obs. & Exposure &  SMS4 Name  \\
($^{h~m~s}$) & ($^{\circ}$~\arcmin~\arcsec) &   (arcsec)   & (10$^{-3}$ ct s$^{-1}$) &                    & (yy-mm-dd) & (yy-mm-dd) &      &   (s)    &             \\
 (1)         &  (2)                         &      (3)     &          (4)            &       (5)          &     (6)    &    (7)     & (8)  &   (9)    &     (10)    \\   
\hline                                                                                                                                 
01 05 22.3   &        $-$45 05 16.9         &      4.7     &       2.0 $\pm$ 0.7     & $6.7\cdot10^{-10}$ & 22-07-26   & 23-02-19  &    8  &   8472   &  B0103$-$453             \\
02 02 13.2   &        $-$76 20 01.0         &      3.8     &      82.8 $\pm$ 5.6     & $ < 10^{-16}$      & 22-09-01   & 23-03-19  &   13  &   3814   & ~B0202$-$765$^\dagger$   \\
02 43 44.5   &        $-$51 12 39.7         &      4.8     &       5.2 $\pm$ 1.1     & $ < 10^{-16}$      & 22-07-31   & 24-01-24  &    6  &   6952   &  B0242$-$514             \\
03 46 30.6   &        $-$34 22 47.4         &      4.0     &       8.8 $\pm$ 1.1     & $ < 10^{-16}$      & 22-12-16   & 23-07-21  &   10  &  11019   &  B0344$-$345             \\
04 29 40.4   &        $-$36 30 52.9         &      4.8     &       7.9 $\pm$ 1.6     & $ < 10^{-16}$      & 22-06-23   & 23-01-09  &    4  &   5009   & ~B0427$-$366$^\dagger$   \\
10 20 04.2   &        $-$42 51 32.5         &      4.5     &      15.0 $\pm$ 2.1     & $ < 10^{-16}$      & 22-12-31   & 23-06-25  &    3  &   4848   &  B1017$-$426             \\
15 30 14.3   &        $-$42 31 53.7         &      3.8     &      23.6 $\pm$ 2.6     & $ < 10^{-16}$      & 22-05-25   & 24-01-10  &    6  &   5005   &  B1526$-$423             \\
18 22 16.1   &        $-$63 59 18.6         &      4.0     &      11.1 $\pm$ 1.2     & $ < 10^{-16}$      & 22-08-01   & 24-02-04  &   13  &  12211   & ~B1817$-$640$^\dagger$   \\
19 57 15.5   &        $-$42 22 23.9         &      5.8     &       1.7 $\pm$ 0.7     & $4.0\cdot10^{-05}$ & 23-06-22   & 23-10-09  &    7  &   5390   &  B1953$-$425             \\
21 43 33.6   &        $-$43 12 50.8         &      4.3     &      11.4 $\pm$ 1.7     & $ < 10^{-16}$      & 22-05-28   & 22-12-24  &   18  &   6288   & ~B2140$-$434$^\dagger$   \\
21 47 22.7   &        $-$81 32 11.3         &      5.2     &       4.0 $\pm$ 1.1     & $8.8\cdot10^{-15}$ & 22-07-23   & 23-07-18  &   16  &   5395   &  B2140$-$817             \\
\hline
   ...       &              ...             &      ...     &            ...          &       ...          & 22-05-22   & 24-02-22  &    2  &    208   & \,\,\,B1754$-$597$^\ast$ \\
\hline
\end{tabular}
\end{center}
\tablecomments{The columns show (1) the Right Ascension and (2) Declination of the X-ray detection; (3) the positional uncertainty $r_c$, at the 90\% confidence level; (4) the 0.3--10 keV count rate, with its uncertainty; (5) the probability $P$ that the signal is a statistical fluctuation of the background; the dates of the first (6) and the latest (7) \sw~observation; (8) the number of stacked \sw~observations; (9) the total XRT exposure time at the coordinates of the SMS4 source; (10) the MRC name of the corresponding SMS4 source.\\
$\dagger$: The X-ray detection of this source was also reported by \cite{2023ApJS..268...32M}.\\
$\ast$: This source, not detected in our campaign due to a very short exposure, is reported in the DR1 {\it Main} catalog \citep{2024arXiv240117274M} as 1eRASS J175906.4$-$594643. See also Table~\ref{tab:dr1}, where all eROSITA-DE sources matching our detections are listed, with a selection of properties from DR1 catalogs.\\
} 
\end{table*}

One of the observations of B1526$-$423 (obsID 03111885013) was affected by bright Earth contamination, due to the loading of optical-UV photons, scattered by the Earth's atmosphere and reflected by the optics of the X-ray telescope onto the XRT detector.  
Following \cite{2011A&A...528A.122P}, we checked the background level in this observation and  found two peaks in two of its five orbits: thus, we filtered the event file of this sequence removing the bad time intervals during which the background rate over the whole detector was higher than 2 ct~s$^{-1}$, also creating the corresponding exposure map.   
We checked that no other observation of this source was affected by this issue, before stacking all the observations.
No such problem was found for other sources in our sample.

For each source we simultaneously uploaded both the event file and the exposure map within {\sc ximage}, and used the {\sc background} command to compute the average background intensity over the whole detector.
We checked the background values in the whole 0.3--10 keV band for all sources with sufficient exposure, spanning a range from 4.1$\times$10$^{-4}$ ct~sqarcmin$^{-1}$~s$^{-1}$ (B2041$-$604) to 7.7$\times$10$^{-4}$ ct~sqarcmin$^{-1}$~s$^{-1}$ (B1814$-$519).
Excluding sources with $\mid b \mid<20^{\circ}$, the upper limit of this range decreases to 7.3$\times$10$^{-4}$ ct~sqarcmin$^{-1}$~s$^{-1}$ (B2032$-$350), and yields results consistent with those reported in \cite{2011A&A...528A.122P}.

We performed a local source detection using the {\sc sosta} command within {\sc ximage}.
This task counts the number of events within a specified box, and corrects these counts for dead time, vignetting, exposure, and Point Spread Function (PSF).
As a result, it provides the source intensity and its significance, in terms of the probability $P$ that the signal is a statistical fluctuation of the background.
If the probability is higher than $P = 10^{-3}$, a count rate 3$\sigma$ upper limit is automatically calculated.

In our use of {\sc sosta}, the side of the box was fixed to 14~pixels ($\sim33\arcsec$), allowing the core of the PSF to be fully included in the box.
Based on our experience in the analysis of XRT images, this choice maximizes the signal-to-noise ratio for faint sources.
Furthermore, we used the qualifier {\it /background\_level} to fix the background intensity at the value previously obtained with the {\sc background} command.
Only for B1754$-$597 was the local background used, since the {\sc back} command in {\sc ximage} had no effect, presumably due to the extremely short available exposure.
For each X-ray image, we used the cursor to precisely center the extraction box on the SMS4 coordinates.

Requiring $P < 10^{-4}$, that is an order of magnitude tighter constraint with respect to the default in {\sc sosta}, to establish an X-ray detection, we distinguish detections (Table~\ref{tab:det}) from non-detections (Table~\ref{tab:und}). 
Considering the 17 observed SMS4 sources, our prescriptions led to 11 X-ray detections. 
To determine the position, and in particular its uncertainty (90\% confidence level), for all these detections, we used the {\sc xrtcentroid} task.

Table~\ref{tab:det} also includes B1754$-$597, the only source for which we were not able to report a detection due to a very short exposure, and that was instead detected by eROSITA and reported in the {\it Main} DR1 catalog (\citealp{2024arXiv240117274M}; see Section~\ref{sec:3.3} for further details on all the sources in our sample detected by eROSITA).

\begin{table*}
\scriptsize
\caption{Results from \sw-XRT observations for the X-ray undetected sources of our sample.}
\label{tab:und}
\begin{center}
\begin{tabular}{cccccc}
\hline
     SMS4 Name     & First Obs. & Latest Obs. & Obs. & Exposure & 3$\sigma$ Upper Limit   \\
                   & (yy-mm-dd) &  (yy-mm-dd) &      &   (s)    & (10$^{-3}$ ct s$^{-1}$) \\ 
          (1)      &    (2)     &    (3)      & (4)  &   (5)    &          (6)            \\
\hline                                                                                                                                 
      B1814$-$519  &  23-02-12  &   24-05-19  &   6  &  3590    &         4.0            \\ 
      B1817$-$391  &  22-07-14  &   24-02-08  &   6  &  2728    &         6.4            \\ 
      B2032$-$350  &  22-09-02  &   23-04-05  &   7  &  5601    &         2.9            \\  
      B2041$-$604  &  22-08-24  &   23-10-19  &  15  &  7979    &         2.2            \\  
      B2331$-$416  &  22-08-19  &   23-04-11  &  15  &  5004    &         4.4            \\ 
\hline
\,\,\,B1754$-$597$^\ast$ &  22-05-22 & 24-02-22 & 2  &   208    &        45.3            \\ 
\hline
\end{tabular}
\end{center}
\tablecomments{The columns show (1) the MRC name of the SMS4 source; the dates of the first (2) and the latest (3) \sw~observation; (4) the number of \sw~observations; (5) the total XRT exposure time; (6) the 0.3--10 keV count rate 3$\sigma$ upper limit at the position of the radio coordinates.\\
$\ast$: This source is also reported in Table~\ref{tab:det}, since it was detected by eROSITA in the 1eRASS (see Table~\ref{tab:dr1}).\\
}
\end{table*}

\cite{2023ApJS..268...32M} reported an analysis of \sw~X-ray observations matching G4Jy-3CRE sources, performed before December 2022  with a minimum exposure of 250~s, yielding a list of 89 G4Jy-3CRE sources observed by \sw. 
Due to their selection criteria, \cite{2023ApJS..268...32M} include only eight of the 17 sources in our sample, and detect only four of the sources given in Table~\ref{tab:det}.

\begin{table*} 
\begin{center}
\scriptsize
\caption{X-ray unabsorbed flux and luminosity for the \sw~detected SMS4 sources of our sample.}
\label{tab:pimms}
\begin{tabular}{cccc|cc|cc}
                      \multicolumn{4}{c}{}                              &  \multicolumn{2}{c}{Power Law Model ($\Gamma=2$)}       & \multicolumn{2}{c}{APEC Model (0.4 Solar, $kT$=3~keV)}   \\  
 \hline
MRC Name             &  $D_L$  &    ${n_{H,\,Gal}}$    &      XRT Count Rate     &     $S_{~0.3-10,unabs}$             &      $L_X$        &      $S_{~0.3-10,unabs}$            &        $L_X$       \\
                     &  (Mpc)  & (10$^{20}$ cm$^{-2}$) & (10$^{-3}$ ct s$^{-1}$) & ($10^{-12}$ erg cm$^{-2}$ s$^{-1}$) &   (erg s$^{-1}$)  & ($10^{-12}$ erg cm$^{-2}$ s$^{-1}$) &  (erg s$^{-1}$)    \\
  (1)                &   (2)   &         (3)           &             (4)         &                  (5)                &        (6)        &              (7)                    &         (8)        \\  
\hline
        B0202$-$765  &  2068.2 &        5.5            &       $82.8\pm5.6$      &                3.35                 & $1.7\cdot10^{45}$ &              2.77                   & $1.4\cdot10^{45}$  \\
        B0242$-$514  &  4911.5 &        2.0            &      ~~$5.2\pm1.1$      &                0.19                 & $5.5\cdot10^{44}$ &              0.15                   & $4.4\cdot10^{44}$  \\
        B0344$-$345  &   233.6 &        1.1            &      ~~$8.8\pm1.1$      &                0.31                 & $2.0\cdot10^{42}$ &              0.28                   & $1.8\cdot10^{42}$  \\
        B0427$-$366  & 11540.7 &        1.3            &      ~~$7.9\pm1.6$      &                0.28                 & $4.5\cdot10^{45}$ &              0.21                   & $3.4\cdot10^{45}$  \\
        B1017$-$426  &  8972.6 &        7.8            &       $15.0\pm2.1$      &                0.64                 & $6.2\cdot10^{45}$ &              0.50                   & $4.8\cdot10^{45}$  \\
        B1817$-$640  &  3997.3 &        6.8            &       $11.1\pm1.2$      &                0.46                 & $8.8\cdot10^{44}$ &              0.37                   & $7.0\cdot10^{44}$  \\
        B2140$-$434  &  3850.5 &        1.5            &       $11.4\pm1.7$      &                0.41                 & $7.3\cdot10^{44}$ &              0.33                   & $5.9\cdot10^{44}$  \\
\hline
        B0103$-$453  &  4294.7 &        1.6            &      ~~$2.0\pm0.7$      &                0.07                 & $1.6\cdot10^{44}$ &              0.06                   & $1.3\cdot10^{44}$  \\
        B1526$-$423  &  2793.4 &        8.7            &       $23.6\pm2.6$      &                1.03                 & $9.6\cdot10^{44}$ &              0.83                   & $7.7\cdot10^{44}$  \\
        B1953$-$425  &   546.1 &        5.2            &      ~~$1.7\pm0.7$      &                0.07                 & $2.4\cdot10^{42}$ &              0.06                   & $2.1\cdot10^{42}$  \\ 
        B2140$-$817  &  3777.6 &        8.3            &      ~~$4.0\pm1.1$      &                0.17                 & $3.0\cdot10^{44}$ &              0.14                   & $2.4\cdot10^{44}$  \\
\hline
\end{tabular}
\end{center}
\tablecomments{The columns show (1) the name of the SMS4 source; (2) the luminosity distance $D_L$, computed using the redshift values given by BH06 and \cite{2024ApJS..271....8G}; (3) the Galactic hydrogen column density in the direction of the X-ray source; (4) the count rate detected by the XRT, as reported in Table~\ref{tab:det}; (5) the unabsorbed flux in the 0.3-10 keV band derived by WebPIMMS, assuming a power-law model with photon index $\Gamma=2$, with (6) the corresponding luminosity in the 0.3--10 keV band; (7) the same as column (5), but assuming an APEC model with a metal abundance corresponding to 0.4 Solar and a plasma temperature $kT=3$~keV, with (8) the corresponding luminosity in the 0.3--10 keV band. Sources are distinguished according to spectroscopic (top) and photometric (bottom) redshift measurements.} 
\end{table*}

We used the count rate values reported in Table~\ref{tab:det} to compute the X-ray flux and luminosity in the 0.3--10 keV band of the 11 XRT detected sources.
This was allowed after confirming all the optical counterparts quoted by BH06 and \cite{2024ApJS..271....8G} (see Section~\ref{sec:4}), and using the redshift values reported in their study.
As shown in Table~\ref{tab:radio}, there are seven sources with spectroscopic redshifts, and photometric redshifts for the remaining four.
For each source, we computed the luminosity distance and used WebPIMMS to convert the XRT count rate into unabsorbed flux in the 0.3--10 keV band.
We used two different spectral models, both absorbed by the Galactic hydrogen column density, to take into account two different emission mechanisms: a power law with photon index $\Gamma=2$ for AGN point-like sources and an APEC model (0.4 Solar abundance, $kT$=3 keV) for thermal emission from a diffuse source.
The results of our analysis are reported in Table~\ref{tab:pimms}.

\subsection{Extent and Shape of the X-ray Emission}
\label{sec:3.1}

After detecting an X-ray source, the extent and surface brightness distribution can provide key details.
For each X-ray detected source, we compared the radial distribution of the detected events with that expected from a point-like source.
We based such a comparison on the analytic PSF model for the \sw-XRT telescope, as derived by the XRT Calibration Team (\url{https://heasarc.gsfc.nasa.gov/docs/heasarc/caldb/swift/docs/xrt/SWIFT-XRT-CALDB-10_v01.pdf}), given by the sum of a Gaussian and a King law.
Since most of the events are found at $\sim1.5$~keV, where the XRT effective area peaks, we adopted the parameters corresponding to the on-axis PSF at this energy value. 
We also note that the dependence of the PSF profile on energy is mild (\citealp{2005SPIE.5898..360M}; \url{https://heasarc.gsfc.nasa.gov/docs/heasarc/caldb/swift/docs/xrt/SWIFT-XRT-CALDB-10_v01.pdf}).

\begin{figure*}
\centering
\includegraphics[width=10cm]{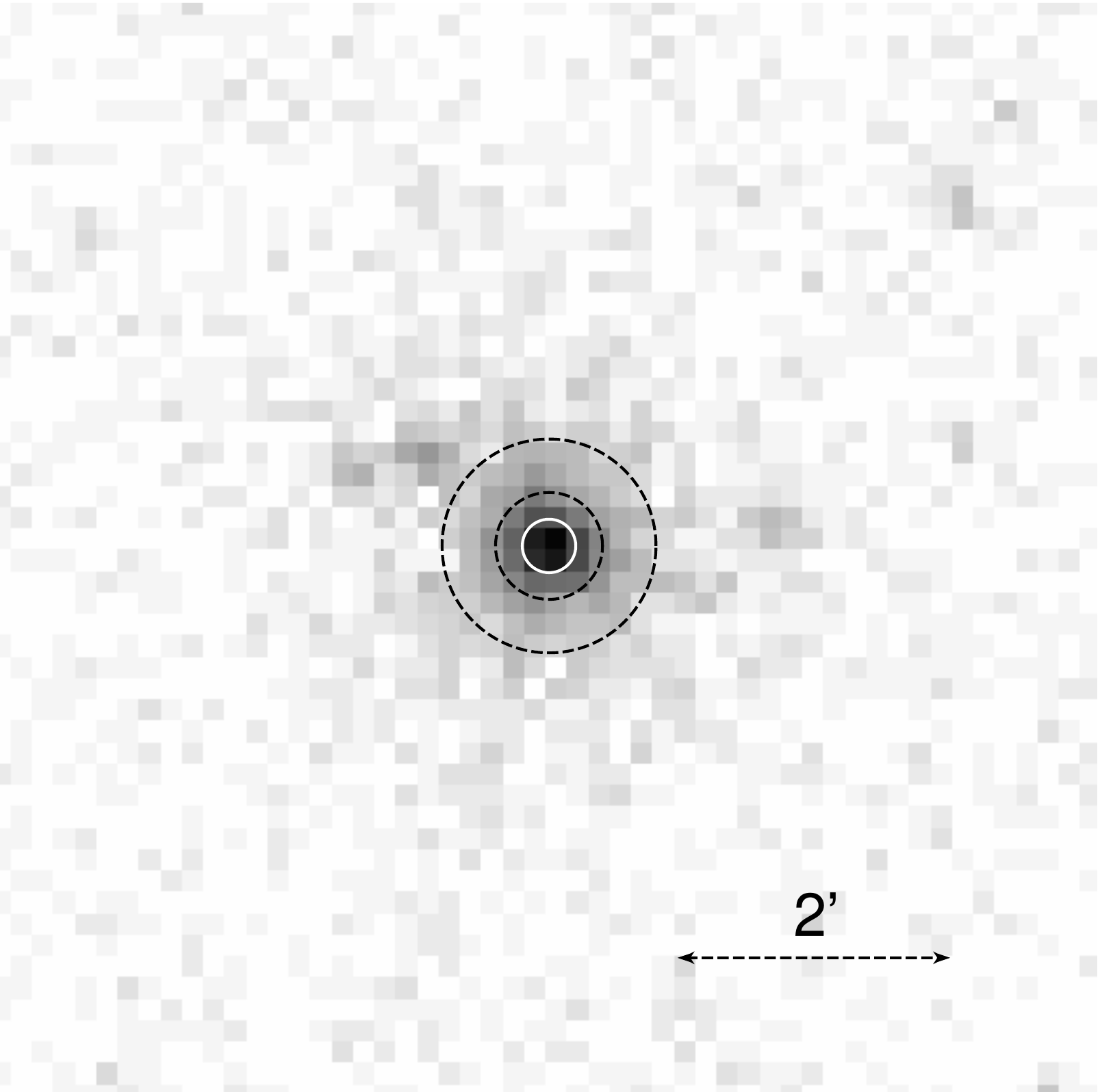} 
\caption{\sw~X-ray map (N is up, and E to the left) in the 0.3--10.0 keV band of MRK~876, a calibration point source with an exposure of 48795~s. The radius of the white circle $C$ is 5 pixels (1 pixel = $\sim2\arcsec.36$), while the inner and the outer radii of the black dashed annulus $A$ are 10 and 20 pixels, respectively. The map is binned by 4x4 pixels.}
\label{fig:mkn876}
\end{figure*}

Our analysis of source extent was carried out in two steps.
First, a ratio test, described in Section~\ref{sec:3.1.1}, is given by the comparison of the number of events in a circle $C$ and a surrounding annulus $A$ centered on the X-ray centroid; these regions are shown for example in Figure~\ref{fig:mkn876}.
Next, for those sources showing clear evidence of extent from the ratio test, we computed their radial profile, as described in Section~\ref{sec:3.1.2}, and compared it with the analytic \sw~PSF model.

To verify our procedure we used MRK~876, a Seyfert\,1 galaxy, that was used by the \sw~team as a point-like source for its first in-flight calibration of the PSF \citep{2005SPIE.5898..360M}\footnote{The document can be also retrieved at \url{https://heasarc.gsfc.nasa.gov/docs/heasarc/caldb/swift/docs/xrt/index.html}} and that has been repeatedly observed by XRT since its launch, achieving a remarkably deep exposure.
We note that Chandra observations, from 2016-2017, of MRK~876 are fully consistent with it being an unresolved source.  
First, MRK~876 shows strong variability (40\%) on time scales of months (see Fig.~8 of \citealp{2022MNRAS.515.3174B}).  
Second, we have merged the 12 Chandra observations and find no structure beyond one single source (14949 counts with 10 background in the energy band 0.6-3.0 keV). 
We find an encircled energy of 95\% (within 2\arcsec, in the energy band 0.6-3.0 keV), which agrees with that of the Chandra PSF calibration (Chandra POG\footnote{\url{https://cxc.cfa.harvard.edu/proposer/POG/}}, see Sections 4.2.3 and 6.6).

\begin{table*} 
\begin{center}
\scriptsize
\caption{Extent of the X-ray emission.}
\label{tab:er}
\begin{tabular}{l|cccc|cccc}
              &         \multicolumn{4}{c}{0.5--4 keV Band}           &     \multicolumn{4}{c}{0.3--10 keV Band}               \\  
\hline
Source Name   &     Background    & $C$  & $A$  &           $ER$      &     Background    & $C$  & $A$  &           $ER$      \\
              & (10$^{-2}$ ct/px) & (ct) & (ct) &                     & (10$^{-2}$ ct/px) & (ct) & (ct) &                     \\
~~~~~~~~(1)   &        (2)        & (3)  & (4)  &           (5)       &        (6)        & (7)  & (8)  &           (9)       \\  
\hline                                                                                                                                 
B0103$-$453   &       0.28        &   2  &  15  &      $0.13\pm0.13$  &       0.62        &   9  &  15  &      $0.60\pm0.35$  \\
B0202$-$765   &       0.17        & 142  &  25  &      $5.68\pm1.61$  &       0.34        & 174  &  27  &      $6.44\pm1.73$  \\   
B0242$-$514   &       0.26        &  13  &   4  &      $3.25\pm2.53$  &       0.58        &  17  &   7  &      $2.43\pm1.51$  \\
B0344$-$345   &       0.44        &  51  &  19  &      $2.68\pm0.99$  &       0.82        &  59  &  25  &      $2.36\pm0.78$  \\
B0427$-$366   &       0.22        &  22  &   8  &      $2.75\pm1.56$  &       0.38        &  25  &   8  &      $3.13\pm1.73$  \\   
B1017$-$426   &       0.21        &  34  &   4  &      $8.50\pm5.71$  &       0.36        &  42  &   8  &      $5.25\pm2.67$  \\
B1526$-$423   &       0.41        &  35  &  82  &      $0.43\pm0.12$  &       0.63        &  38  &  95  &      $0.40\pm0.11$  \\
B1817$-$640   &       0.67        &  48  &  14  &      $3.43\pm1.41$  &       1.22        &  66  &  27  &      $2.44\pm0.77$  \\   
B1953$-$425   &       0.26        &   4  &   4  &      $1.00\pm1.00$  &       0.45        &   4  &   3  &      $1.33\pm1.44$  \\
B2140$-$434   &       0.27        &  26  &   7  &      $3.71\pm2.13$  &       0.52        &  36  &  10  &      $3.60\pm1.74$  \\   
B2140$-$817   &       0.24        &   8  &   3  &      $2.67\pm2.48$  &       0.49        &   9  &   6  &      $1.50\pm1.11$  \\
\hline \hline                                              \multicolumn{9} {c} {Point-like Source Test}      \\ \hline
MRK~876 (90~s)    &       0.04        &   11 &   3  &      $3.67\pm3.22$  &       0.05        &   14 &   4  &      $3.50\pm2.69$  \\
MRK~876 (276~s)   &       0.02        &   31 &   6  &      $5.17\pm3.04$  &       0.02        &   34 &   8  &      $4.25\pm2.23$  \\
MRK~876 (1000~s)  &       0.03        &  130 &  26  &      $5.00\pm1.42$  &       0.07        &  178 &  36  &      $4.94\pm1.19$  \\
MRK~876 (48795~s) &       1.17        & 3559 & 585  &      $6.08\pm0.35$  &       2.10        & 4831 & 806  &      $5.99\pm0.30$  \\ 
\hline                                              
\end{tabular}
\end{center}
\tablecomments{The columns show (1) the MRC name of the SMS4 source; (2) and (6) the measured number of background counts, per pixel, where 1 pixel = 2.36\arcsec~on a side; (3) and (7) the background-corrected number of counts $C$ within a circle with a radius of 5 pixels; (4) and (8) the background-corrected number of counts $A$ within an annulus with inner and outer radii of 10 and 20 pixels, respectively; (5) and (9) the extent ratio $ER$, with its 1$\sigma$ uncertainty. Results for the point-like source MRK~876, using observations with different exposures (reported in parentheses), are shown for comparison. The significance of source extent can be evaluated by comparing the measured $ER$ value, columns 5 and 9, to that for the model PSF that yields $ER=5.73$ for an unresolved source.} 
\end{table*}

\subsubsection{Ratio Test for Source Extent} 
\label{sec:3.1.1}

Following \am, we used as extraction regions for our analysis (see Figure~\ref{fig:mkn876}) a circle $C$ with a radius of 5~pixels ($\sim12\arcsec$) for the source and an annulus $A$ with inner and outer radii of 10 and 20 pixels, respectively, for the background.
Both regions are centered at the coordinates of each X-ray centroid, given in Table~\ref{tab:det}.
We verified that the annulus did not include any spurious X-ray sources.
Unlike Paper I, which used only the broad band (0.3-10.0 keV), we also carried out our analysis in the 0.5--4 keV band to focus more on thermal emission, appropriate to groups and clusters.  
After extracting the counts from the circles $C$ and the annuli $A$, we corrected them for the background contribution, computed as reported in Section~\ref{sec:4}.

The fraction $ER=C/A$ expected from a point-like source, computed from the analytic PSF model, is equal to 5.73.
Thus, a deviation from this value reveals that the distribution of events differs from that expected from a point-like source, making $ER$ a marker of extended X-ray emission. 
The $ER$ values of the X-ray detected sources, for both the 0.5--4~keV and the 0.3--10~keV bands, are reported in Table~\ref{tab:er}.

According to the results in Table~\ref{tab:er}, B0103$-$453 and B1526$-$423 show clear evidence of extended emission, with $ER$ values lower than 5.73 at very high statistical significance; for B1526$-$423, $ER < 5.73$ by more than 40$\sigma$ in both considered energy bands and for B0103$-$453 by more than 30$\sigma$ (in the narrow band), based in the difference between the measured and expected values of $ER$.
Extended emission is also suggested for B0344$-$345, although with lower statistical significance, but higher than $3\sigma$ in both energy bands. 
The extent for B1953$-$425 is based on a very low number of counts.
The remaining sources are consistent with being unresolved.

For comparison, we applied the same method to MRK~876, having a mean count rate of $\sim0.2$ ct s$^{-1}$ \citep{2005SPIE.5898..360M}.
Among the more than three hundred observations present in the \sw~archive, we searched for a few observations with adequate exposures to obtain a net number of counts comparable with what we found for most of our sources.
Furthermore, we also stacked the two observations (obsID~00050300004 and obsID~00050302001) with the longest exposures, for a total of 48795~s.
As shown in Table~\ref{tab:er}, starting from the observation with the shortest exposure (obsID~00095146191, 90~s) and moving to higher exposures (obsID~00095724014, 276~s; obsID~00095724031, 1000~s), to conclude with the stack of the two longest exposures, the uncertainty range of the extent ratio gradually shrinks around the expected value $ER = 5.73$, for both the 0.5--4 keV and the 0.3--10 keV bands.

\subsubsection{Radial Profile Analysis of Extended Sources}
\label{sec:3.1.2}

To characterize in greater detail the soft, diffuse X-ray emission, possibly extending to angular distances larger than $\approx$1\arcmin~(see Section~\ref{sec:3.1.1}) from the X-ray centroid, we computed the X-ray radial profiles for sources with $ER < 5.73$ by more than $\simeq 3\sigma$ -  B0103$-$453, B0344$-$345, B1526$-$423 - with sufficient S/N.   
For comparison, we did the same for B1017$-$426, one of two point-like sources in our sample, as well as for MRK~876.

\begin{figure*}
\gridline{\fig{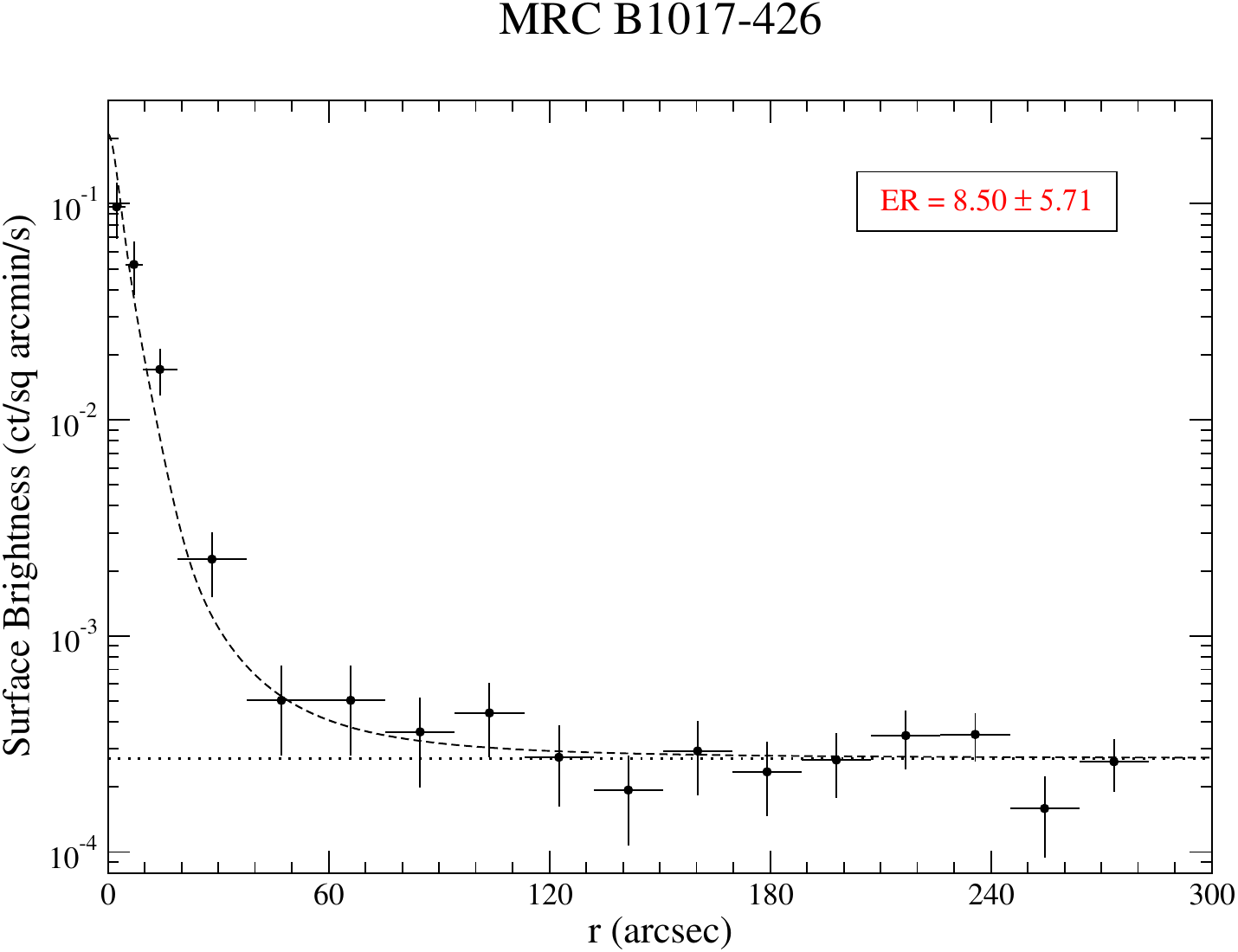}{0.45\textwidth}{($z$=1.28; 1$\arcmin$ = 0.50 Mpc)} \fig{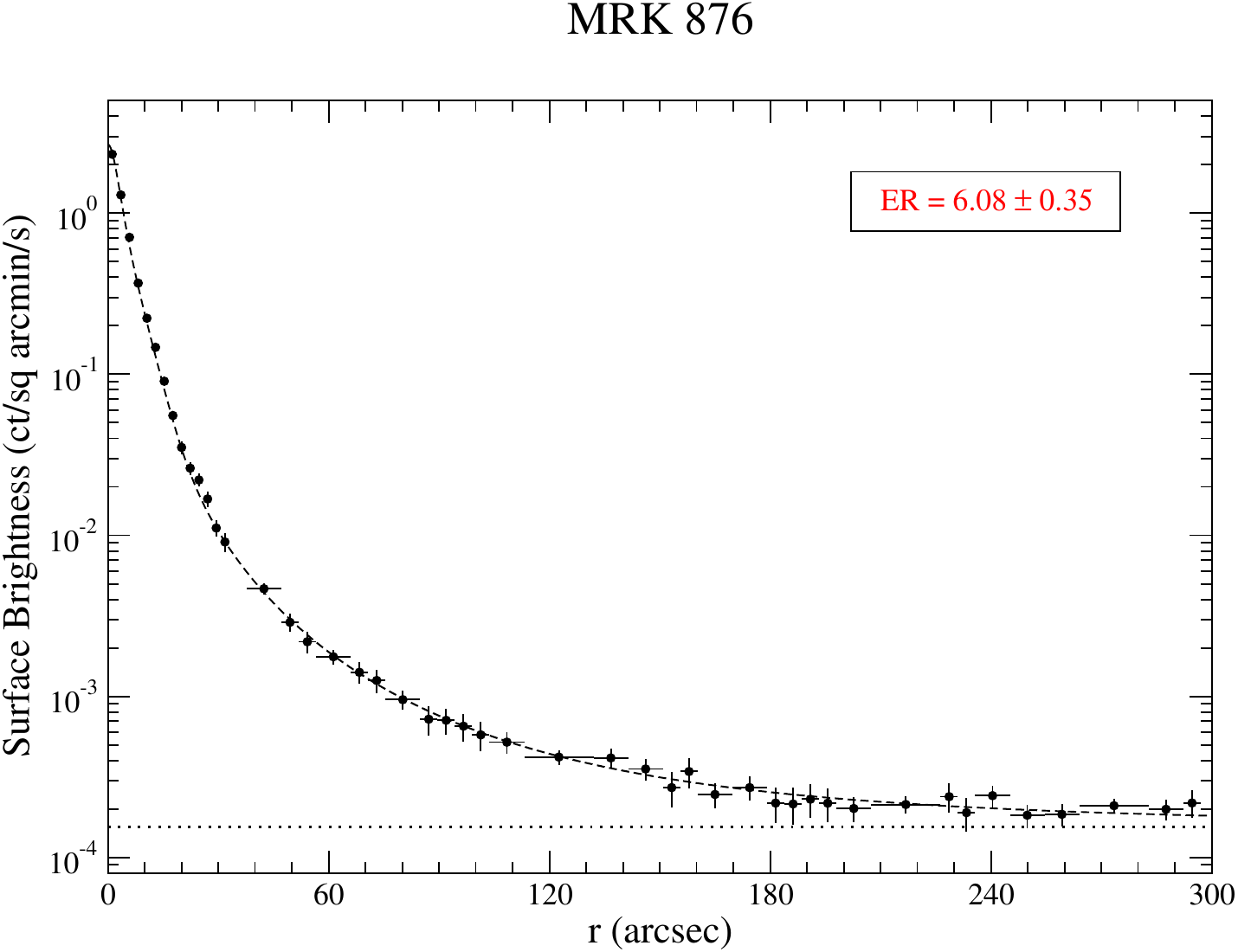}{0.45\textwidth}{($z$=0.12019; 1$\arcmin$ = 0.13 Mpc)} }
\caption{Radial profile in the 0.5--4 keV band of a point-like source in our sample, compared with MRK~876. For each source, the $ER$ value is given, together with the spectroscopic redshift $z$ and the corresponding scale factor. The dotted horizontal line represents the mean background level computed over the whole image, while the dashed line corresponds to the fit of data with the analytic model of the PSF. As discussed in the text, the radial profiles for both sources are consistent with the \sw~PSF model.}
\label{fig:radprofpoint}
\end{figure*}

\begin{figure*}
\gridline{\fig{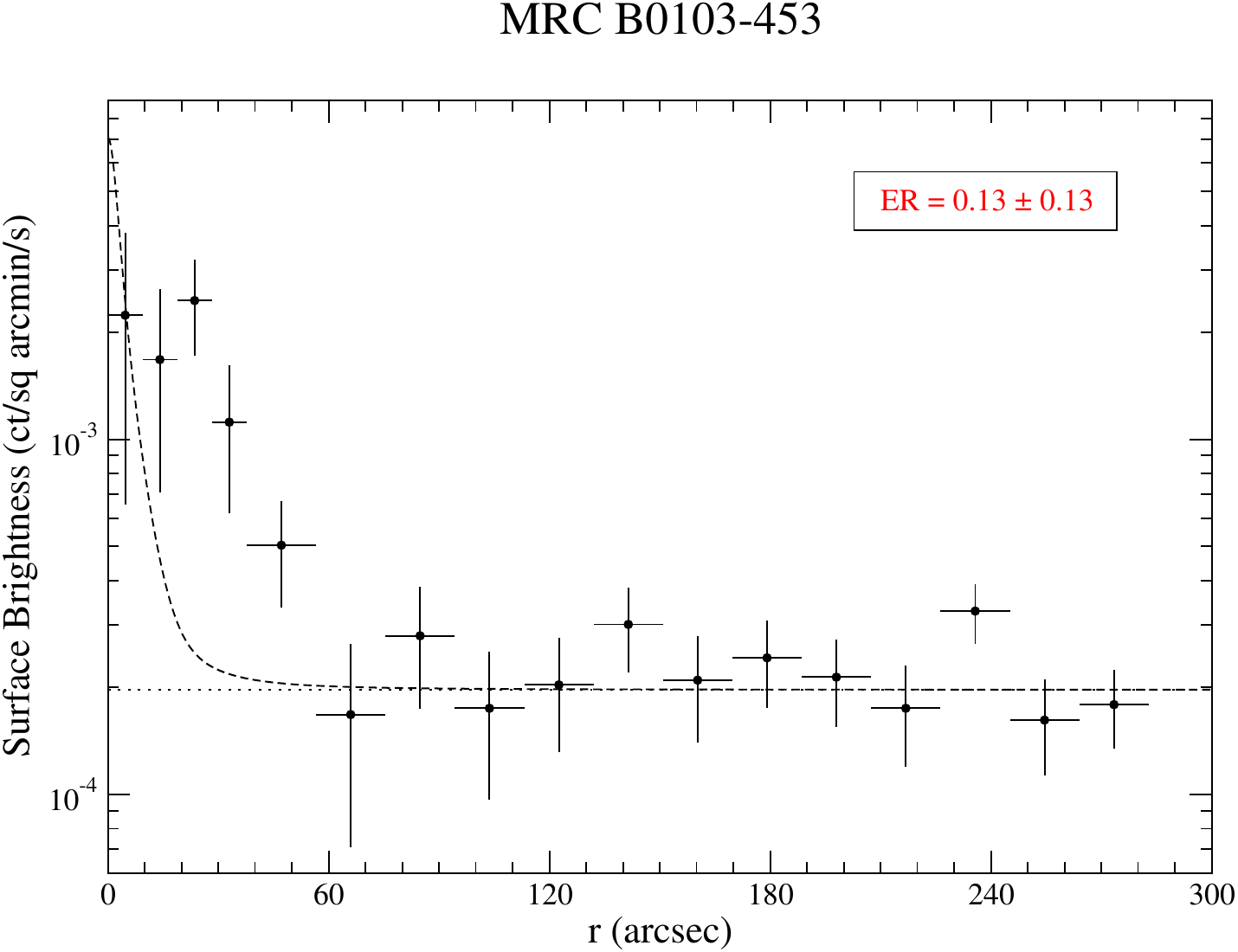}{0.45\textwidth}{($z$=0.71; 1$\arcmin$ = 0.43 Mpc)} \fig{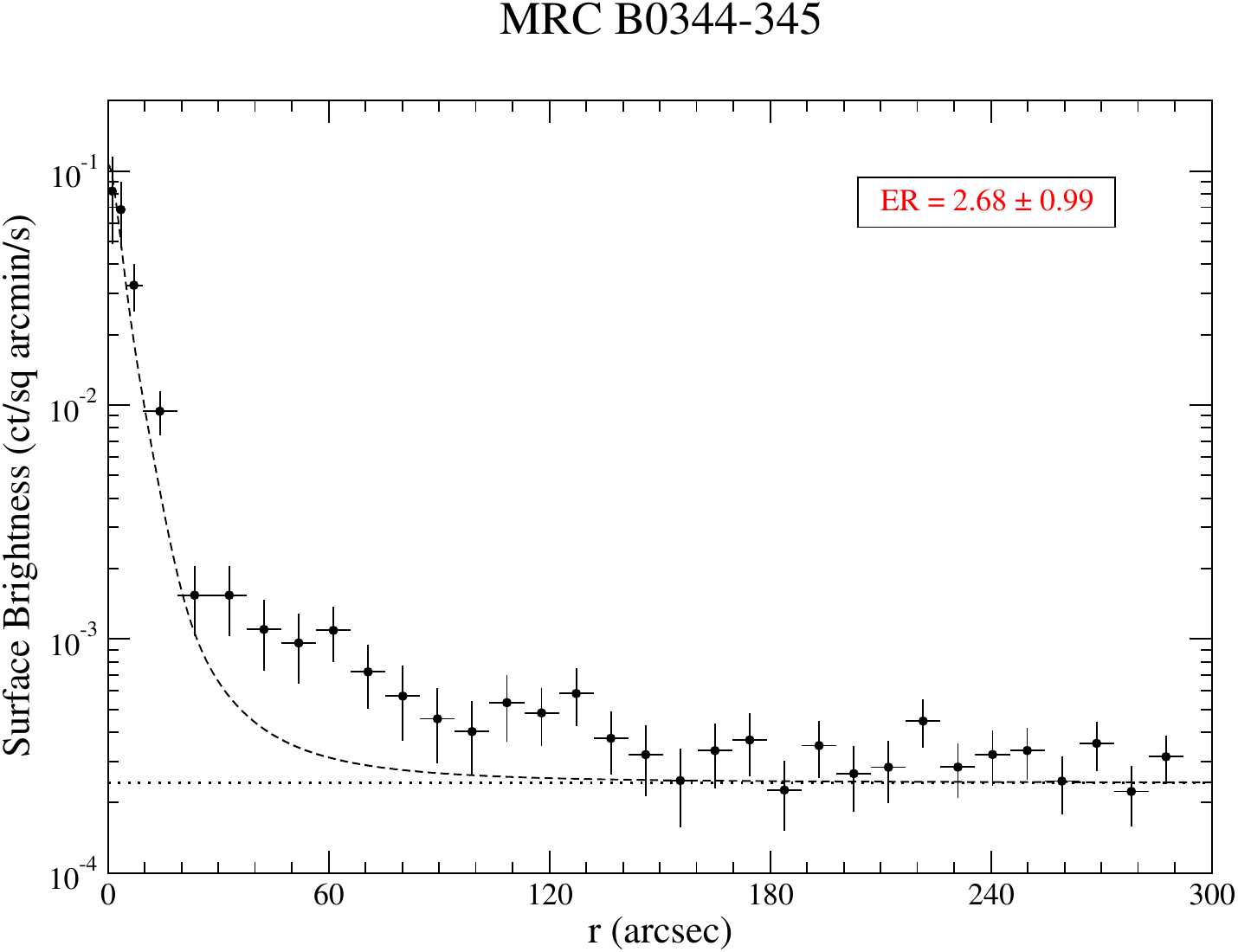}{0.45\textwidth}{($z$=0.0538; 1$\arcmin$ = 61.2 kpc)}}
\gridline{\fig{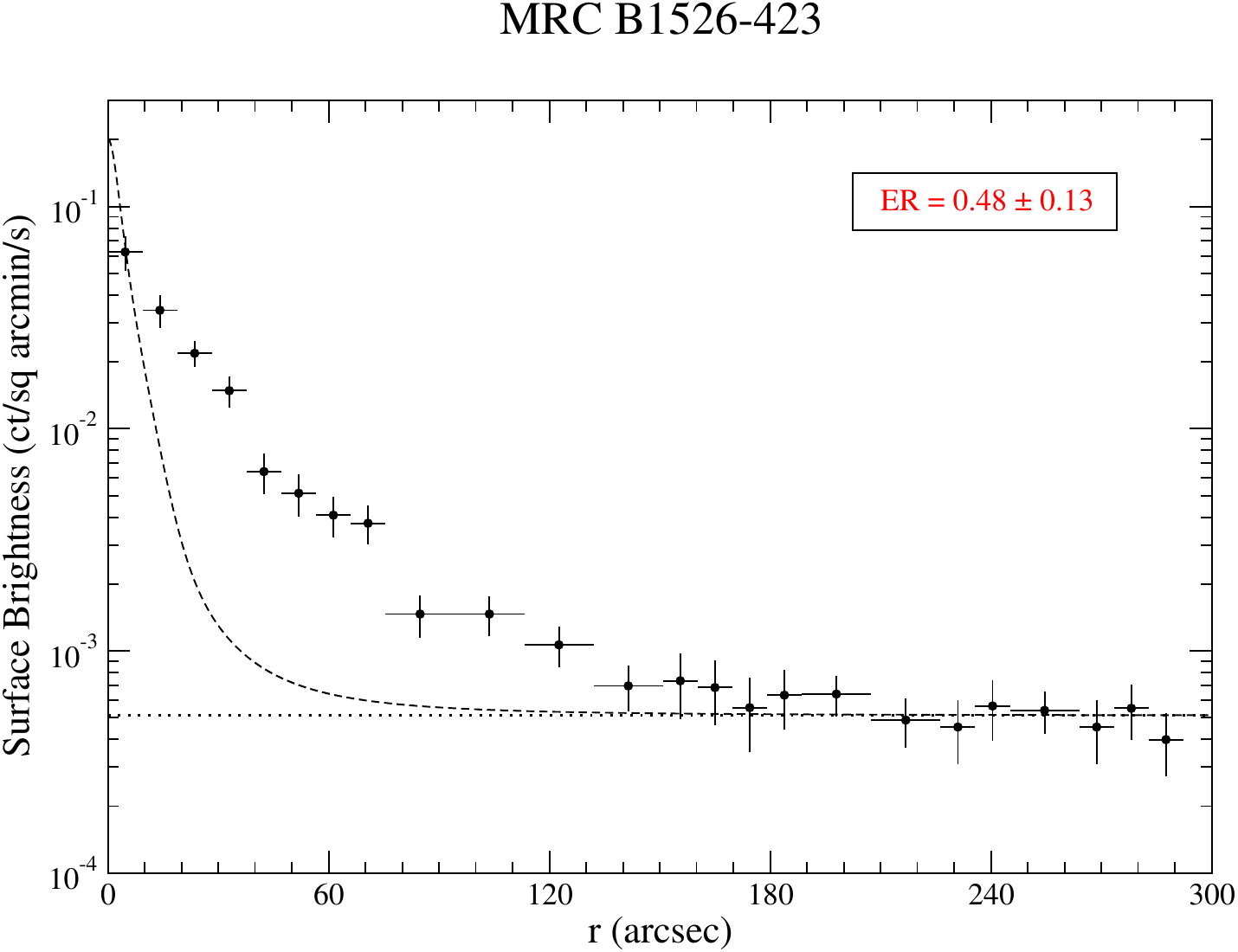}{0.45\textwidth}{($z$=0.50; 1$\arcmin$ = 0.36 Mpc)}}
\caption{Radial profiles in the 0.5--4 keV band of extended SMS4 sources in our sample. For each source, the $ER$ value is given. For all sources, the redshift $z$ is given with the corresponding scale factor. The redshift is spectroscopic for B0344$-$345, and is photometric for the other two sources. The dotted horizontal line represents the mean background level computed over the whole image, while the dashed line corresponds to the fit of data with the analytic model of the PSF. As is apparent, the radial profiles confirm the extents for these three sources.}
\label{fig:radprofext}
\end{figure*}

To compute the X-ray radial profile, we applied {\sc ximage} to the stacked event file in the 0.5--4 keV band and used the {\sc psf} command, taking into account the corresponding exposure map and performing the computation up to 5\arcmin~from the coordinates of the X-ray centroid.
Before using the {\sc psf} command, we searched for spurious point-like sources within 5\arcmin~using the {\sc detect} command and leaving the requested $S/N$ ratio set to the default value $S/N$=2. 
We removed these spurious sources using the {\sc remove\_sources} command, taking into account the exposure map and replacing for each pixel the detected number of counts with the mean background value computed over the whole detector.
For the {\sc psf} command we did not correct data for background, setting its value to zero.
The data were then fitted with the analytic model representing the PSF of the XRT telescope previously described, including at this step a constant value equal to the mean background computed over the whole image.

Before describing results for the extended sources, we show, in Figure~\ref{fig:radprofpoint}, radial profiles for two point-like sources, B1017$-$426 and MRK~876, to fully validate the detailed shape of the \sw~PSF for comparison with the extended sources in the discussion below.
For MRK~876, we removed seven unrelated detections before building its radial profile with {\sc XIMAGE}, while no unrelated sources were found around B1017$-$426.
For each panel, a dotted horizontal line represents the mean background level computed over the whole image, while a dashed line corresponds to the fit of data with the analytic model of the PSF.
As shown in Fig.~\ref{fig:radprofpoint}, the radial profiles obtained for B1017$-$426 and MRK~876 agree with that expected from a point-like source ($\chi^2=17.0$ for $\nu=16$ degrees of freedom for B1017$-$426, and $\chi^2=31.7$ for $\nu=46$ degrees of freedom for MRK~876). 
Hence, we have validated the shape of the independently derived \sw~PSF and can confidently compare this analytic shape to the radial profiles of the sources in our sample.

The radial profiles of the three extended sources in our sample, given in Figure~\ref{fig:radprofext}, show both the X-ray extent and relationship to a point source.
Before building these profiles, we removed three unrelated sources for B0344$-$345, and one unrelated source for both B1526$-$423 and B1953$-$425. 
The X-ray surface brightness of B0103$-$453 appears to be constant up to 30\arcsec~and then decreases to the background level at $\sim1\arcmin$.  
The radial profiles of B0344$-$345 and B1526$-$423 show X-ray emission extending to  $\sim2$\arcmin, but B0344$-$345 shows a sharp drop within $\sim30\arcsec$ and then a steady decline, while for B1526$-$423, the decline is much smoother.
A possible interpretation for these features is that the emission of B0344$-$345 might be due to the combined contribution from a brighter compact object (the core of the radio galaxy) surrounded by more tenuous emission.
For B1526$-$423, the more uniform profile, lacking a strong central peak, suggests a relatively strong diffuse X-ray component with comparatively weak X-ray emission from any central AGN.

\begin{figure*}[thbp]
\centerline{
\includegraphics[width=0.9\textwidth]{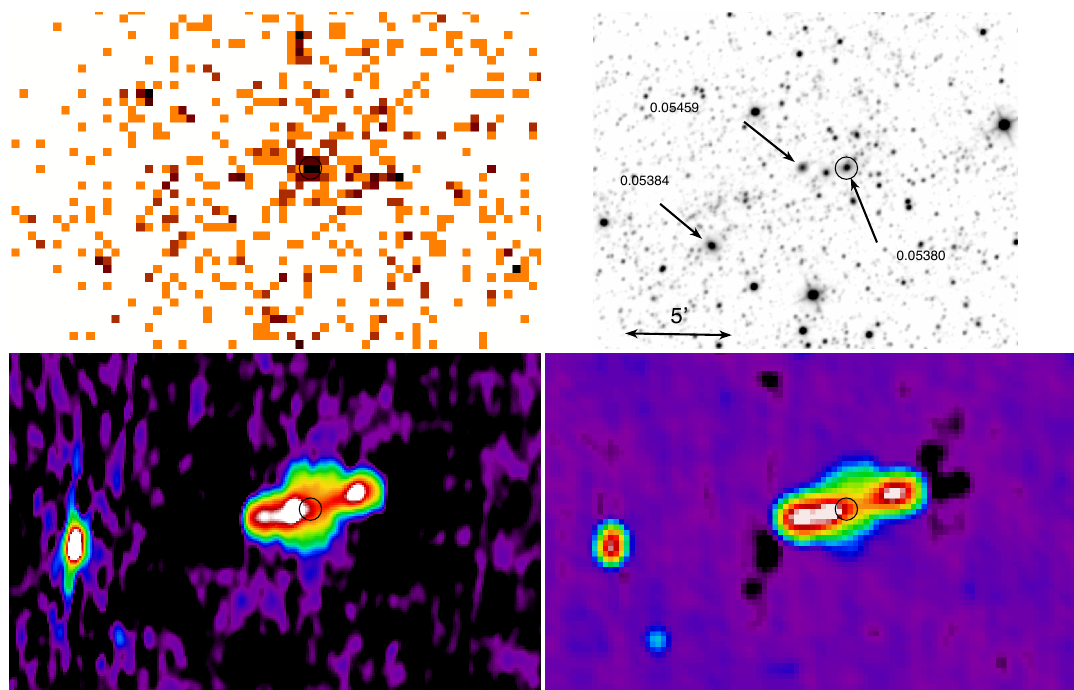}
}
\caption{Images of B0344$-$345 (matched in scale; N is up and E to the left), (a: top left) the 11 ks \sw~image (1-4 keV); (b: top right) WISE $3.4~\mu$m image; (c: bottom left) GMRT survey field (153~MHz) observation; (d: bottom right) NVSS (1.4 GHz) image. The \sw~data show emission from a compact source, presumably an AGN, as well as diffuse emission, extending about 2\arcmin. In the IR image, the three galaxies with measured redshifts (and their redshift values) are indicated. The X-ray source ID is indicated with a black circle and is included in all four images. 
}          
\label{fig:0344}
\end{figure*}

\begin{figure*}
\gridline{\fig{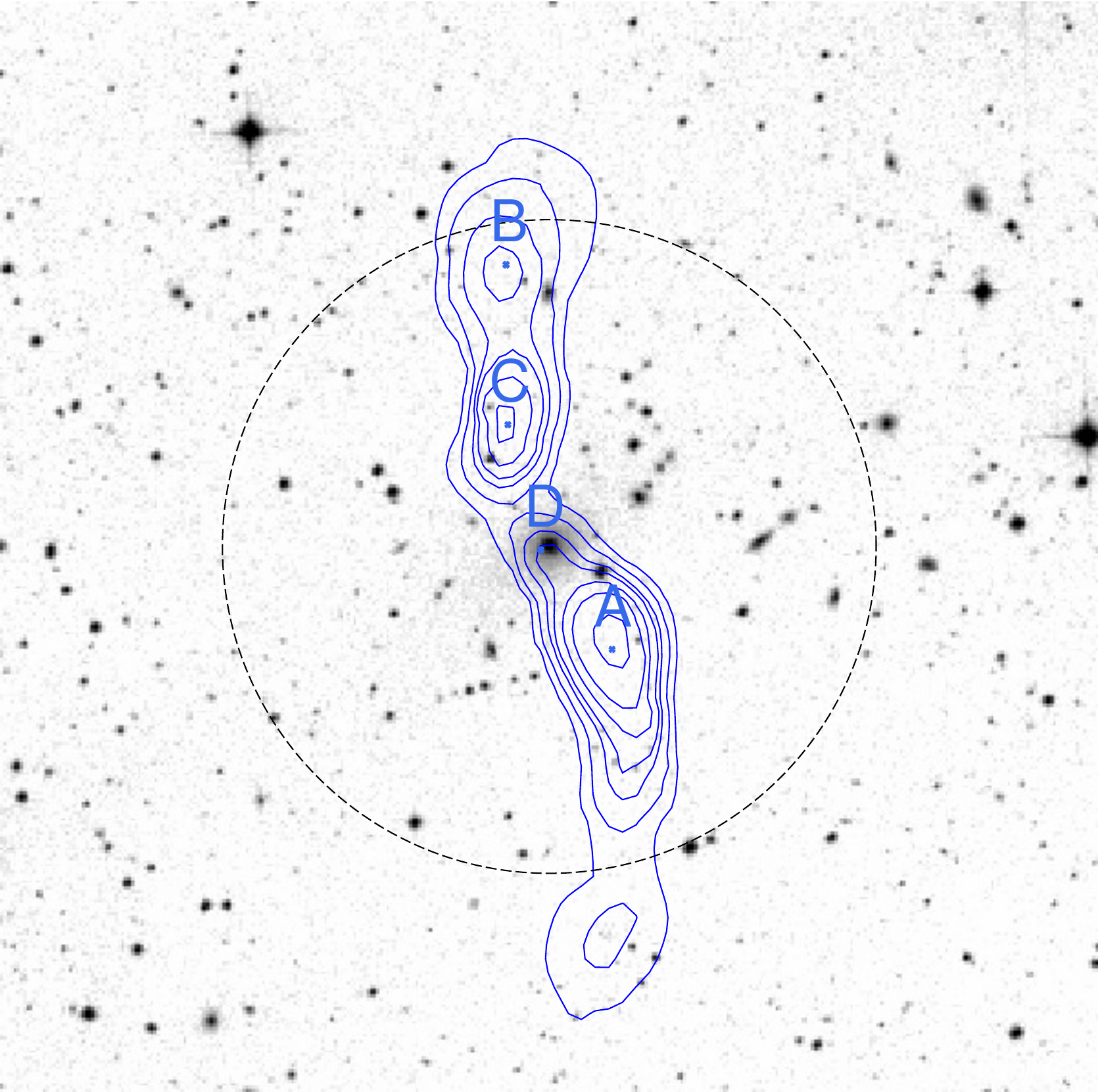}{0.26\textwidth}{} \fig{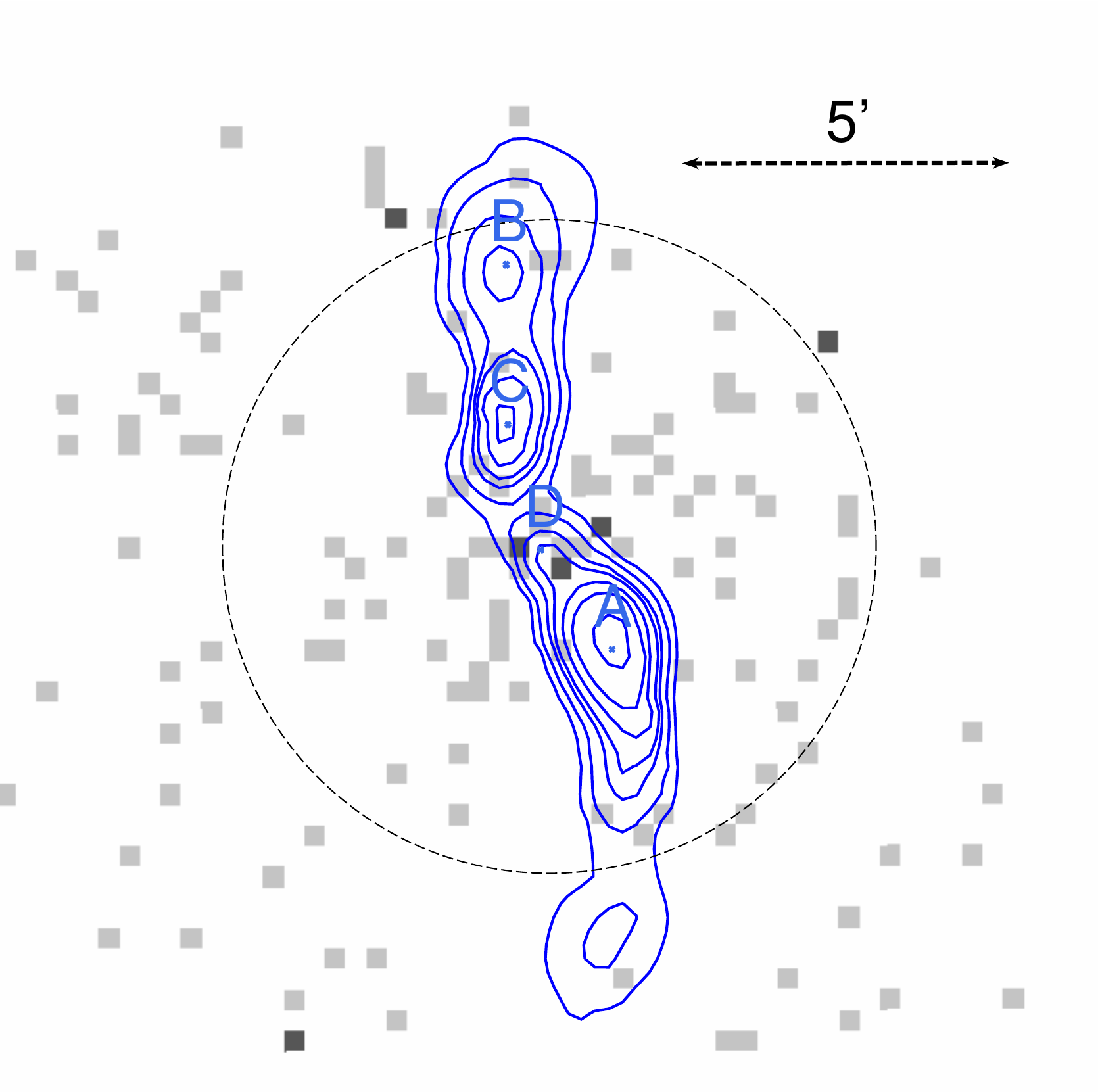}{0.26\textwidth}{} \fig{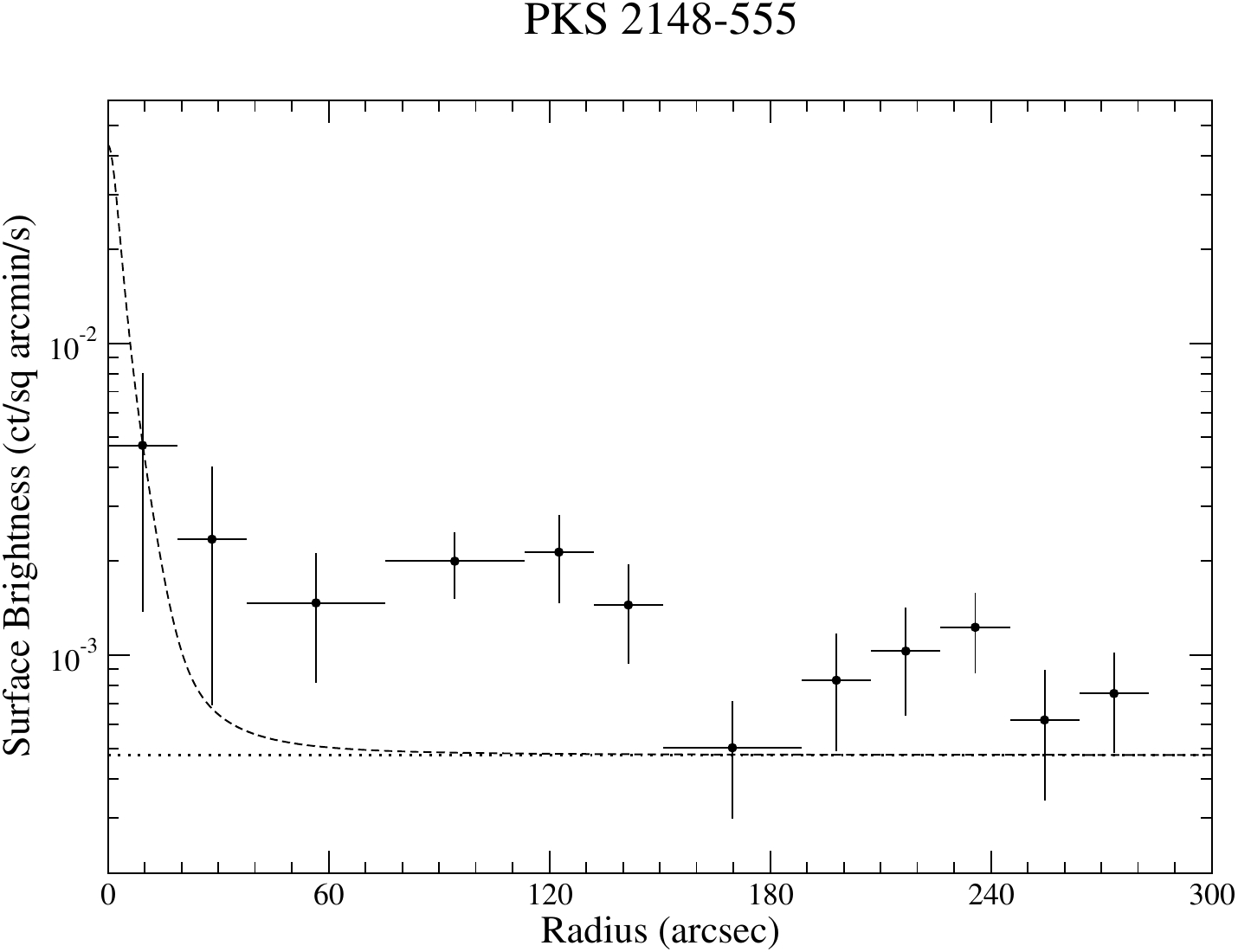}{0.35\textwidth}{($z$=0.038783; 1$\arcmin$ = 44.9 kpc)}}
\caption{The extended X-ray source PKS~2148$-$555 (first presented in \am). Optical map from DSS2 (left), 0.5--4 keV X-ray map, binned by 8x8 pixels, from \sw-XRT (center), and the corresponding 0.5--4 keV radial profile (right). In both the optical and the X-ray maps, N is up and E is to the left. The radial profile is computed up to 5\arcmin~from the X-ray centroid, corresponding to the dashed circles (note 1\arcmin = 44.9~kpc). The dotted horizontal line represents the mean background level computed over the whole image, while the dashed line corresponds to the fit of data with the analytic model of the PSF.}
\label{fig:pks2148}
\end{figure*}

Of the three just described extended sources, the most interesting is B0344$-$345 with its large X-ray extent; the second large extended X-ray source B1526$-$423 is difficult to study, since it lies close to the Galactic plane. 
The \sw, WISE, GMRT, and NVSS images of B0344$-$345 are shown in Fig.~\ref{fig:0344}. 
The \sw~image shows emission extending more than 2\arcmin~(about 120~kpc) around the central peaked X-ray emission (see also Fig.~\ref{fig:radprofext} for the radial profile; the mass of the associated SMBH is estimated to be $M_{BH} \approx 2.3 \times 10^8$ \msun~by \citealp{2003A&A...399..869B}). 
As Fig.~\ref{fig:0344} shows, the radio emission is primarily extended E-W with two hot spots as well as a bright knot on the eastern side. 
The optical image (Fig.~\ref{fig:0344}b) shows the locations of three galaxies with measured redshifts (indicated in the figure) with a full range of less than 250~\kms.

According to the results reported in Table~\ref{tab:pimms}, for B0344$-$345, the total X-ray luminosity (the central peak and the extended emission) is about $2\times10^{42}$~\ergss, typical of a galaxy group. 
Using the correlation of X-ray luminosity with temperature for thermal gas in dark matter potentials, from groups to clusters (e.g., see \citealp{2016MNRAS.463..820Z} and \citealp{2020ApJ...892..102L}), for a luminosity of $2\times10^{42}$~\ergss, we would expect a cool temperature, $\sim 1$ keV (as noted above, there were too few source counts to fit a detailed spectrum).
For such low temperature systems, poor groups or individual galaxies, with shallow potentials, compared to rich clusters, the effects of ``radio mode" feedback could be significant (for discussion of radio mode feedback, in simulations as well as from galaxies to clusters, see, e.g.,  \citealp{2006MNRAS.365...11C,2019SSRv..215....5W,2009AIPC.1201..198N,2022hxga.book....5H,2002MNRAS.332..729C}).

Two graphic examples of the influence of radio mode feedback are M~87 (e.g., \citealp{2007ApJ...665.1057F}) and Fornax A, aka NGC~1316 (see \citealp{2003ApJ...586..826K}; in the outskirts of the Fornax cluster) whose thermal gas atmospheres range from 2~keV (for M~87) to~0.5~keV (for Fornax~A). 
Fornax~A, with its X-ray bright central AGN and thermal extended emission \citep{2010ApJ...721.1702L} is an apt comparison to B0344$-$345. 
The AGN in Fornax~A has undergone a $\sim5\times10^{58}$ erg outburst (\citealp{2010ApJ...721.1702L}; see also \citealp{2021A&A...656A..45M} who detected outflowing gas at velocities of 2000\kms) that has strongly perturbed its hot, X-ray emitting atmosphere.
Hence, deep X-ray observations of the extended, low luminosity emission around B0344$-$345 could provide interesting constraints on the SMBH activity and the impact of any outbursts on the surrounding thermal gas.

In \am, where we had not presented X-ray radial profiles, one of the most interesting extended sources was PKS~2148$-$555 (see Figure~\ref{fig:pks2148}), and we expand its analysis here. 
In Fig.~\ref{fig:pks2148}, we provide the source X-ray radial profile that shows emission extending to at least $\sim3$\arcmin~ ($\sim100$~kpc; and possibly farther). 
PKS~2148$-$555 is the BCG of the galaxy cluster A3816 (\citealp{1991AJ....102.1917B} and \citealp{2002MNRAS.331..717L}) which is itself a member of a supercluster of Abell clusters, along with A3862 and A3869 (\citealp{2001AJ....122.2222E}). 
PKS~2148$-$555 is an FR~I, as one might expect for a system in a rich environment (\citealp{2011AJ....141...88W}; see also \citealp{2015A&A...576A.101M,2018MNRAS.478.3848M,2019MNRAS.489.2488K}). 
The PKS~2148$-$555 jets exhibit numerous knots and gradual curvature as the jet traverses the ICM (see Figs. 5, 6, 9 in \citealp{2002MNRAS.331..717L}).  
As the cluster BCG and the extent of the radio jets, one should expect interesting interactions of the jet with the ambient X-ray emitting ICM in deeper X-ray images.

\subsection{Hardness Ratio and Spectral Analysis} 
\label{sec:3.2}

We used the {\sc counts} command within {\sc ximage} to extract the counts within circles with a radius of 10~pixels (23\arcsec.6), centered at the position of our X-ray detections.
As adopted in \am, we used the two contiguous 0.3--3 keV and 3--10 keV energy bands to distinguish soft ($S$) and hard ($H$) X-ray photons.
Then, we computed the hardness ratio as $HR = (H - S) / (H + S)$.
As shown in Table~\ref{tab:hr}, the only source that shows some evidence of hard emission, despite a considerable uncertainty, is B0103$-$453.
For most of the remaining sources, excluding B1817$-$640 and B2140$-$434, $HR\apprle-0.25$.

\begin{table*} 
\begin{center}
\scriptsize
\caption{Hardness ratio of the X-ray detections.}
\label{tab:hr}
\begin{tabular}{cccc}
\hline
SMS4 Name   & $S$ (ct)  & $H$ (ct) &   $HR$      \\
   (1)      &   (2)     &  (3)     &      (4)    \\  
\hline                                                                                           
B0103$-$453 &     8     &   14  &~~$0.27\pm0.36$ \\
B0202$-$765 &   216     &   29  & $-0.68\pm0.04$ \\
B0242$-$514 &    18     &   10  & $-0.29\pm0.21$ \\
B0344$-$345 &    59     &   23  & $-0.44\pm0.10$ \\
B0427$-$366 &    27     &    7  & $-0.59\pm0.13$ \\
B1017$-$426 &    50     &    7  & $-0.75\pm0.07$ \\
B1526$-$423 &    91     &   22  & $-0.61\pm0.07$ \\
B1817$-$640 &    64     &   45  & $-0.17\pm0.12$ \\
B1953$-$425 &     5     &    3  & $-0.25\pm0.41$ \\
B2140$-$434 &    34     &   24  & $-0.17\pm0.16$ \\
B2140$-$817 &    11     &    6  & $-0.29\pm0.27$ \\
\hline
\end{tabular}
\end{center}
\tablecomments{The columns show (1) the MRC name of the SMS4 source; (2) the measured number of counts $S$ in the soft (0.3--3 keV) band within a circle of 10~pixels (23\arcsec.6) centered at the coordinates of the X-ray centroid, reported in Table~\ref{tab:det}; (3) the measured number of counts $H$ in the hard (3--10 keV) band within the same circle; (4) the hardness ratio $HR$, with its 1$\sigma$ uncertainty.}
\end{table*}

For three sources (B0202$-$765, B1526$-$423, B1817$-$640) we had sufficient counts to perform a full spectral analysis with {\sc xspec}.  
We note that efforts to separate a compact source from surrounding thermal emission cannot be fully successful, due to scattered flux by the \sw-XRT optics.  
Hence, caution is required in evaluating the spectral fits.
For B0344$-$345, we searched within a circle of 10~pixels (23\arcsec.6), that covers $\sim$80\% of the XRT PSF, to focus on the core of the radio galaxy and avoid contributions from the diffuse emission shown in Figure~\ref{fig:radprofext}: despite considerable \sw~exposure on this source, the modest count rate led to obtain only 81 counts, a value that is too low to expect precise constraints.

\begin{itemize}
\item {\bf B0202$-$765} For this point-like source, we used the {\sc xrtproducts} task to extract source events from a circle with a radius of 20 pixels ($\sim47\arcsec$), thus covering $\sim$90\% of the XRT PSF, centered at the X-ray source coordinates, and background events from a circle with a radius of 50~pixels ($\sim2\arcmin$), to avoid spurious sources.
The number of events in the spectrum files corrected for the background contribution is 264, and we grouped the spectral file requiring at least 20 events per bin.

We fitted the spectrum with an absorbed power law, taking into account the spectroscopic redshift and fixing the hydrogen column density to the corresponding Galactic value (\citealp{2016A&A...594A.116H}).
We obtained $\Gamma=1.96\pm0.11$ and $S_{~0.3-10,unabs}=2.95^{+0.29}_{-0.20}\times10^{-12}$ erg~cm$^{-2}$~s$^{-1}$, with $\chi^2=12.05$ (11 d.o.f.).
Uncertainties on both $\Gamma$ and $S_{~0.3-10,unabs}$ are given at a 1$\sigma$ confidence level.
Leaving the $n_H$ parameter free to vary, we found no evidence of intrinsic absorption.

\item {\bf B1817$-$640} Similarly to B0202$-$765, also for B1817$-$640 we used circles with radii of $\sim$47\arcsec~and $\sim$2\arcmin~to extract events for the source and the background, respectively.
We grouped the spectral file, with 141 events corrected for the background contribution, requiring at least 20 events per bin.

Fitting the spectrum with an absorbed power law, taking into account the spectroscopic redshift and fixing the Hydrogen column density to the corresponding Galactic value, we obtained a very poor fit ($\chi^2=21.5$, with 5 d.o.f.).
However, leaving the $n_H$ parameter free, we obtained a significant improvement in the $\chi^2$ value ($\chi^2=7.19$, with 4 d.o.f.) with an increase in the Hydrogen column density value $n_H=5.33^{+2.56}_{-2.00}\times10^{22}$ cm$^{-2}$, in excess of the Galactic absorption $n_{H,Gal}=6.79\times10^{20}$ cm$^{-2}$.
Moreover, we were able to find $\Gamma=1.67^{+0.57}_{-0.50}$ and $S_{~0.3-10,unabs}=1.56^{+0.03}_{-0.54}\times10^{-12}$ erg~cm$^{-2}$~s$^{-1}$.
All the uncertainties are given at a 1$\sigma$ confidence level.

\item {\bf B1526$-$423} For this extended source (see Fig.~\ref{fig:radprofext}), we used for the source extraction region a circle with a radius of 2\arcmin, and another circle with a radius of 3.1\arcmin~for the background.
Also in this case, we used the {\sc xrtproducts} task to extract the products for the spectral analysis, this time specifying the {\it extended=yes} option in the task.
The number of the spectral counts obtained was 371, and we grouped the data requiring at least 30 events per bin.

We fitted the spectrum with an absorbed APEC model (\citealp{2001ApJ...556L..91S}), fixing the hydrogen column density to the Galactic value, the metal abundance at 0.4 Solar, and using the photometric redshift $z$=0.50. 
We obtained a plasma temperature $kT=10.7^{+5.6}_{-2.8}$~keV and an unabsorbed flux $S_{~0.3-10,unabs}=2.99^{+0.16}_{-0.27}\times10^{-12}$ erg cm$^{-2}$ s$^{-1}$, implying a luminosity $L_{~0.3-10}=2.62^{+0.15}_{-0.20}\times10^{45}$ erg s$^{-1}$, with $\chi^2=11.02$ (10~d.o.f).
The uncertainties on $kT$, flux, and luminosity are given at a 1$\sigma$ confidence level.
The larger region used for the spectral analysis of B1526$-$423 yielded a flux more than three times larger than the value reported in Table~\ref{tab:pimms}, derived for a point-like source using WebPIMMS.
\end{itemize}


\subsection{Comparison with other X-ray Catalogs} 
\label{sec:3.3}

The Living \sw-XRT Point Source catalog (LSXPS, \citealp{2023MNRAS.518..174E}) is a dynamic catalog that is updated almost in real time as soon as new \sw~observations are archived.
In addition to the whole catalog, two subsets ({\it clean} and {\it ultra-clean}) are found in the LSXPS. 
In January 2024, data from the first eROSITA All-Sky Survey (eRASS1) were released \citep{2024arXiv240117274M}.
Three catalogs ({\it Main, Hard, and Supplementary}) are found in the First Data Release (DR1).
Here we compare the results of our \sw~campaign with the LSXPS and the eROSITA DR1 catalogs.

\begin{table*} 
\begin{center}
\scriptsize
\caption{Positional match with other X-ray catalogs.}
\label{tab:xcat}
\begin{tabular}{c|ccc|cc|cc}
       \multicolumn{4}{c}{}                  &  \multicolumn{2}{c}{LSXPS}  &  \multicolumn{2}{c}{DR1} \\   
\hline
SMS4 Source & $r_{c}$  & $r_{l}$  & $r_{e}$  & $d_{cl}$ & $r_{cl}$ &       $d_{ce}$       & $r_{ce}$ \\
            & (arcsec) & (arcsec) & (arcsec) & (arcsec) & (arcsec) &       (arcsec)       & (arcsec) \\
   (1)      &   (2)    &   (3)    &   (4)    &   (5)    &   (6)    &         (7)          &   (8)    \\  
\hline                                                                                                                                 
B0103$-$453 &   4.7    &   5.5    &   ...    &   2.2    &   7.2    &         ...          &   ...    \\
B0202$-$765 &   3.8    &   3.7    &   1.5    &   0.8    &   5.3    &         1.3          &   4.1    \\
B0242$-$514 &   4.8    &   6.6    &   8.5    &   3.3    &   8.2    &         5.0          &   9.9    \\
B0344$-$345 &   4.0    &   2.6    &   5.2    &   1.2    &   4.8    &         2.4          &   6.6    \\
B0427$-$366 &   4.8    &   4.7    &   5.6    &   2.4    &   6.7    &         5.4          &   7.4    \\
B1017$-$426 &   4.5    &   4.4    &   4.0    &   0.5    &   6.3    &         2.9          &   6.0    \\
B1526$-$423 &   3.8    &   5.4    &   6.2    &   3.1    &   6.6    &         6.1          &   7.3    \\
B1754$-$597 &   ...    &   ...    &   8.9    &   ...    &   ...    &         ...          &   ...    \\
B1817$-$640 &   4.0    &   4.5    &   7.4    &   0.7    &   6.0    &         2.6          &   8.4    \\
B1953$-$425 &   5.8    &   ...    &   ...    &   ...    &   ...    &         ...          &   ...    \\  
B2140$-$434 &   4.3    &   4.4    &   6.8    &   1.8    &   6.2    &         9.1          &   8.0    \\
B2140$-$817 &   5.2    &   6.6    &  10.8    &   8.8    &   8.4    &         3.7          &  12.0    \\
\hline
\end{tabular}
\end{center}
\tablecomments{The columns show (1) the MRC name of the SMS4 source; (2) the positional uncertainty (90\% c.l.) of the \sw~detection from our campaign $r_{c}$; (3) the positional uncertainty (90\% c.l.) of the LSXPS source $r_{l}$; (4) the positional uncertainty (90\% c.l.) of the eROSITA-DE DR1 source $r_{e}$; (5) the angular separation $d_{cl}$ between our detection and the LSXPS source; (6) the matching radius $r_{cl}$ between our detection and the LSXPS source; (7) the angular separation $d_{ce}$ between our detection and the eROSITA-DE DR1 source; (8) the matching radius $r_{ce}$ between our detection and the eROSITA-DE DR1 source.}
\end{table*}

We searched for matching X-ray sources in the LSXPS (\url{https://www.swift.ac.uk/LSXPS/}) within 3\arcmin~from the coordinates of our X-ray detections: we found at least one LSXPS source for all our detections, excluding B1953$-$425.
As expected, we did not find any LSXPS source for all the undetected SMS4 sources listed in Table~\ref{tab:und}.

For eight of ten sources, the primary LSXPS source belongs to the {\it ultra-clean} sample, and is typically within $\sim$3\arcsec.5~from the coordinates of our detections, that is comparable with the smaller positional uncertainty of our detections (see Table~\ref{tab:det}).   
Of the remaining two sources, 1) the extended emission of B1526$-$423, is incorrectly reported as four LSXPS sources within 80\arcsec~and 2) for B2140$-$817, the LSXPS position is found at 8\arcsec.8 from our position and likely incorrect in LSXPS.

For all our detected sources, we computed the angular separations $d_{cl}$ between our detections and the LSXPS sources, as well as the matching radius $r_{cl}=(r_{c}^2+r_{l}^2)^{1/2}$, where $r_{c}$ and $r_{l}$ are the positional uncertainties of corresponding sources, respectively. 
As shown in Table~\ref{tab:xcat}, we found that $d_{cl}<r_{cl}$ in all cases, excluding the noted above B2140$-$817.  
For the extended source B1526$-$423, we include in Table~\ref{tab:xcat} only one of the two sources from the {\it ultra-clean} sample.

\begin{table*} 
\begin{center}
\scriptsize
\caption{eROSITA-DE DR1 counterparts to SMS4 sources in our sample.}
\label{tab:dr1}
\begin{tabular}{ccccccc}
\hline
SMS4 Source &    1eRASS Name     & $r_{e}$  & Exposure & Detection likelihood &  Count Rate   &         0.2--2.3 keV Flux           \\
            &                    & (arcsec) &   (s)    &                      & (ct s$^{-1}$) & ($10^{-12}$ erg cm$^{-2}$ s$^{-1}$) \\
   (1)      &       (2)          &    (3)   &   (4)    &        (5)           &      (6)      &                 (7)                 \\  
\hline                                                                                                                                 
B0202$-$765 & J020213.5$-$762001 &    1.5   &   198    &       1549.95        &     1.84      &            1.72                     \\
B0242$-$514 & J024344.1$-$511235 &    8.5   &   206    &          8.63        &     0.04      &            0.03                     \\
B0344$-$345 & J034630.4$-$342245 &    5.2   &   257    &         62.45        &     0.11      &            0.10                     \\
B0427$-$366 & J042939.9$-$363053 &    5.6   &   193    &         86.36        &     0.18      &            0.17                     \\
B1017$-$426 & J102003.9$-$425130 &    4.0   &   125    &         86.07        &     0.24      &            0.22                     \\
B1526$-$423 & J153014.1$-$423147 &    6.2   &   113    &        242.77        &     1.39      &            1.29                     \\
~~B1754$-$597$^\ast$ & J175906.4$-$594643 &    8.9   &   129    &          9.94        &     0.07      &            0.07                     \\ 
B1817$-$640 & J182216.4$-$635917 &    7.4   &   106    &         11.25        &     0.07      &            0.06                     \\
B2140$-$434 & J214334.2$-$431244 &    6.8   &    94    &         23.63        &     0.13      &            0.12                     \\
B2140$-$817 & J214724.3$-$813212 &   10.8   &   159    &          8.81        &     0.03      &            0.03                     \\
\hline
            &                    &          &          &                      &               &        2.3--5.0 keV Flux            \\
            &                    &          &          &                      &               & ($10^{-12}$ erg cm$^{-2}$ s$^{-1}$) \\
\hline
B0202$-$765 & J020213.5$-$762001 &    1.5   &   203    &       1576.47        &     1.89      &            2.27                     \\
B1526$-$423 & J153014.1$-$423147 &    6.0   &   114    &        289.37        &     1.53      &            2.56                     \\
\hline
\end{tabular}
\end{center}
\tablecomments{The columns show (1) the MRC name of the SMS4 source; (2) the IAU name of the eROSITA-DE DR1 source, with (3) the positional uncertainty (90\% c.l.) $r_{e}$; (4) the exposure at the position of the source ({\sc ape\_exp\_1} parameter in the DR1 catalog); (5) the detection likelihood; (6) the count rate and (7) the flux of the 1eRASS source, in the specified bands. Top: values derived from the {\it Main} catalog in the 0.2--2.3 keV band. Bottom: values derived from the {\it Hard} catalog in the 2.3--5.0 keV band.\\
$\ast$: This source was not detected in our campaign, due to a very short exposure.\\
}
\end{table*}

We searched in the eROSITA data archive (eRODat, \url{https://erosita.mpe.mpg.de/dr1/erodat/}) for 1eRASS sources matching 17 of the 18 sources in our original proposal (one of the sources, B2032$-$350, is not covered in the footprint of the released data). 
These 17 sources include B1827$-$360, that as of May~2024 had not been observed by \sw. 
We used the coordinates reported in Table~\ref{tab:det} for our eleven X-ray detections and those in Table~\ref{tab:radio}, corresponding to the G4Jy radio centroids, for the remaining sources.

As a result, we found a source in the {\it Main} (0.2--2.3 keV) catalog for nine of the eleven \sw~detections.
For the two sources (B0202$-$765 and B1526$-$423) with the highest rates detected from both \sw~and eROSITA, a source was found also in the {\it Hard} (2.3--5.0 keV) catalog.
The two missing sources are those for which we detected the lowest rate values, equal to (B0103$-$453) or lower (B1953$-$425) than $2\times10^{-3}$ ct~s$^{-1}$.
The latter source is the same that was missing in the LSXPS itself, but that we were able to detect with our forced photometry.
For these two sources, nothing relevant was found in the {\it Supplementary} catalog, with lower confidence detections.

For the six remaining sources, a detection in the {\it Main} Catalog was found only for B1754$-$597, the only source in Table~\ref{tab:und} for which the lack of an X-ray detection in the available \sw~data was most probably due just to the short X-ray exposure. 
For the remaining five, including B1827$-$360, nothing relevant was found even in the {\it Supplementary} catalog.

Some details for the eROSITA sources, taken from the eRODat web site, are reported in Table~\ref{tab:dr1}.
Among others, a significant parameter provided in the DR1 catalogs is the extent likelihood and the extext in arcseconds, but we did not report them in Table~\ref{tab:dr1} since they were found to be 0 for all sources, excluding B1526$-$423.

Similarly to our analysis for the LSXPS catalog, we computed $d_{ce}$ between our detections and corresponding eROSITA sources, as well as the matching radius $r_{ce}=(r_{c}^2+r_{e}^2)^{1/2}$, where $r_{e}$ is the positional uncertainty (90\% c.l.) of eROSITA-DE DR1 sources.
These values, given in Table~\ref{tab:xcat}, show that, overall, the consistency is excellent; only for B1240-434 is $d_{ce} > r_{ce}$, by a very small amount.

\section{Multi-frequency analysis} 
\label{sec:4}

In this Section, we first discuss the matching of the SMS4 radio sources with the detected X-ray sources, and establish the radio/X-ray association when the radio and the X-ray positional uncertainties overlap. 
Using comparable values for the positional uncertainties of both radio and X-ray sources, we showed in \am~that the probability of chance coincidence between an X-ray detection and an SMS4 source is negligible. 
We then discuss the WISE and optical identifications, mainly utilizing the X-ray position. 
The association with infrared or optical sources is allowed only if their coordinates lie within the corresponding X-ray positional uncertainty.

As also reported in \am, all 12 sources in our sample, detected in the X-rays, were classified by BH06 as radio galaxies (eight objects), quasars (three objects) and a quasar candidate (B0427$-$366).
In the optical band, emission from the AGN itself, but likely also from the host galaxy, can be detected.
In type~II AGNs, the nuclear optical emission might be absorbed and obscured, but the infrared is more likely to reach the observer.
In our efforts to localize the core of the AGN, following the same criterion already used in \am, with our multifrequency analysis we require a detection in both the infrared and optical bands.

As a preliminary step to match our X-ray detections with infrared and optical counterparts, we used the SkyView Virtual Observatory to retrieve infrared maps in the $W1$~filter (3.4~$\mu$m) from the AllWISE Data Release \citep{2014yCat.2328....0C} Images Atlas, and optical maps in the $r$~filter (0.62~$\mu$m) from the Space Telescope Science Institute (STScI) $2^{nd}$ Digitized Sky Survey (DSS2).
These maps are shown in Fig.~\ref{fig:iroptcand}.
We used TOPCAT to cross-match our list of X-ray detections with sources from selected radio, infrared, and optical catalogs.


\subsection{Associating X-ray Sources with Radio Sources}
\label{sec:4.1}
To reliably associate our 12 X-ray detected sources, listed in Table~\ref{tab:det}, with corresponding radio sources, we matched the source positions at these two energy ranges.
We used G4Jy (W20) as the main matching radio catalog for the sources, that were all surveyed by GLEAM.
Following W20, the typical rms positional uncertainties for G4Jy sources are $\sigma_{\alpha,S}$ $\approx$ 1\arcsec.5, $\sigma_{\delta,S}$ $\approx$ 1\arcsec.7, when their brightness-weighted centroids were computed after a cross-correlation with SUMSS data, while they are $\sigma_{\alpha,N}$ $\approx$ 0\arcsec.5, $\sigma_{\delta,N}$ $\approx$ 0\arcsec.6, when the same operation was based on NVSS data.

To match the radio and X-ray positions, we conservatively used a circular confidence region sufficiently large to ensure, with high probability, that the radio source lies within the X-ray positional uncertainty region.
For this circular region, we used a radius $r_X$ corresponding to the error radius, at 90\% confidence level, for sources at the X-ray limit of sensitivity, to which we added the largest uncertainty $r_r$ for the radio position, at the same confidence level as for the X-rays.
For the X-ray band, we find $r_X=5\arcsec.8$ from Table~\ref{tab:det}, that is the positional uncertainty of B1953$-$425.
For the radio band, we take the larger of $\sigma_{\alpha,S}$ and $\sigma_{\delta,S}$, and multiply it by a factor of 1.645, derived from the Normal distribution, obtaining $r_r=2\arcsec.8$.

Using a circle with $R_{90}=r_r+r_X=8\arcsec.6$, we verified that the combined 90\% confidence positional uncertainty widely overlaps - and in most cases totally includes - the corresponding G4Jy positional uncertainty for all \sw-XRT detections, with the exception of B0103$-$453 and B0344$-$345.
As reported in Table~\ref{tab:radio}, these two FR~II radio galaxies are the two sources in our sample with the largest angular sizes, exceeding 2\arcmin~(B0103$-$453) and even 4\arcmin~(B0344$-$345); their complex radio structure is also shown in Figure~\ref{fig:xrtmaps}.
For both radio sources, we also noticed a mismatch between the G4Jy centroid and the coordinates of the SUMSS component that is closer to the core of the radio galaxy: as shown in Figure~\ref{fig:iroptcand}, for both SMS4 sources our \sw~X-ray detection, and the corresponding infrared/optical counterpart that we found, lies in between the G4Jy and the SUMSS source positions.
We conclude that the \sw~X-ray observations are able to accurately determine the position of the AGN, for both B0103$-$453 and B0344$-$345, and overcome any uncertainties arising from the extended nature of the radio emission which can be significantly affected by jets, knots, and lobes, especially at lower frequencies where the radio angular resolution is sometimes limited.

We repeated an analogous computation for B1754$-$597, detected only by eROSITA.  
Unlike the \sw~ X-ray positions, the eROSITA position with its uncertainty of 8\arcsec.9 does not match the radio position.
Using $R_{95}=13\arcsec.9$, the combined 95\% confidence positional uncertainty, does overlap the radio positional uncertainty.


\begin{table*} 
\begin{center}
\scriptsize
\caption{Our candidate counterparts for X-ray detected SMS4 sources.}
\label{tab:ir_opt}
\begin{tabular}{ccccccc}
\hline
MRC Name             &    Infrared Source     & $W1$  & $W1-W2$ & Optical Source & $\Delta$ \\
                     &                        & (mag) &  (mag)  &                & (arcsec) \\
   (1)               &          (2)           &  (3)  &   (4)   &       (5)      &    (6)   \\  
\hline                                                                                                                                            
B0103$-$453          & J010522.21$-$450517.2  & 14.71 &   1.30  &   S2S4040102   &  0.17   \\
B0202$-$765          & J020213.68$-$762003.1* & 12.45 &   1.16  &   S0X7028858   &  0.05   \\
B0242$-$514          & J024344.45$-$511238.6* & 15.04 &   0.37  &   S2TQ011216   &  0.77   \\
B0344$-$345          & J034630.57$-$342246.2* & 12.31 &   0.21  &   S33I021144   &  0.01   \\ 
B0427$-$366          & J042940.12$-$363053.6* & 14.58 &   1.07  &   S2GB002483   &  0.49   \\ 
B1017$-$426          & J102003.93$-$425130.0  & 14.00 &   1.37  &   S5Q2037849   &  0.15   \\ 
B1526$-$423          & J153014.29$-$423151.7  & 13.95 &   0.18  &   S9CT375122   &  0.48   \\ 
B1817$-$640          & J182216.18$-$635918.4* & 14.90 &   0.79  &   SAQ2103623   &  0.21   \\ 
B1953$-$425          & J195715.25$-$422219.8* & 15.36 &   1.68  &   SCEL082842   &  0.10   \\ 
B2140$-$434          & J214333.40$-$431253.0* & 12.48 &   1.17  &   SBZ5022318   &  0.06   \\ 
B2140$-$817          & J214724.62$-$813213.0* & 14.48 &   0.91  &   SAGB001072   &  0.75   \\ 
\hline                                                                                       
~B1754$-$597$^{\dagger} $ & J175907.08$-$594649.7 & 14.49 & 1.44 &   S7FD049277   &  0.35   \\                   
\hline                                                                                       
\end{tabular}
\end{center}
\tablecomments{\scriptsize{The columns show (1) the MRC name of the SMS4 source; (2) the AllWISE \citep{2014yCat.2328....0C} infrared source identification; (3) the AllWISE magnitude in the $W1$ filter; (4) the AllWISE $W1-W2$ infrared color; (5) the GSC~2.4.2 \citep{2021yCat.1353....0L} optical source identification; (6) the angular separation $\Delta$ between the given infrared and optical sources.\\
$*$: this infrared source was previously associated by W20. \\
$\dagger$: as discussed in the text, there are two viable counterparts for this source, detected by eROSITA. The given source IDs, lying within the X-ray positional uncertainty (see Fig.~\ref{fig:iroptcand}h) are those which include, as optical counterpart, the one consistent with BH06.\\
}}
\end{table*}

\subsection{Cross Matches with Infrared and Optical Catalogs}
\label{sec:4.2}

In the infrared band, the main catalog that we adopted in our cross match with the X-ray detected sources is AllWISE, and the typical positional uncertainties for our sources are in the range of 0.03--0.09 arcseconds. 
As in our previous paper, we took into account the infrared color of our WISE candidates, comparing $W1-W2$ with the threshold ($W1-W2\geq0.8$ mag) established by \cite{2012ApJ...753...30S} in their simple criterion for selecting AGNs.
In the optical band, the main catalog that we used for source identification is the 2$^{nd}$ Generation Guide Star Catalog (GSC~2.4.2) by \cite{2021yCat.1353....0L}, which has a typical positional uncertainty in the 0.1--0.3 arcseconds range for our sources.

\begin{figure*}
\gridline{\fig{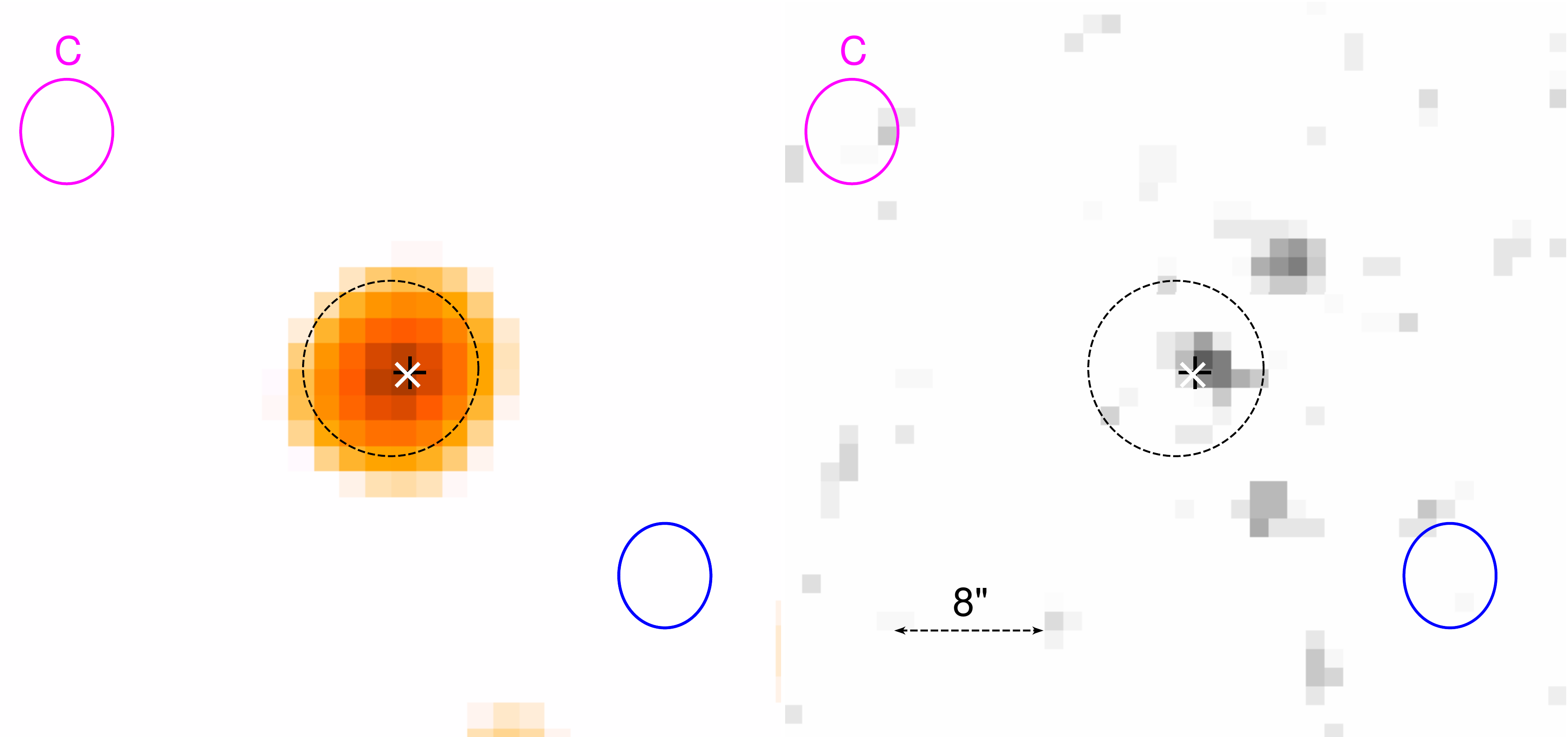}{0.49\textwidth}{(a - MRC~B0103$-$453)} \fig{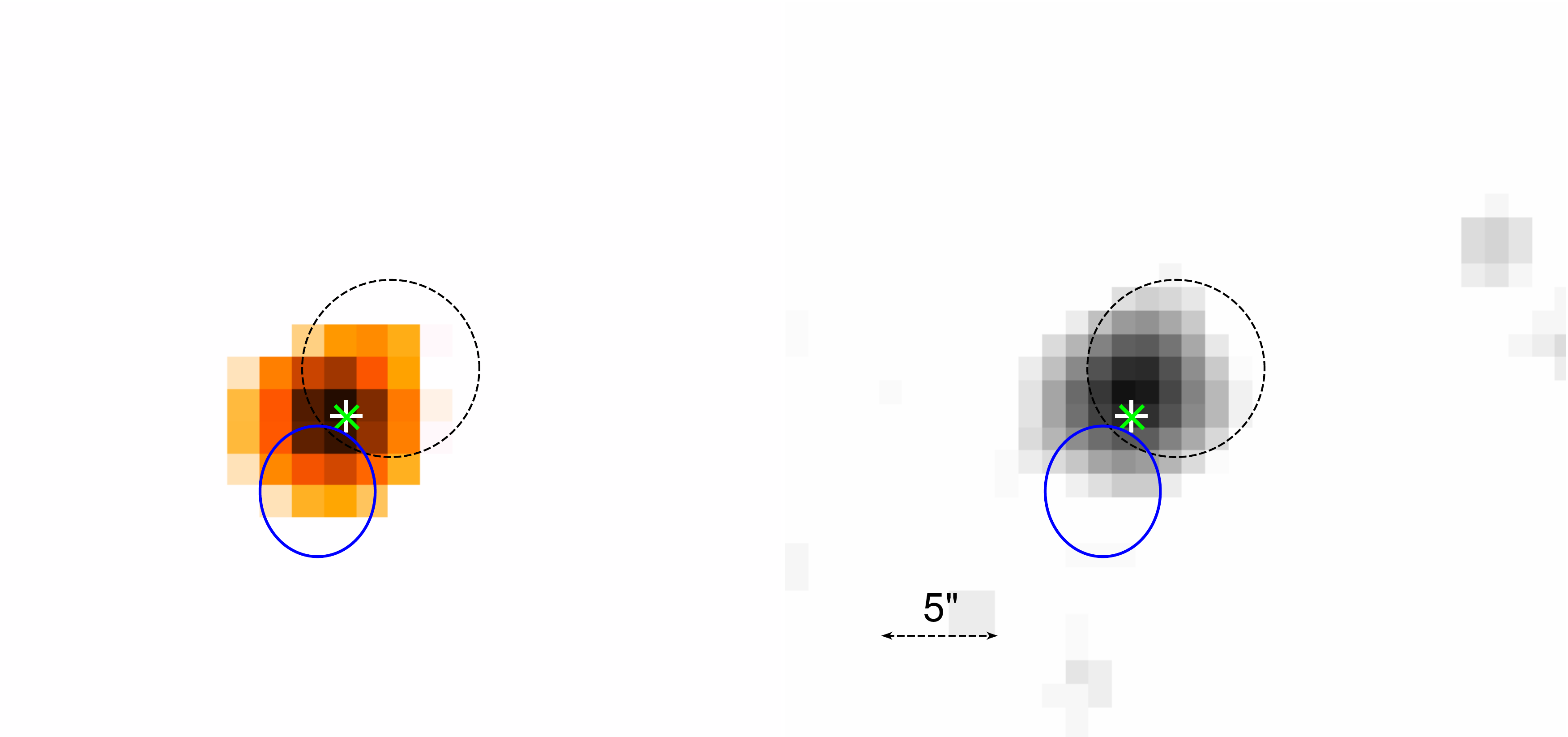}{0.49\textwidth}{(b - MRC~B0202$-$765)}}
\gridline{\fig{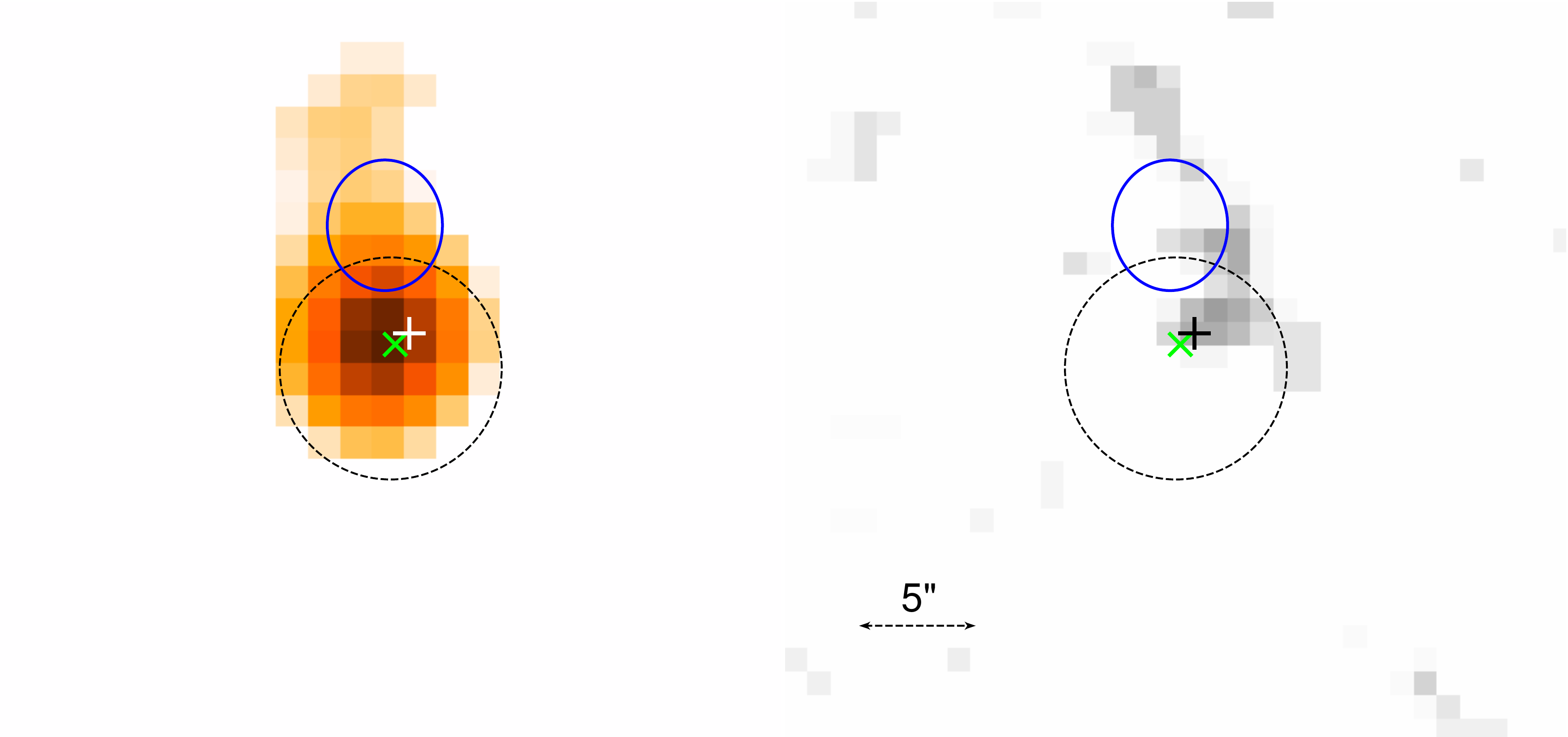}{0.49\textwidth}{(c - MRC~B0242$-$514)} \fig{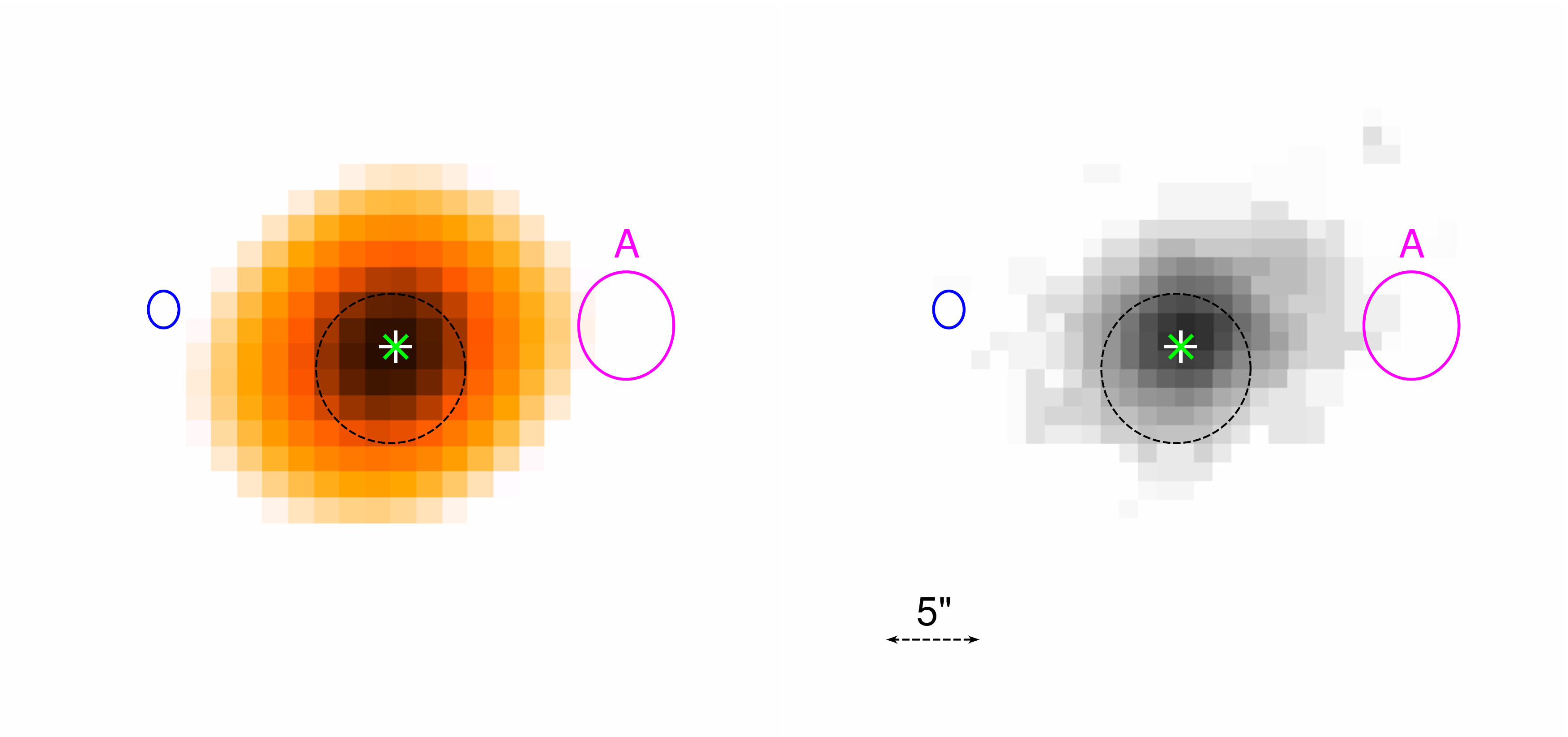}{0.49\textwidth}{(d - MRC~B0344$-$345)}}
\gridline{\fig{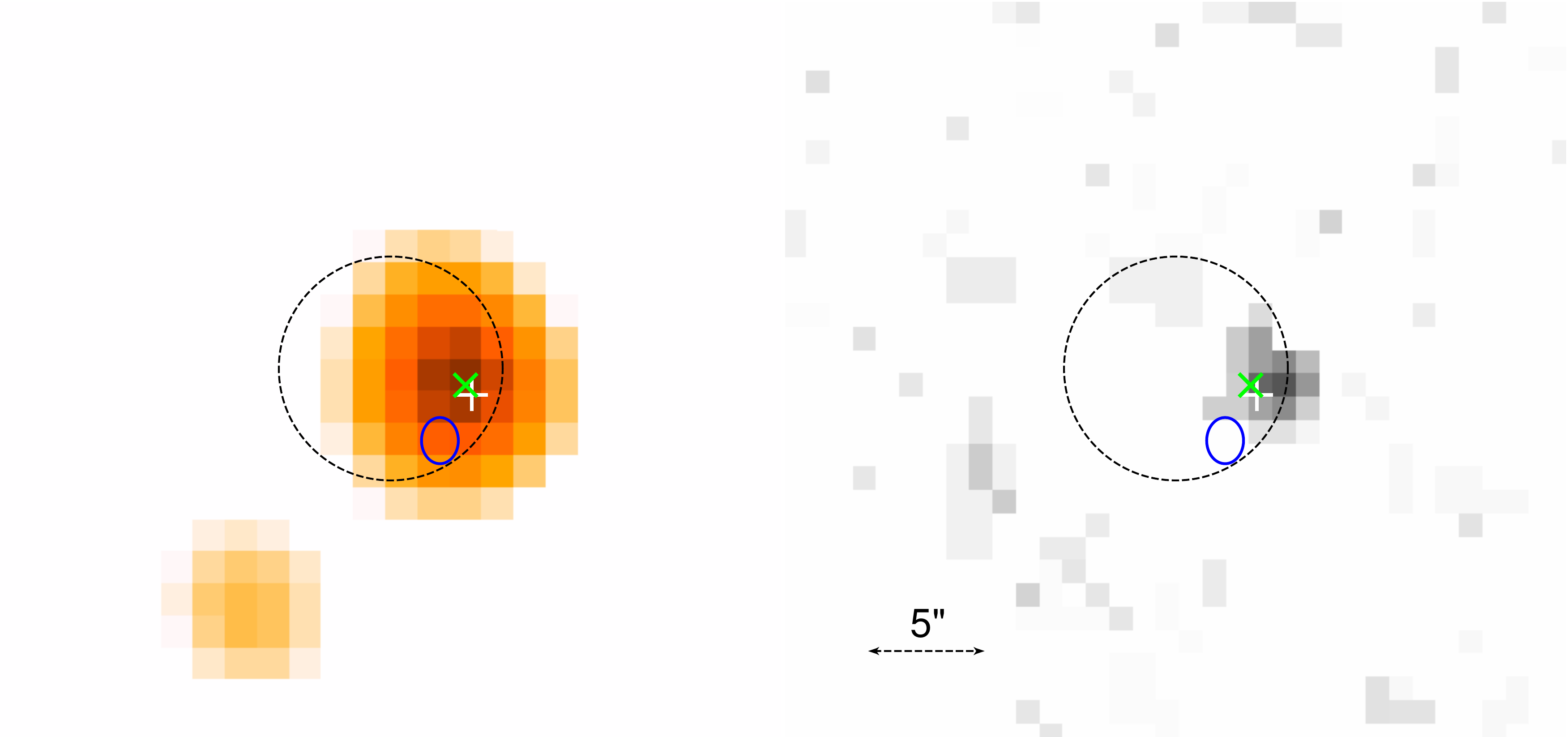}{0.49\textwidth}{(e - MRC~B0427$-$366)} \fig{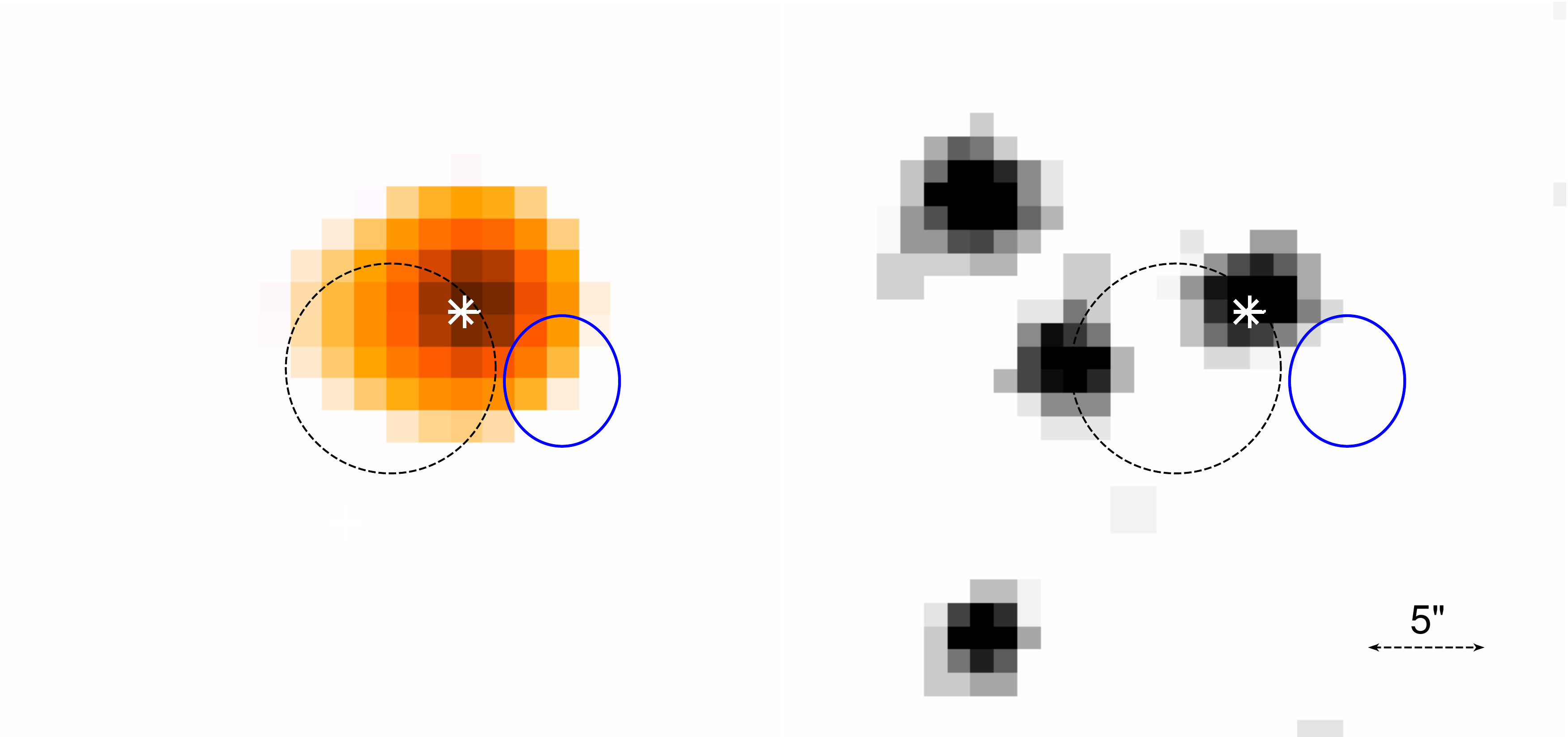}{0.49\textwidth}{(f - MRC~B1017$-$426)}}
\caption{Infrared and optical maps, (matched in scale for each source; N is up and E is to the left), of the eleven SMS4 sources detected by \sw-XRT in our sample and also of B1754$-$597, detected by eROSITA. Infrared maps (left side of each panel) in the $W1$ filter (3.4 $\mu$m) are from AllWISE, while optical maps (right side; 1~pixel=1\arcsec) in the $r$ filter are from DSS2. Blue ellipses mark the positional uncertainty of G4Jy sources; ellipses are magenta for SUMSS components, also shown for the FR~II radio galaxies B0103$-$453 and B0344$-$345, with letters corresponding to those assigned in Figure~\ref{fig:xrtmaps}. Dashed circles mark the positional uncertainty of the X-ray sources; the circle is black for our \sw-XRT sources, and red for the 1eRASS source corresponding to B1754$-$597. Positional uncertainties are given at 90\% confidence level for both the radio and the X-ray sources. Crosses (x) and plus signs (+) mark infrared sources from AllWISE and optical sources from GSC~2.4.2. Black or white are equally used for AllWISE and GSC~2.4.2 sources to improve the visibility with respect to the map in the background. Green is used for AllWISE counterparts also associated by W20, and for the optical counterpart of B1754$-$597 suggested by BH06.}
\label{fig:iroptcand}
\end{figure*}

\setcounter{figure}{6}
\begin{figure*}
\gridline{\fig{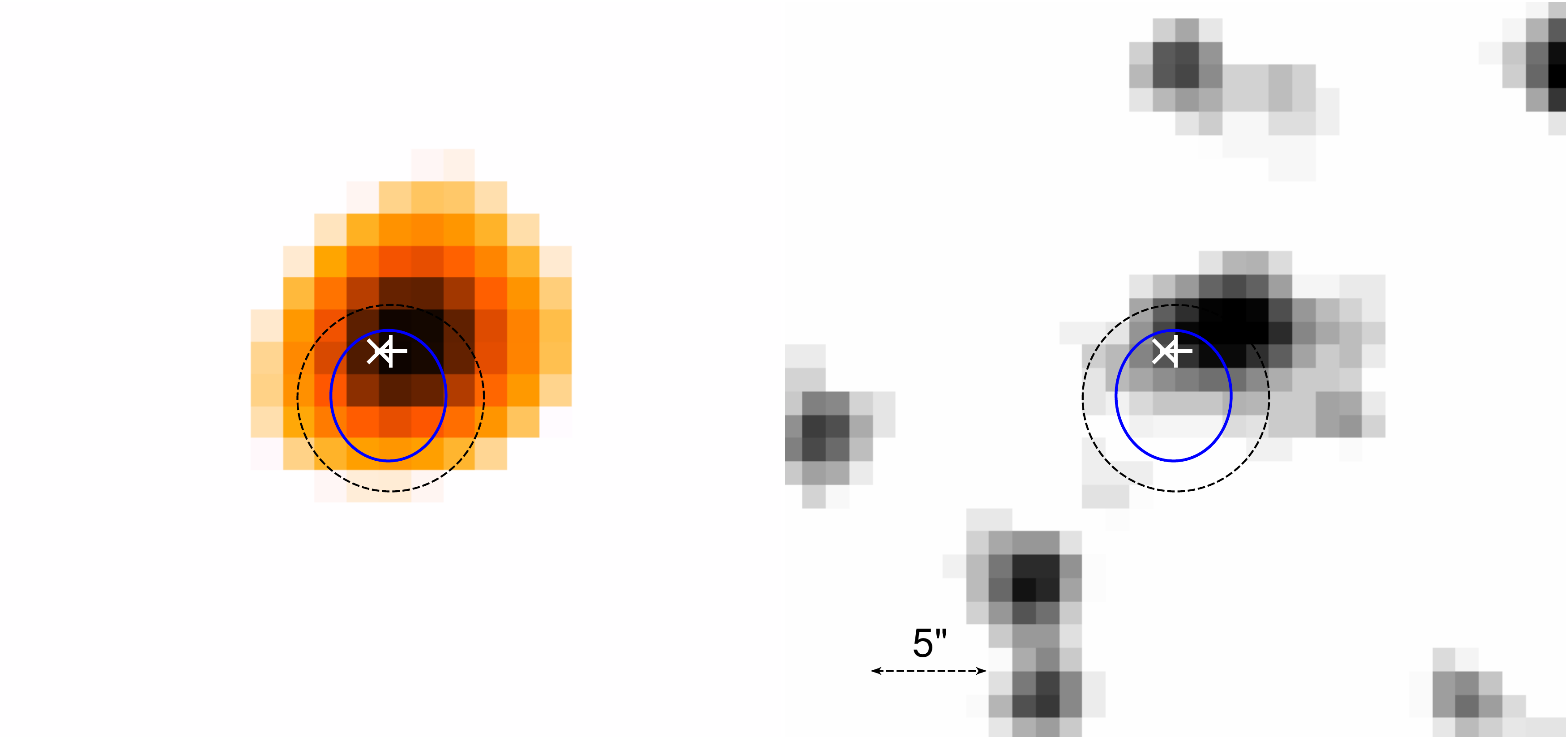}{0.49\textwidth}{(g - MRC~B1526$-$423)} \fig{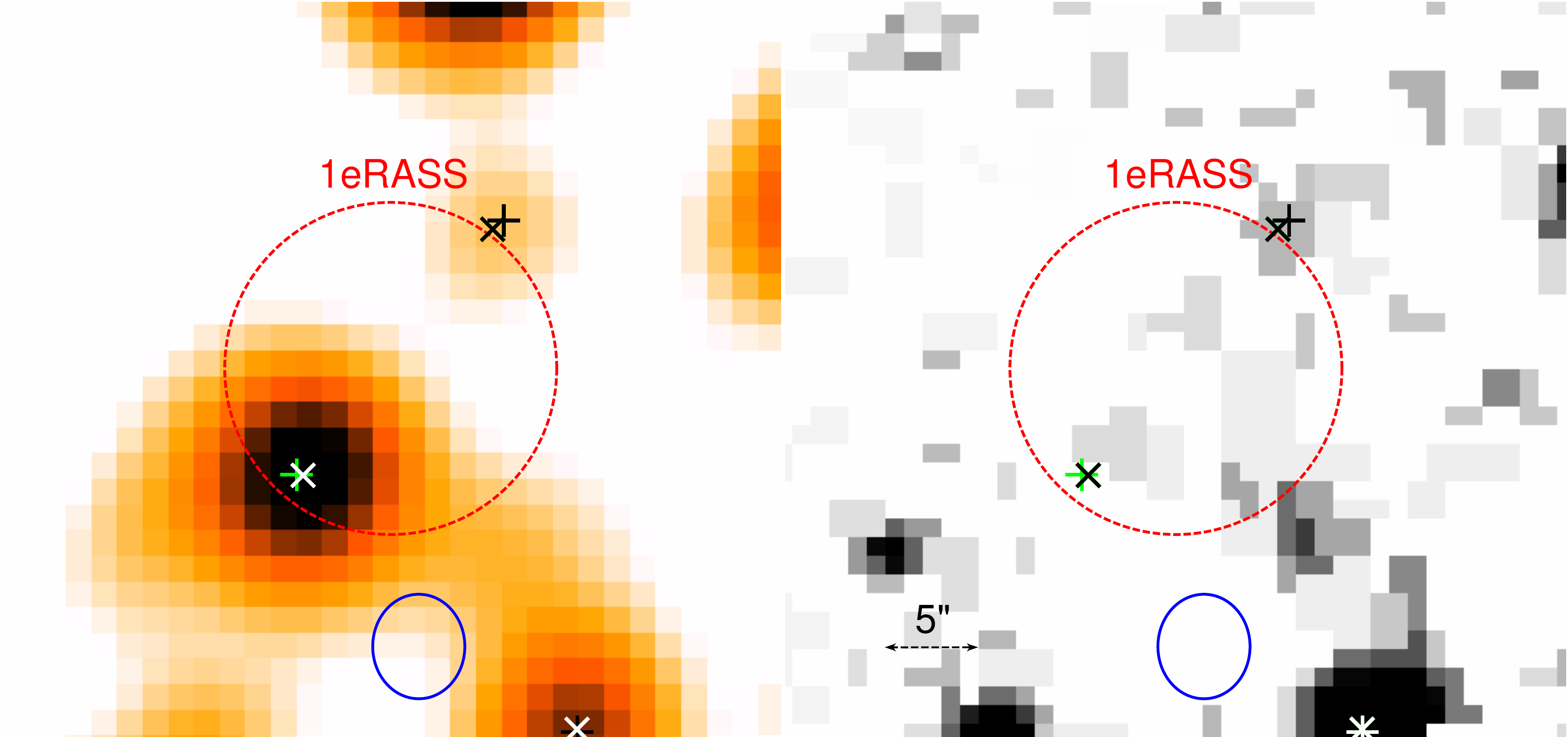}{0.49\textwidth}{(h - MRC~B1754$-$597)}}
\gridline{\fig{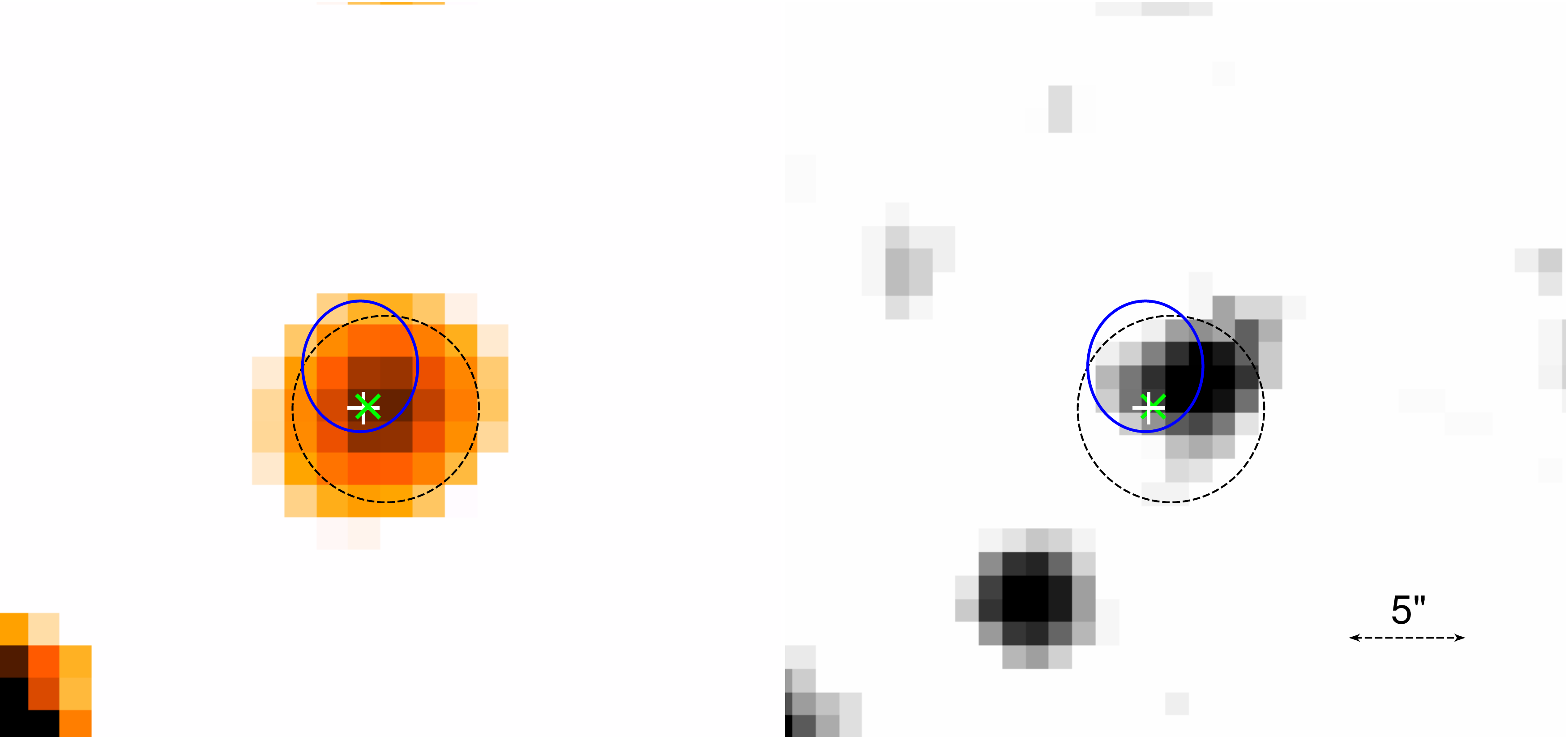}{0.49\textwidth}{(i - MRC~B1817$-$640)} \fig{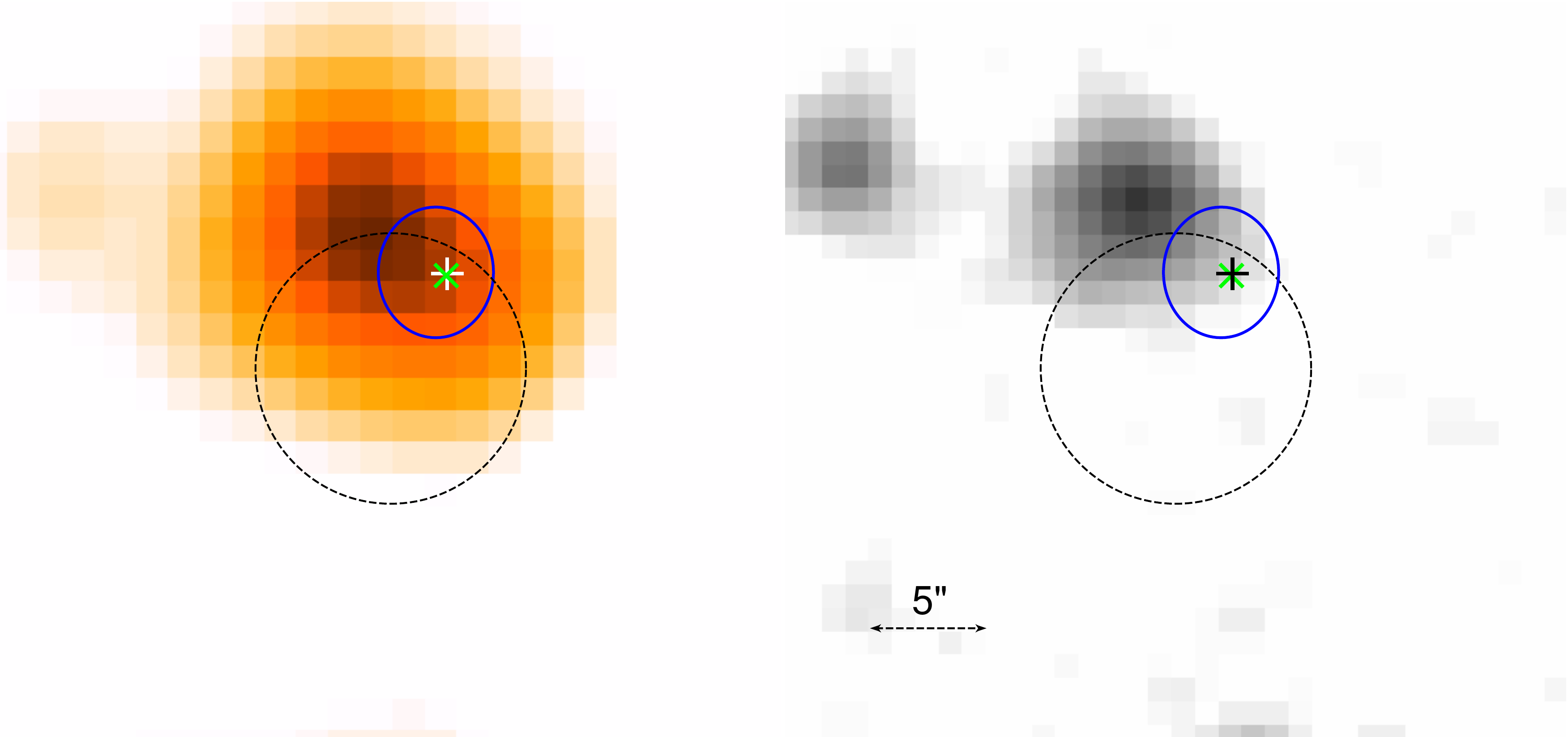}{0.49\textwidth}{(j - MRC~B1953$-$425)}}
\gridline{\fig{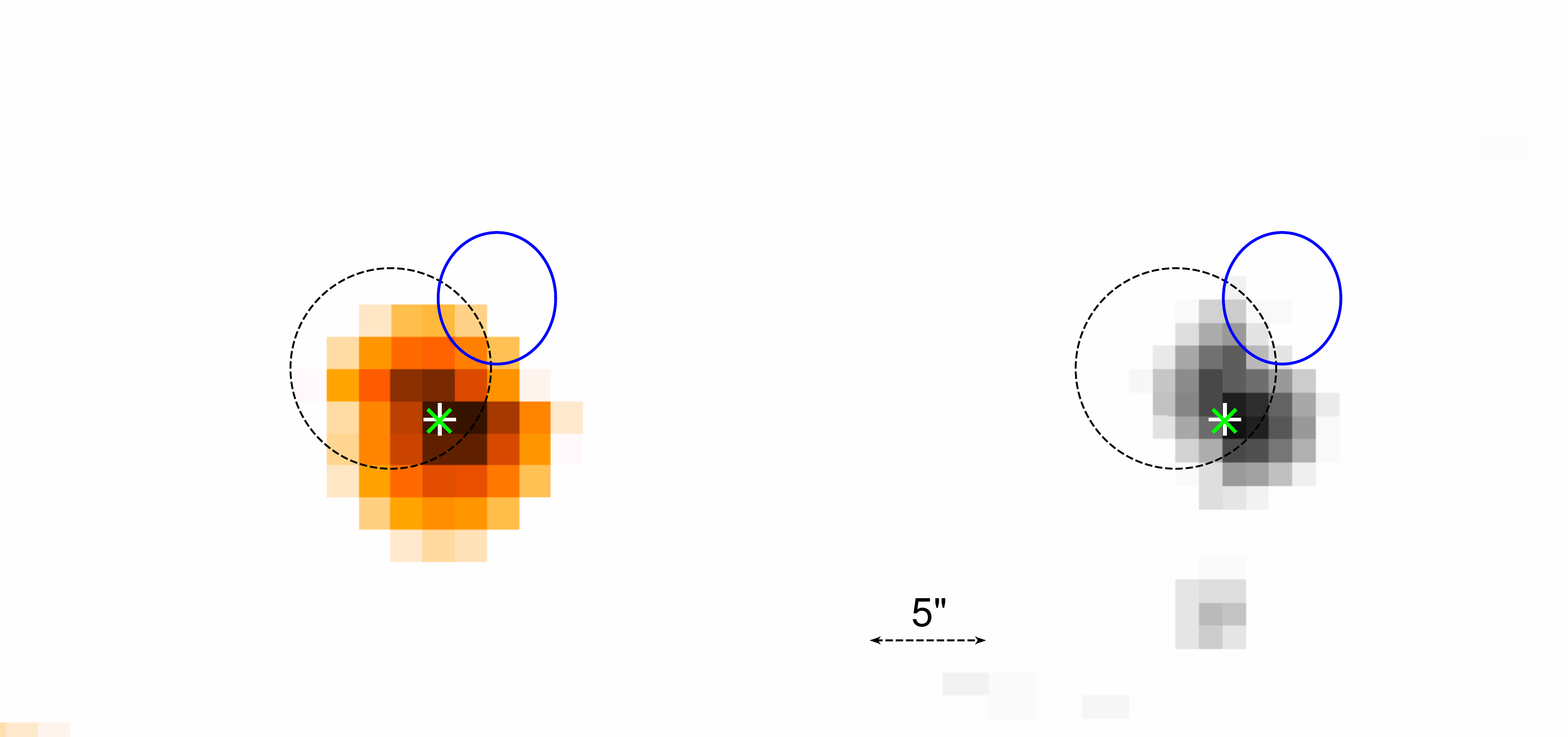}{0.49\textwidth}{(k - MRC~B2140$-$434)} \fig{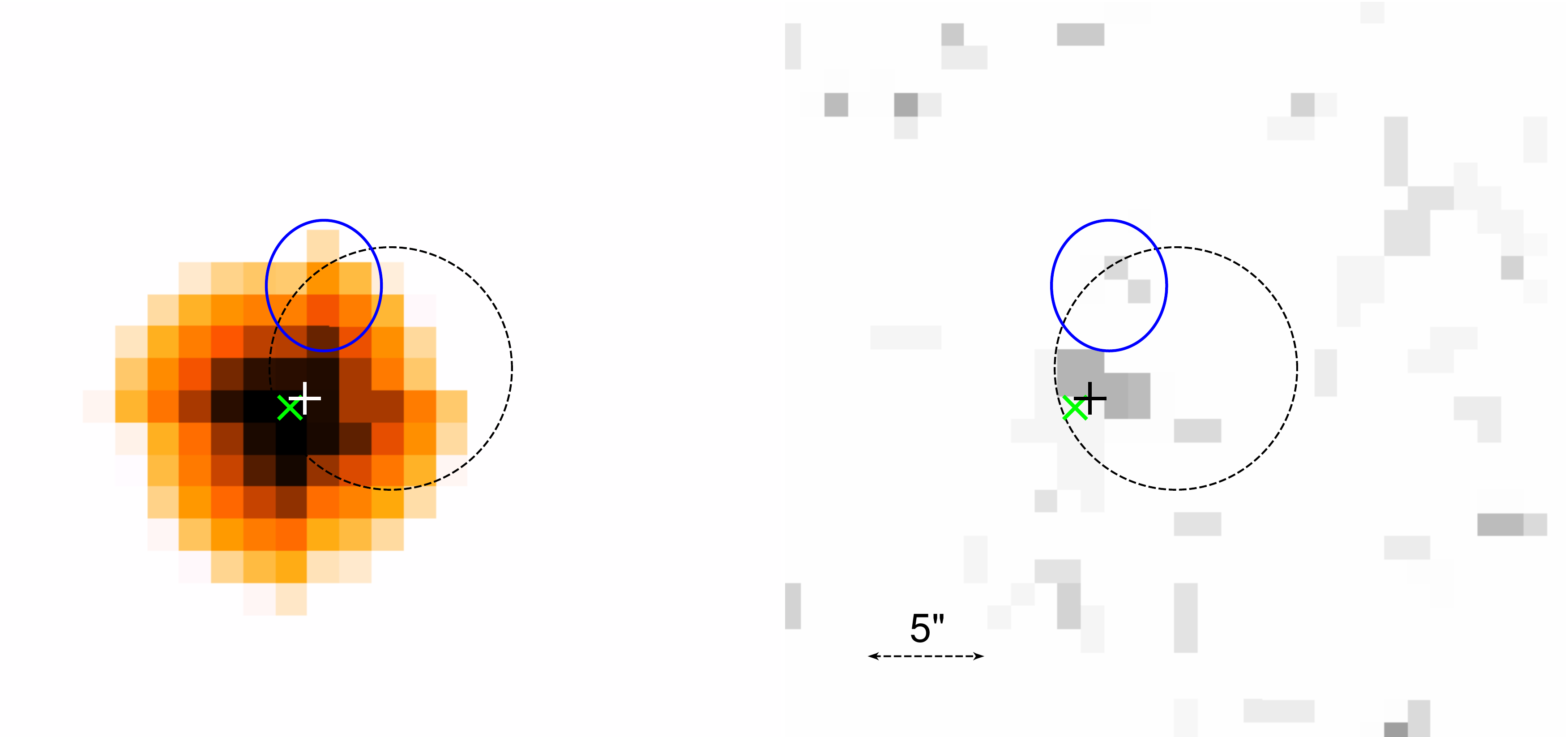}{0.49\textwidth}{(l - MRC~B2140$-$817)}}
\caption{{\it (continued)}}
\end{figure*}

For all the 11 SMS4 sources detected by \sw, we find one candidate counterpart, detected in both the infrared and the optical bands.
These candidates are unique, and lie within the boundaries of the X-ray positional uncertainty.
While we are confident in the identification of the counterparts, not all have IR AGN colors. 
B0242$-$514, B0344$-$345, and B1526$-$423 do not lie within the AGN color range defined above; this is anomalous particularly for B0344$-$345 ($W1-W2=0.21$ mag), since its AGN nature is quite evident from its FR~II radio structure (see Fig.~1). 
Hence, following the same classification scheme adopted in \am, we define all these SMS4 sources, with unique IDs, as class~A but now distinguish the three SMS4 sources (B0242$-$514, B0344$-$345, and B1526$-$423) lacking AGN colors as class~A2, with the remaining sources as class~A1.

A different situation is found for the eROSITA detected source B1754$-$597, for which we found two candidates: one within the X-ray uncertainty region, with AGN infrared colors, and another one outside, but close to its boundaries (see Figure~\ref{fig:iroptcand}), whose infrared counterpart is AllWISE~J175905.73$-$594636.5, with $W1-W2=0.20$ mag.
This occurrence defines B1754$-$597 as a class~B source.

All our candidate counterparts are listed in Table~\ref{tab:ir_opt}; for B1754$-$597, only the candidate within the uncertainty region and with AGN infrared colors is reported.
As shown in column (6), the angular distance between the infrared and the optical sources is always less than 0\arcsec.8, and only in five cases higher than 0\arcsec.25, attesting to the fact that they can be both attributed to the same astrophysical source.


\newpage
\subsection{Comparison with Earlier Studies}
\label{sec:4.3}
 
W20 associated an infrared counterpart from AllWISE to eight of the twelve SMS4 sources - including B1754$-$597 - for which X-ray emission was detected.
More recently, AllWISE infrared counterparts for B1017$-$426 and B1526$-$423 were reported by \cite{2023ApJS..265...32M}.
As a result, with our approach, we confirm all the infrared counterparts previously addressed in the literature, and we provide for the first time AllWISE counterparts for B0103$-$453 and B1754$-$597.

Optical counterparts were associated by BH06 to each of the twelve X-ray detected sources, using either the plates from the UK Schmidt Southern Sky Survey or R-band CCD images from a dedicated campaign at the AAT.
As shown in Table~\ref{tab:opt}, we find excellent agreement between the coordinates reported in the GSC~2.4.2 catalog and those reported by BH06, with angular distances in most cases less than 1\arcsec~(only B0202$-$765 and B1754$-$597 exceed 1\arcsec, but are $<2$\arcsec) leading us to conclude that both optical sources refer to the same astrophysical objects.
Thus, we independently confirm all BH06 counterparts, and we list the GSC~2.4.2 sources in Table~\ref{tab:ir_opt} for the sake of simplicity.

Comparing our optical candidates with those given more recently in \cite{2023ApJS..265...32M} (see their Table~3) and \cite{2024ApJS..271....8G} (see their Table~1), we provide additional candidates for B0103$-$453 and B1754$-$597, the same sources mentioned earlier for the infrared band.
In particular, we verified that the counterpart provided by \cite{2024ApJS..271....8G} for B0242$-$514 matches the counterpart that we give in Table~\ref{tab:ir_opt}, with an angular separation of 1\arcsec.3.
We also note that, for B1953$-$425, the angular separation between our optical counterpart and that given by \cite{2023ApJS..265...32M} exceeds 5\arcsec.5, and therefore might refer to a different source.

\begin{table*} 
\begin{center}
\scriptsize
\caption{Comparison between optical counterparts.}
\label{tab:opt}
\begin{tabular}{c|cc|ccc}
\hline
            & \multicolumn{2}{c}{GSC~2.4.2 Source}        &     \multicolumn{3}{c}{\cite{2006AJ....131..114B}}     \\
  MRC Name  & R.A. (J2000) &        Decl. (J2000)         & R.A. (J2000) &          Decl. (J2000)       & $\Delta$ \\
            & ($^{h~m~s}$) & ($^{\circ}$~\arcmin~\arcsec) & ($^{h~m~s}$) & ($^{\circ}$~\arcmin~\arcsec) & (arcsec) \\
   (1)      &     (2)      &              (3)             &      (4)     &             (5)              &   (6)    \\  
\hline                                                                                                                                            
B0103$-$453 & 01 05 22.20  &        $-$45 05 17.1         & 01 05 22.12  &        $-$45 05 17.1         &   0.88   \\
B0202$-$765 & 02 02 13.69  &        $-$76 20 03.1         & 02 02 13.41  &        $-$76 20 04.7         &   1.92   \\
B0242$-$514 & 02 43 44.39  &        $-$51 12 38.2         & 02 43 44.40  &        $-$51 12 37.9         &   0.31   \\
B0344$-$345 & 03 46 30.58  &        $-$34 22 46.2         & 03 46 30.52  &        $-$34 22 45.8         &   0.85   \\
B0427$-$366 & 04 29 40.11  &        $-$36 30 54.1         & 04 29 40.05  &        $-$36 30 54.3         &   0.72   \\
B1017$-$426 & 10 20 03.92  &        $-$42 51 30.0         & 10 20 03.85  &        $-$42 51 30.2         &   0.79   \\
B1526$-$423 & 15 30 14.25  &        $-$42 31 51.7         & 15 30 14.25  &        $-$42 31 51.5         &   0.21   \\
B1817$-$640 & 18 22 16.21  &        $-$63 59 18.5         & 18 22 16.21  &        $-$63 59 18.7         &   0.16   \\
B1953$-$425 & 19 57 15.26  &        $-$42 22 19.8         & 19 57 15.31  &        $-$42 22 19.1         &   0.92   \\ 
B2140$-$434 & 21 43 33.41  &        $-$43 12 53.0         & 21 43 33.36  &        $-$43 12 52.2         &   0.94   \\
B2140$-$817 & 21 47 24.34  &        $-$81 32 12.6         & 21 47 24.52  &        $-$81 32 12.9         &   0.48   \\
\hline                                                                                       
~B1754$-$597$^{\ast}$ & 17 59 07.13 & $-$59 46 49.7        & 17 59 07.25  &        $-$59 46 48.9         &   1.22   \\   
\hline                                                                                       
\end{tabular}
\end{center}
\tablecomments{\scriptsize{The columns show (1) the MRC name of the SMS4 source; (2) the Right Ascension (J2000) and (3) the Declination (J2000) of the GSC~2.4.2 source that we give as optical counterpart (see Table~\ref{tab:ir_opt}); (4) the Right Ascension (J2000) and (5) the Declination (J2000) of the optical counterpart given by \cite{2006AJ....131..114B} (see their Table~7); (6) the angular separation $\Delta$ with respect to the corresponding GSC~2.4.2 source.\\
$\ast$: this source was detected in the X-rays by eROSITA.}\\ 
}
\end{table*}


\newpage
\section{SMS4 Sources Still Unobserved with Narrow FOV Instruments}
\label{sec:5}
Since 2015, we successfully proposed two observational campaigns to improve the X-ray coverage of the SMS4 sample of 137 bright radio sources using \sw~snapshot observations. 
As we began our program, 57 sources had existing \cha, \xmm, or \sw~data (seven only by \sw), leaving 80 sources that lacked pointed moderate/high resolution observations. 
In \am~we discussed 31 sources, including seven targets previously observed by \sw. 
Thus, with 17 radio sources presented in the current paper, our \sw~campaigns observed half (41) of the sources that were not previously observed in X-rays.

\begin{table*} 
\begin{center}
\scriptsize
\caption{eROSITA-DE DR1 likely counterparts to SMS4 Sources Unobserved by Narrow FOV Instruments.}
\label{tab:dr1_18}
\begin{tabular}{cccccccc}
\hline
SMS4 Source  & G4Jy ID &   1eRASS Name     &  $r_e$   & Exposure & Detection likelihood &  Count Rate   &         0.2--2.3 keV Flux           \\
             &         &                   & (arcsec) &   (s)    &                      & (ct s$^{-1}$) & ($10^{-12}$ erg cm$^{-2}$ s$^{-1}$) \\
   (1)       &  (2)    &       (3)         &    (4)   &   (5)    &        (6)           &      (7)      &                 (8)                 \\  
\hline                                                                                                                                 
B0003$-$833  &    12   & J000617.3$-$830608 &   8.1    &   184   &       12.76          &     0.04      &                 0.04                \\
B0036$-$392  &    70   & J003826.5$-$385946 &   3.8    &   100   &       83.15          &     0.28      &                 0.26                \\
B0119$-$634  &   145   & J012140.3$-$630900 &   1.6    &   182   &      584.87          &     0.87	    &                 0.81                \\
B0251$-$675  &   301   & J025155.6$-$671759 &   1.6    &   278   &      803.04          &     0.80      &                 0.75                \\
B0646$-$398  &   629   & J064811.1$-$395711 &   5.8    &   180   &       39.81          &     0.10      &                 0.09                \\
B0658$-$656  &   636   & J065813.1$-$654453 &   2.3    &   866   &      281.35          &     0.13      &                 0.12                \\
B0719$-$553  &   653   & J072015.4$-$552521 &   4.6    &   301   &      104.90          &     0.17      &                 0.16                \\
B1123$-$351  &   917   & J112552.6$-$352341 &   7.1    &   145   &       23.00          &     0.11      &                 0.10                \\
B1136$-$320  &   934   & J113916.8$-$322229 &  12.2    &   138   &       10.00	        &     0.05      &                 0.05                \\
B1215$-$457  &   986   & J121806.2$-$460028 &   2.3    &   183   &      354.33          &     0.57      &                 0.53                \\
B1421$-$382  &  1157   & J142416.4$-$382648 &   0.8    &   173   &     2564.57	        &     3.51      &                 3.26                \\
B1756$-$663  &  1455   & J180118.1$-$662305 &   8.2    &   148   &        7.15	        &     0.05      &                 0.05                \\
B1840$-$404  &  1505   & J184428.8$-$402159 &   7.7    &    78   &       41.14	        &     0.27      &                 0.25                \\
B2150$-$520  &  1740   & J215408.3$-$515013 &   8.1    &    91   &        7.44	        &     0.06      &                 0.06                \\
B2201$-$555  &  1756   & J220504.8$-$551732 &   6.3    &   100   &       16.90	        &     0.09      &                 0.08                \\
B2226$-$411  &  1781   & J222918.6$-$405125 &   6.3    &    93   &       40.04	        &     0.19      &                 0.18                \\
B2252$-$530  &  1797   & J225548.9$-$524547 &  10.7    &    99   &        6.38	        &     0.04      &                 0.04                \\
B2332$-$668  &  1843   & J233510.7$-$663658 &   2.5    &   151   &      307.68	        &     0.61      &                 0.57                \\
\hline
\end{tabular}
\end{center}
\tablecomments{The columns show (1) the MRC name of the SMS4 source; (2) the corresponding G4Jy ID; (3) the IAU name of the eROSITA-DE DR1 source, with (4) the positional uncertainty (90\% c.l.) $r_e$; (5) the exposure at the position of the source ({\sc ape\_exp\_1} parameter in the DR1 catalog); (6) the detection likelihood; (7) the count rate and (8) the flux of the 1eRASS source, in the specified band.}
\end{table*}

Considering that one source, B2223$-$528, was observed by Chandra in November 2022, 38 sources from the original sample still lack dedicated, pointed observations. 
We searched in the DR1 {\it Main} Catalog for these 38 SMS4 sources, using the coordinates of the corresponding G4Jy sources. 
We followed the same procedure described in Section~\ref{sec:4.1} to compute the matching radius $R$ between the radio and the X-ray sources. 
As $R_X$, we adopted the highest positional uncertainty (90\% c.l.) of the 1eRASS sources with the shortest angular distance from the coordinates of each G4Jy source. 
Using the X-ray positional uncertainty found for B1136$-$320, $R_X=12\arcsec.2$. 
Adopting for $R_r$ the same value reported in Section~\ref{sec:4.1}, we obtain $R$=15\arcsec: with this value, we find 18 SMS4 sources so far detected by eROSITA.
The list of these SMS4 sources, reported in Table~\ref{tab:dr1_18}, provides the corresponding 1eRASS source, that is the likely X-ray counterpart from the eROSITA DR1 Catalog.

Finally, considering 4 sources (B2049$-$368, B2115$-$305, B2226$-$386, and B2259$-$375) lying outside the eROSITA-DE footprint, 16 SMS4 sources are undetected by eROSITA in the DR1.


\section{Summary} 
\label{sec:6}

In the current paper, we present a final set of 17 objects from our sample\footnote{One of the 17 was not detected by \sw, but is found in the eROSITA survey. 
One additional source from our \sw~proposal, B1827$-$360, remains unobserved.} of radio sources, selected in \am as having high flux density at 181~MHz, and observed since May~2022 as part of a second \sw~observational campaign (PI~Maselli). 
As a whole, considering both \sw~ campaigns, the number of SMS4 sources in the \sw~archive increased by 41 sources (30\% of the SMS4 sample).

Following \am, for each source we carried out a local source detection to derive source intensity and significance (see Section~\ref{sec:3}). 
We detected 11 sources (see Table~\ref{tab:det}), four of which were also detected by \cite{2023ApJS..268...32M}.
For all the \sw~detected sources we derived their X-ray properties, including extent and hardness ratio.

To investigate source extent, we carried out a simple ratio test (see Section~\ref{sec:3.1.1} and Table~\ref{tab:er} for details), that uses the ratio of source counts in a circle with radius $\sim12$\arcsec~divided by the source counts in the annulus $\sim24$-48\arcsec. 
This ratio was then compared to that expected for an unresolved source, derived from the \sw~PSF. 
We found two sources, B0103$-$453 and B1526$-$423, to be clearly extended and, at lower significance, B0344$-$345 to be likely extended.
The remaining sources are consistent with point-like emission.

In addition to the analysis performed in \am, for the three sources mentioned above, we generated their radial profiles (see Section~\ref{sec:3.1.2}, and Fig.~\ref{fig:radprofext}) to characterize their extent up to 5\arcmin~from the X-ray centroid. 
The X-ray emission for B0344$-$345 and B1526$-$423 extends to at least 2\arcmin~(see Fig.~\ref{fig:radprofext}). 
For B0103$-$453, despite lower S/N compared to the two sources above, X-ray emission extends to at least 1\arcmin~(see Fig.~\ref{fig:radprofext}).

Since we did not characterize source extent in Paper~I, we present in Fig.~\ref{fig:pks2148} the radial profile of the very extended emission for PKS~2148$-$555, along with a radio map superposed on the DSS2 image.
As a result, in both our \sw~campaigns, described in \am~and in the current paper, we were able to characterize the extended nature of six sources. 
For comparison, from eROSITA~DR1, only B1526$-$423 and PKS~2148$-$555 are given as extended in the DR1 catalogs, and with a much smaller extent than found in our analysis.

Hardness ratio analysis (see Section~\ref{sec:3.2}) reveals that the two point-like sources, B0202$-$765 and B1017$-$426, as well as a few other sources with robust statistics such as B0427$-$366 and B1526$-$423, are characterized by soft X-ray emission.
The $HR$ value of one source, B0103$-$453, an extended source, instead suggests hard emission, but a higher S/N would be needed to strengthen conclusions for this source, as well as for the remaining sources.
Despite a count rate exceeding 10$^{-2}$~ct~s$^{-1}$ for five of our X-ray detections (B0202$-$765, B1017$-$426, B1526$-$423, B1817$-$640, and B2140$-$434), none of them is found in any catalog of hard X-ray sources.

We performed a spectral analysis (see Section~\ref{sec:3.2} for details) for three sources with adequate S/N, adopting a power law model for B0202$-$765 and B1817-640, and a thermal plasma emission model for the extended source B1526$-$423. 
For B0202$-$765, we obtained a photon index $\Gamma=1.96^{+0.19}_{-0.19}$ and an unabsorbed flux $S_{~0.3-10,unabs}=2.95^{+0.39}_{-0.36}\times10^{-12}$ erg~cm$^{-2}$~s$^{-1}$, with no evidence of intrinsic absorption. 
A column density in excess to the Galactic value was instead found for B1817$-$640, with $n_H=5.33^{+2.56}_{-2.00}\times10^{22}$ cm$^{-2}$.
For B1526$-$423, we found a plasma temperature $kT=10.7^{+5.6}_{-2.8}$~keV but, as we noted earlier (see Section~\ref{sec:3.2}), there may be some contamination from scattered photons from a central AGN.
The luminosity that we found for this source is $L_{~0.3-10}=2.62^{+0.15}_{-0.20}\times10^{45}$ erg~s$^{-1}$.

In late January 2024, data products from 1eRASS were published \citep{2024arXiv240117274M}, enabling a first comparison of our \sw~snapshots with results obtained by eROSITA. 
The eROSITA DR1 catalogs listed 10 of the 17 SMS4 sources of our sample, with nine from the \sw~detected sources and one (B1754$-$597) not detected by \sw.  
Also, the one source in our sample that was not observed with \sw, B1827$-$360, was not detected by eROSITA.

In Section~\ref{sec:4}, we matched the twelve X-ray detections (eleven from \sw~and one from DR1) with the AllWISE and the GSC~2.4.2 catalogs to search for infrared and optical counterparts.
As in \am, we required a detection in both the infrared and the optical bands to establish a counterpart at lower frequencies for our X-ray detections. 
As a result, we were able to establish a single candidate counterpart for eleven of the twelve  X-ray detections.
Only for the source detected by eROSITA did we find a second, possible (but less likely) candidate.
Comparing our results with the counterparts previously proposed by W20 and BH06, our analysis, relying on X-ray source positions, independently confirms the eight infrared counterparts provided by W20 and all twelve optical counterparts provided by BH06.

For all four sources for which W20 did not provide an infrared counterpart in their G4Jy sample, we identify an AllWISE source; the same infrared source was given by \cite{2023ApJS..265...32M} for B1017$-$426 and B1526$-$423, and in addition we provide infrared counterparts for B0103$-$453 and B1754$-$597. 
As regards the optical band, we provide a counterpart for these same two sources (B0103$-$453 and B1754$-$597) in addition to those reported by \cite{2023ApJS..265...32M} and \cite{2024ApJS..271....8G}, and also suggest a different optical counterpart for B1953$-$425 with respect to \cite{2023ApJS..265...32M}.

As reported in Section~\ref{sec:5}, among the 38 SMS4 sources still lacking pointed X-ray observations, 18 are already detected in the eROSITA DR1 catalogs, 16 are in the DE footprint but not yet detected, and four lie outside the DE footprint.

In conclusion, this second paper completes our analysis, from our two \sw~campaigns, of the \sw-XRT observations of bright radio sources in the southern hemisphere, comparable to the 3C sample in the north.
As described above, these include notably interesting sources - bright across the spectrum from radio to X-ray - that remain poorly studied but have promise for detailed investigations.


\begin{acknowledgments}
The authors thank the referee for a careful reading of the paper and for very helpful comments that improved the paper.
They also thank the \sw~PI, Brad Cenko, his deputies, and the Science Operations Team for performing the requested observations. 
W.F., C.J., and R.K. acknowledge support from the Smithsonian Institution and the Chandra High Resolution Camera Project through NASA contract NAS8-03060.
A.M. acknowledges financial support from the ASI-INAF agreement No. 2022-14-HH.0. 
This work has been partially supported by the ASI-INAF program I/004/11/4.
This research has made use of archival data, software or online services provided by the ASI Space Science Data Center (SSDC); the High Energy Astrophysics Science Archive Research Center (HEASARC) provided by NASA’s Goddard Space Flight Center; the SIMBAD database operated at CDS, Strasbourg, France; the NASA/IPAC Extragalactic Database (NED) operated by the Jet Propulsion Laboratory, California Institute of Technology, under contract with the National Aeronautics and Space Administration; the NASA/IPAC Infrared Science Archive, which is funded by the National Aeronautics and Space Administration and operated by the California Institute of Technology. 
\end{acknowledgments}

%

\vspace{5mm}
\facilities{\sw~(XRT), SkyView Virtual Observatory (https://skyview.gsfc.nasa.gov/current/cgi/query.pl), IRSA.} 





\bibliography{aas}{}
\bibliographystyle{aasjournal}



\end{document}